\RequirePackage{amsmath}
\documentclass[12pt,a4paper,final]{iopart}
\usepackage{iopams}  
\usepackage{graphicx}
\usepackage[breaklinks=true,colorlinks=true,linkcolor=blue,urlcolor=blue,citecolor=blue]{hyperref}

\usepackage[stable]{footmisc}
\usepackage{array}
\usepackage{verbatim}
\usepackage{hyperref}
\usepackage{color}
\usepackage{mathtools}

\usepackage{graphics}
\usepackage{epsfig}
\usepackage{mathrsfs,amsmath}
\usepackage{amsthm} % Theorem Formatting
\usepackage{amssymb}	% Math symbols such as \mathbb

\usepackage{graphicx}% Include figure files
\usepackage{dcolumn}% Align table columns on decimal point

\usepackage{subfig,enumerate}
\usepackage{kbordermatrix} % include package @ document preamble
\usepackage{multirow}
\usepackage{bigdelim}

\usepackage{blkarray}
\usepackage{multirow}
\usepackage{tikz,xstring,etoolbox,catchfile}
\usepackage{enumerate}
\usepackage{enumitem}

\newcommand{\red}{\color{red}}
\newcommand{\blu}{\color{blue}}

\newcommand{\bla}{\color{black}}
\newcommand{\green}{\color{blue!10!black!60!green}}

\newcommand{\brown}{\color{red!80!green!50!black!100!}}

\newcommand{\violet}{\color{violet}}

\newcommand{\defcol}{\color{blue!80!black}} 
\newcommand{\ampcol}{\color{red!80!black}} 
\newcommand{\concol}{\bla}
\newcommand{\errcol}{\bla}  

\newcommand{\Htogcol}{\color{black}}
\newcommand{\Htogacol}{\ampcol}
\newcommand{\Htogdcol}{\defcol}

\newcommand{\conxcol}{\red}
\newcommand{\conycol}{\blu}
\newcommand{\conzcol}{\green}

\newcommand{\Utot}{U\bla}

\newcommand{\Uc}{\concol U_c\bla}
\newcommand{\UcD}{\concol U_c^\dagger\bla}
\newcommand{\Uerr}{\errcol \tilde{U}\bla}

\newcommand{\Hc}{\concol {H}_c\bla}
\newcommand{\Herr}{\errcol {H}_0\bla}
\newcommand{\Htog}{\Htogcol\tilde{H}_0\bla}
\newcommand{\TimeOrderingOperator}{\mathcal{T}}

\newcommand{\Htoga}{\Htogacol\tilde{H}^{(\Omega)}_0\bla}
\newcommand{\Htogd}{\Htogdcol\tilde{H}^{(z)}_0\bla}
\newcommand{\Hd}{\defcol H_0^{(z)}\bla}
\newcommand{\Ha}{\ampcol H_0^{(\Omega)}\bla}

\newcommand{\EV}{\boldsymbol{a}}

\newcommand{\EVa}{\ampcol\boldsymbol{a}^{(\Omega)}_1\bla}
\newcommand{\EVd}{\defcol\boldsymbol{a}^{(z)}_1\bla}

\newcommand{\sigvec}{\boldsymbol{\sigma}}

\newcommand{\sigphiL}{\hat{\sigma}_{\phi_l}}

\newcommand{\Id}{\mathbf{I}}
\newcommand{\nvec}{\hat{\boldsymbol{n}}}

\newcommand{\Unitary}{\mathcal{U}}

\newcommand{\sigx}{\hat{\sigma}_x}
\newcommand{\sigy}{\hat{\sigma}_y}
\newcommand{\sigz}{\hat{\sigma}_z}

\newcommand{\sig}{\hat{\sigma}}

\newcommand{\HistMat}{\mathbf{\Lambda}^{(l-1)}}

\newcommand{\ProjVec}{\vec{\mathbf{\mathbb{T}}}^{(l)} }

\newcommand{\Ba}{\ampcol\beta_{_{\Omega}}\bla}
\newcommand{\Bd}{\defcol\beta_z\bla}
\newcommand{\Sa}{\ampcol S_{_{\Omega}}\bla}
\newcommand{\Sd}{\defcol S_z\bla}

\newcommand{\Ra}{\ampcol\mathbf{R}^{(\Omega)}\bla}
\newcommand{\Rd}{\defcol\mathbf{R}^{(z)}\bla}
\newcommand{\Fa}{\ampcol F_{_{\Omega}}\bla}
\newcommand{\Fd}{\defcol F_z\bla}

\newcommand{\RPlxx}{\conxcol R^{P_l}_{xx}\bla(\omega)}
\newcommand{\RPlxy}{\conxcol R^{P_l}_{xy}\bla(\omega)}
\newcommand{\RPlxz}{\conxcol R^{P_l}_{xz}\bla(\omega)}
\newcommand{\RPlyx}{\conycol R^{P_l}_{yx}\bla(\omega)}
\newcommand{\RPlyy}{\conycol R^{P_l}_{yy}\bla(\omega)}
\newcommand{\RPlyz}{\conycol R^{P_l}_{yz}\bla(\omega)}
\newcommand{\RPlzx}{\conzcol R^{P_l}_{zx}\bla(\omega)}
\newcommand{\RPlzy}{\conzcol R^{P_l}_{zy}\bla(\omega)}
\newcommand{\RPlzz}{\conzcol R^{P_l}_{zz}\bla(\omega)}

\newcommand{\VofL}{\violet V_l\bla(\omega)}
\newcommand{\BofL}{\brown B_l\bla(\omega)}

\newcommand{\CtrlSpace}{\mathfrak{C}}

\newcommand{\CPSMatElem}{\boldsymbol{\Gamma}}
\newcommand{\tauMIN}{\tau_{_\MinDim}}

\newcommand{\PAL}{\text{PAL}}

\newcommand{\Walsh}{w}

\newcommand{\MinDim}{\mathcal{M}}
\newcommand{\bP}{\boldsymbol{P}}
\newcommand{\Pk}{P^{(k)}}

\newcommand{\fampP}{q}
\newcommand{\fampH}{\tilde{\fampP}}

\newcommand{\WAMampP}{X}
\newcommand{\WAMampH}{\tilde{\WAMampP}}

\newcommand{\Rabivec}{\vec{\boldsymbol{\Omega}}}
\newcommand{\WAMFRabiPlus}{X_+}
\newcommand{\WAMFRabiMin}{X_-}

\newcommand{\WPMFPhasePlus}{Y_+}
\newcommand{\WPMFPhaseMin}{Y_-}

\newcommand{\WPMampP}{Y}
\newcommand{\WPMampH}{\tilde{\WPMampP}}
\newcommand{\RabiWPMF}{\Omega_0}

\newcommand{\tauSK}{\tau_{_\text{SK1}}}
\newcommand{\OmegaSK}{\Omega_{_\text{SK1}}}
\newcommand{\phiSK}{\phi_{_\text{SK1}}}

\newcommand{\RabiWSK}{\Omega_0}

\newcommand{\sinc}{\text{sinc}}
\newcommand{\FT}{\mathscr{F}}
\newcommand{\ket}[1]{\left| #1 \right>} % for Dirac bras
\renewcommand{\kbldelim}{[} % change default array delimiters to parentheses
\renewcommand{\kbrdelim}{]}
\usepackage{appendix}

\begin{document}

%__________________________________________________________________________________________

%							     TITLE/AUTHORS
%__________________________________________________________________________________________

%\author{}%
%\author{}%

%\author{}%
%\email{}

%\affiliation{}
%\affiliation{}
%\date{today}%
%\begin{center}

\title[Walsh-synthesized noise-filtering quantum logic]{Walsh-synthesized noise-filtering quantum logic}

\author{H. Ball$^{1,2}$, M.J. Biercuk$^{1,2}$}

\address{$^1$ARC Centre for Engineered Quantum Systems, School of Physics, The
University of Sydney, NSW 2006 Australia}
\address{$^2$National Measurement Institute, West Lindfield, NSW 2070 Australia}

\eads{\mailto{michael.biercuk@sydney.edu.au}}

%__________________________________________________________________________________________

%							     ABSTRACT 
%__________________________________________________________________________________________

\begin{abstract}

We study a novel class of open-loop control protocols constructed to perform arbitrary nontrivial single-qubit logic operations robust against time-dependent non-Markovian noise.  Amplitude and phase modulation protocols are crafted leveraging insights from functional synthesis and the basis set of Walsh functions.  We employ the experimentally validated generalized filter-transfer function formalism in order to find optimized control protocols for target operations in SU(2) by defining a cost function for the filter-transfer function to be minimized through the applied modulation.   Our work details the various techniques by which we define and then optimize the filter-synthesis process in the Walsh basis, including the definition of specific analytic design rules which serve to efficiently constrain the available synthesis space.  This approach yields modulated-gate constructions consisting of chains of discrete pulse-segments of arbitrary form, whose modulation envelopes possess intrinsic compatibility with digital logic and clocking.  We derive novel families of Walsh-modulated noise filters designed to suppress dephasing and coherent amplitude-damping noise, and describe how well-known sequences derived in NMR also fall within the Walsh-synthesis framework.  Finally, our work considers the effects of realistic experimental constraints such as limited modulation bandwidth on achievable filter performance.

\end{abstract}

\vspace{2pc}
\submitto{\NJP}
%\maketitle

%__________________________________________________________________________________________

%							TABLE OF CONTENTS
%__________________________________________________________________________________________

\tableofcontents\newpage
\pagestyle{plain}
\pagenumbering{arabic}

%__________________________________________________________________________________________

%							  SECTION: INTRODUCTION 
%__________________________________________________________________________________________

\section{Introduction}\label{Sec:Intro} 

In realistic laboratory settings, decoherence in quantum systems is dominated by \emph{time-dependent} non-Markovian noise processes with long correlations, frequently characterized by low-frequency dominated noise power spectra~\cite{Clarke2004, Faoro2004, Bylander2011, Harmon2007, BiercukQIC2009}. These may arise either from environmental fluctuations or - in the important case of \emph{driven} quantum systems - from noise in the control device itself~\cite{rutman1978}.  In either case, the result is a reduction in the fidelity of a target control operation, including both memory and nontrivial operations. These phenomena present a major challenge as quantum devices move from proof of principle demonstrations to realistic applications, where performance demands on the quantum devices are frequently extreme.  Accordingly, finding ways to control quantum systems efficiently and effectively in the presence of noise is a central task in quantum control theory~\cite{Tarn80,Tarn2003,James2007,James2009}.

A range of techniques relying on both open- and closed-loop control have been devised to address this challenge~\cite{Viola1999,Altafini2013, QECLidar2013} at various levels in a layered architecture for quantum computing~\cite{JonesPRX2012}. In particular, open-loop dynamical error suppression strategies (without the need for measurement or feedback) such as dynamic decoupling (DD)~\cite{BiercukNature2009, LongStorage, RBDD, DavidsonDD}, dynamically corrected gates (DCGs)~\cite{khodjasteh2009dcg, Khodjasteh2010, DasSarmaGate, Suter_Interleaved, HansonInterleaved, RBInterleaved, UhrigPRA2012}, and composite pulsing~\cite{Vandersypen2004,MerrillArXv2012,Kabytayev2014}, have emerged as \emph{resource-efficient} approaches for physical-layer decoherence control.  They are joined by a new class of continuously modulated (``always-on'') dynamical decoupling and dynamically protected gate schemes~\cite{ViolaEDD, JonesNJP2012,CaiNJP2012, FanchiniPRA2007, XuPRL2012, BermudezPRA2012,ChaudhryPRA2012, LemmerNJP2013} inspired by well established techniques in NMR.  

These schemes all address the question of decoherence mitigation, but looking across their breadth, have both benefitted and suffered from reliance on a wide range of theoretical techniques.  Unfortunately the analytic tools for crafting control protocols employed in any particular setting do not necessarily translate equivalently between approaches, nor do the methods generally employed for evaluating efficacy easily translate to experimentally measured characteristics of the environment.  This is a major challenge for experimentalists or systems designers attempting to determine which of the many open-loop control schemes to employ in a particular experiment.  As an example, the powerful group theoretic insights and consideration of time-varying environments that permit the construction of error-robust, bounded-strength SU(2) operations for quantum information in Viola's DCG framework are quite different from the geometric considerations and quasi-static noise assumptions widely employed in NMR composite pulsing.  This issue has been highlighted recently as new work has revealed striking differences between the time-domain noise sensitivity of control protocols as compared to longstanding notions of error cancellation in the Magnus expansion~\cite{SoareNatPhys2014, ViolaFFFArXve2014}.

A unified and experimentally relevant framework for \emph{devising} and \emph{evaluating} error-suppressing gates in realistic noise environments is therefore needed to secure the role of dynamical error suppression in systematic designs of quantum technologies including fault-tolerant quantum computers.  Kurizki provided a promising path towards this end with his seminal work framing the problem of finding decoherence-suppressing control protocols by considering appropriate frequency-domain modification of the system-environment coupling~\cite{KurizkiPRL2001, KurizkiPRL2004}.  Residual errors could be calculated through overlap integrals of the power spectrum describing the environmental noise, and functions capturing the frequency-domain response of any applied control.  This framework -- effectively a quantum generalization of transfer functions widely used in control engineering~\cite{TransferFunction} -- provides a simple heuristic approach to understanding the performance of an arbitrary control protocol in an arbitrary noise environment.   Stated simply, effective error-suppressing control protocols ``filter'' the noise over a user-defined band, therefore mitigating decoherence in the quantum system~\cite{BiercukJPB2011}.  

Early demonstrations of this framework applied to the simple case of implementing the protected identity operator to qubits by dynamical decoupling~\cite{Martinis2003, Kuopanportti2008, BiercukNature2009,UysPRL2009,UhrigPRL2007,CywinskiPRB2008}, where the filter functions could be calculated for pure dephasing in a straightforward manner using concepts of \emph{linear control}~\cite{BiercukJPB2011}.  Expanding significantly beyond this work, the challenge of crafting generalized analytic forms for the transfer functions describing arbitrary single-qubit control compatible with universal non-commuting noise (a problem in \emph{nonlinear} control) has recently been addressed theoretically~\cite{GreenPRL2012, GreenNJP2013, Kabytayev2014, ViolaFFFArXve2014}, and validated in experiment~\cite{SoareNatPhys2014}.  Further theoretical extensions of filter-transfer functions to two-qubit gates highlight the breadth of applicability of this approach to quantum control~\cite{HayesPRL2014, HayesPRA2011, GreenArXve2014}.

%has taken on a in importance as recent theoretical efforts have seen growth in the prominence of dynamical decoupling schemes involving continuously-modulated (``always-on'') control fields, moving away from the unphysical assumptions of instantaneous, unbounded pulsing in traditional DD. This includes schemes for extending single-qubit coherence times, improving performance of single-qubit~\cite{}, and two-qubit~\cite{} quantum logic operations in a variety of noise environments.

Beyond its simple intuitive nature, the power of the filter transfer function approach comes from the fact that it can in principle be applied to studying dynamic-error-suppression control protocols derived through any manner of analytic approach.   It permits the application of well tested engineering concepts for control systems design; the complex physics associated with quantum dynamics in time-dependent environments with non-commuting noise and control Hamiltonians is relegated to the calculation of the generalized filter transfer functions themselves, and once derived these may be deployed in block-diagram systems analyses~\cite{TransferFunction} .

With these significant advances and the promise of applying the suite of insights from control theory to the quantum regime, the noise-filtering approach to quantum control has leapt to the fore, providing a unifying framework applicable over a wide parameter range of interest to real experimental settings.   Nonetheless, outstanding challenges remain in how to leverage the generalized filter-transfer-function framework~\cite{GreenPRL2012, GreenNJP2013} to systematically craft effective error-suppressing gate constructions while also heeding realistic system constraints imposed by hardware systems.  For instance, the presence of finite timing precision and limited classical communication bandwidth between the physical (quantum) layer and a classical controller~\cite{JonesPRX2012} impose new constraints not generally captured when solely considering quantum dynamical evolution of an individual state. 

We address this challenge, introducing a quantum control toolkit permitting the realization of physical-layer error-suppressing control protocols that are simultaneously effective in suppressing error and compatible with a variety of major hardware restrictions.  We leverage the generalized filter-transfer function formalism as a unifying theoretical construct, and employ techniques from functional analysis in order to realize appropriate modulation protocols applied to a near-resonant carrier frequency for enacting high-fidelity quantum control operations on single qubits~\cite{OwrutskyPRA2012,JonesNJP2012}.  Our work identifies the Walsh functions -- square-wave analogues of the sines and cosines -- as natural building blocks for constructing the modulation protocols designed to filter time-varying noise over a user defined band while enacting a nontrivial qubit rotation. The Walsh functions are defined in a uniform piecewise-constant fashion, building intrinsic compatibility with discrete clocking~\cite{Hodgson2010} and classical digital logic, and have previously been identified as providing a powerful mathematical framework in the context of quantum control sequencing~\cite{HayesPRA2011}.  Moreover, they may be arbitrarily combined using Fourier-like synthesis using techniques for arbitrary waveform generation well established in digital signal processing.  

We treat a \emph{Walsh-modulated} driven qubit system weakly interacting with both dephasing and coherent amplitude-damping noise processes.  The task of finding Walsh-synthesized modulation patterns that produce effective filters is reduced to minimizing a cost function measuring the extent to which noise over a user-defined spectral band is filtered by the applied control. The performance of resulting control protocols is completely characterized by their \emph{Walsh spectra}, facilitating intuitive analytic design rules based on symmetry and spectral properties of the Walsh basis.   Our work details the various techniques and mathematical constructs through which we define and then optimize the filter-synthesis process in the Walsh basis, and considers the effects of realistic experimental constraints such as limited modulation bandwidth.  

With these insights, we derive novel families of Walsh-modulated noise filters designed to suppress dephasing and coherent amplitude-damping noise, and describe their properties.   Modulation protocols are tailored to a particular operation on SU(2), but are otherwise largely \emph{model-robust} (being tailored to suppress noise over a frequency band rather than to a specific time-domain noise signal), and \emph{portable} between different qubit technologies.  Combined with the discovery, presented here, that several prominent composite pulse protocols derived in NMR actually fall within the Walsh-synthesis basis -- mirroring similar insights in the context of dynamical decoupling~\cite{HayesPRA2011} -- this work positions the Walsh functions as a natural basis for crafting physical-layer error suppression strategies for scalable quantum technologies.

The remainder of this paper is organized as follows. In Sec. \ref{Sec:PhysicalSetting} we describe our model quantum system by defining relevant control and noise Hamiltonians. In Sec. \ref{Sec:BuildingNoiseFilters}  we review the generalized filter-transfer function formalism used to derive a spectral representation of the operational infidelity. Notation for defining and parameterizing the control space is introduced and explicit expressions for computing corresponding filter functions are presented. Sec. \ref{Sec:CharacteristicsOfNoiseFilters} provides a formal definition of a filter cost function used for optimizing operational fidelity over the control space and deriving useful filters. Performance characteristics of these filters are discussed and interpreted, with care taken to differentiate filter order from Magnus order. In Sec. \ref{Sec:FilterDesignByWalshSynthesis} physically motivated constraints on the control space are established by synthesizing control waveforms as superpositions of functions in the Walsh basis, bounding the dimensionality of the filter-optimization task.  Two useful representations of the Walsh basis -- Paley ordering and the Hadamard representation -- are introduced. We then develop a range of analytic filter-design rules for efficient filter construction based on the symmetry and spectral properties of the Walsh functions.  In Secs. \ref{Sec:WAMFs} -  \ref{Sec:UWMFs} we apply the above framework to derive several novel families of noise filters implementing nontrivial logic gates. These include filters for dephasing and coherent amplitude-damping noise in addition to concatenated filters for universal noise. In Sec. \ref{Sec:EffectOfBandwidthLimitsOnWalshFilters} we study how relaxing the assumption of perfect square pulses reduces the performance of filters optimized in the Walsh basis, and demonstrate that these filter properties may be recovered in general by simply re-optimizing under the assumption of non-square pulses. We then close with a brief summary and outlook, followed by a number of appendices containing detailed derivations of relevant quantities used in the main text.

\section{Physical Setting}\label{Sec:PhysicalSetting}%\label{Sec:ControlSetting NoiseModel and FFFormalism}

We begin by establishing the Hamiltonian framework for the control and noise interactions treated in this paper. This is necessary background in order to study noise filtering via Walsh-synthesized control fields implementing logic gates. We consider a model quantum system consisting of an ensemble of identically prepared noninteracting qubits immersed in a weakly interacting noise bath and driven by an external control device. Working in the interaction picture with respect to the qubit splitting, state transformations are represented as unitary rotations of the Bloch vector. In this interaction picture the generalized time-dependent Hamiltonian is written
\begin{align}\label{TotalHamiltonian}
H(t)=\Hc(t)+ \Herr(t)
\end{align}
where $\Hc(t)$ describes perfect control of the qubit state, e.g. via an ideal external driving field, and the noise Hamiltonian $\Herr(t)$ captures undesirable interactions with a time-varying non-Markovian noise environment. The full qubit dynamics are governed by the Schrodinger equation $i\dot{U}(t,0) = H(t)U(t,0)$ where the time-evolution operator $U(t,0)$ transforms an initial state $\ket{\psi(0)}$ to the final $U(\tau,0)\ket{\psi(0)}$ after an interaction of duration $\tau$.

In the absence of noise the total Hamiltonian reduces to $H(t) = \Hc(t)$, in which case time-evolution is determined purely by control operations according to $i\dot{U}_c(t,0) = H_c(t)U_c(t,0)$. An \emph{intended} evolution path under ideal control is therefore described by the \emph{control propagator} $\Uc(t,0) = \TimeOrderingOperator\exp\big(-i\int_0^t\Hc(t')dt'\big)$, with $\TimeOrderingOperator$ denoting the time-ordering operator. For a single qubit the time-dependent control Hamiltonian may in general be written $\Hc(t) = \Omega(t)\nvec(t)\cdot\sigvec/2$. Here $\nvec(t)\cdot\sigvec\equiv n_x\sigx+n_y\sigy+n_z\sigz$ is the rotation generator, $\nvec(t)\in\mathbb{R}$ is a unit vector defining the instantaneous axis of rotation, and $\Omega(t)$ is the instantaneous rate of rotation (Rabi rate) for the Bloch vector. 
%Control Schrodinger
%\begin{align}\label{ControlSchrodinger}
%\color{red}
%i\frac{d}{dt}U_c(t,0) = H_c(t)U_c(t,0)
%\end{align}
%\green Representing the qubit state on the Bloch sphere, state manipulation maps to a rotation of the Bloch vector in $\mathbb{R}^3$ associated with the unitary operator $U(\theta, \sig_{\rvec}): = \exp\big(\frac{-i\sigvec\cdot\rvec\theta}{2}\big)$, reflecting the homeomorphism between $SU(2)$ and $SO(3)$. In effect, the spin operator $\sig_{\rvec} := \rvec\cdot\sigvec$ generates a rotation though an angle $\theta$ about an axis defined by the unit vector $\rvec\in\mathbb{R}^3$. 
%				   	---------------------------------
%					    sub SECTION: Noise Interaction Model
%				   	---------------------------------
%\subsection{Noise Interaction Model}\label{subSec:NoiseInteractionModel}

Environmental interactions are modeled semi-classically, with stochastic noise processes expressed in terms of time-dependent fluctuating classical noise fields.  We consider time-dependent dephasing (detuning) and coherent amplitude-damping processes, captured respectively through (stochastic) rotations about $\sig_{z}$ and about the instantaneous direction of control $\nvec(t)\cdot\sigvec$. The universal noise Hamiltonian thus takes the form $\Herr(t) = \Hd(t) + \Ha(t)$ where $\Hd(t)$ and $\Ha(t)$ denote noise interactions in the dephasing and amplitude noise quadratures respectively. 
%Hamiltonian, and rotations generated by  the instantaneous control Hamiltonian $\Hc(t)$
%the instantaneous spin operator $\sig_\phi:=\cos(\phi)\sig_x+\sin(\phi)\sig_y$. 
%\ampcol
Dephasing noise thus contributes the additive term  
\begin{align}\label{DephasingNoiseHamiltonian}
\Hd(t) = \Bd(t)\sig_z
\end{align}
where $\Bd(t)$ describes a time-varying noise field. Coherent amplitude-damping noise contributes the multuplicative term
\begin{align}\label{AmplitudeNoiseHamiltonian}
\Ha(t) &= \frac{\Ba(t)\Omega(t)}{2}\nvec(t)\cdot\sigvec = \Ba(t)\Hc(t).
\end{align}
Including this term is equivalent to making the substitution $\Omega(t)\longrightarrow\Omega(t)(1+\Ba(t))$ in the control Hamiltonian, where $\Ba(t)$ describes a (multiplicative) noise source in the amplitude of the driving field. Inclusion of this term in the noise Hamiltonian enables us to go beyond previous studies where attention has been restricted to dephasing processes. This novel approach is important for most realistic experimental situations where correctable non-Markovian amplitude-damping errors arise from noise in the control system itself (for example, fluctuations in the strength of the driving field).

%where $\Bd(t)$ describes a time-varying noise field. During each pulse we also make the substitution $\Omega_l\longrightarrow\Omega_l(1+\Ba(t))$ where $\Ba(t)$ describes a (multiplicative) noise source in the amplitude of the driving field. Thus the amplitude noise Hamiltonian takes the form XXX
%\begin{align}\label{AmplitudeNoiseHamiltonian}
%\Ha &= \Ba(t)\sum_{l=1}^{n} G^{(l)}(t) \frac{\Omega_l}{2} \sigphiL
%\end{align}

%\noindent and generates errors in the intended rotation angle coaxial with the target rotation axis $\sigphiL$, corresponding to a coherent amplitude-damping process. 

In our model both  $\Bd(t)$ and  $\Ba(t)$ are assumed to be classical random variables with zero mean and non-Markvovian power spectra. We also assume they are \emph{wide sense stationary} and \emph{independent}\footnote{The assumption of independence is reasonable, for instance, in the case of a driving field where random fluctuations in frequency and amplitude arise from different physical processes. A general model including correlations between noise processes is possible, however, following the approach outlined by \emph{Green et al.}~\cite{GreenNJP2013}.}. The former implies the autocorrelation functions $\langle\beta_i(t_1)\beta_i(t_2)\rangle$,  $i\in\{z,\Omega\}$, depend only on the time \emph{difference} $t_1-t_2$. The latter implies the cross-correlation functions vanish. That is, $\langle\beta_i(t_1)\beta_j(t_2)\rangle = 0$ where $i,j\in\{z,\Omega|i\ne j\}$. The angle brackets denote a time average of the random variables.  Finally, our model permits access to a wide range of parameter regimes, from quasistatic (noise slow compared to $\Hc(t)$) to the limit in which the noise fluctuates on timescales comparable to or faster than $\Hc(t)$.

These noise Hamiltonians generate uncontrolled rotations in the qubit dynamics, leading to errors in the evolution path (and hence the final state) relative to the target transformation intended under $\Hc(t)$. An estimate for this error is derived in the next section using the generalized filter-transfer function formalism.

%__________________________________________________________________________________________

%		      SECTION: Calculating operational fidelity & the first-order approximation
%__________________________________________________________________________________________

%				-------------------------------------------------
%				sub:SECTION: Calculating operational fidelity & the first-order approximation
%				-------------------------------------------------

\section{Building Noise Filters}\label{Sec:BuildingNoiseFilters}
Overall, our objective is to craft control protocols such that the deleterious effects of time-dependent noise on the intended evolution of an arbitrary qubit transformation are suppressed -- \emph{filtered} by the control.  Accordingly, we require a measure for the operational fidelity in the presence of both noise and the relevant control. For this we employ the method developed by \emph{Green et al.}\cite{GreenNJP2013}. In this framework the error contributed by the noise fields over the duration of the control is approximated, to first order, via a truncated Magnus expansion. Each noise field then contributes a term to the gate infidelity in the spectral domain expressed as an overlap integral between the noise power spectrum and an appropriate generalized filter-transfer function. We we describe this in detail below. 

\subsection{Calculating Operational Fidelity}
In the absence of noise interactions, state evolution is determined by $i\dot{U}_c(t)=\Hc(t)U_c(t)$ with $U_c(t)$ the ideal evolution operator describing the target operation. Including the effects of noise, however, time evolution is determined by the operator $U(t)$ satisfying $i\dot{U}(t) = (\Hc(t)+\Hd(t)+\Ha(t))U(t)$. Our measure for operational fidelity is given by $\mathcal{F}_{av}(\tau) = \frac{1}{4}\langle |\text{Tr}(\UcD(\tau)U(\tau))|^2\rangle$, effectively measuring the extent to which the intended and realized operators ```overlap'', as captured by the Hilbert-Schmidt inner product ~\cite{Schumacher1996}. Computing the evolution dynamics, however, is very challenging since the control and noise Hamiltonians do not commute at different times; sequential application of the resulting time-dependent, non-commuting operations gives rise to both dephasing and depolarization errors, mandating approximation methods. 

Our error model assumes non-dissipative qubit evolution with both control and noise interactions resulting in unitary rotations. Hence we approximate the evolution operator as a unitary using a Magnus expansion~\cite{Blanes2009,Magnus1954}. This involves moving to a frame co-rotating with the control known as the \emph{toggling frame}, originally appearing in the development of average Hamiltonian theory ~\cite{Waugh1968}. This approach allows us to separate the part of the system evolution due solely to the control from the part affected by environmental coupling, and is standard in the study of coherent control in NMR~\cite{Waugh1968, Ernst1987} and quantum information.%~\cite{}.\bla This procedure is the basis of the generalized filter-transfer function formalism~\cite{GreenNJP2013}. 

Defining the \emph{error propagator} $\Uerr(t)\equiv\Uc^\dagger(t)\Utot(t)$, the total evolution operator is written $\Utot(t) = \Uc(t)\Uerr(t)$. In this case the realized evolution operator approaches the target operation as $\Uerr(\tau) \rightarrow \Id$, establishng the condition for suppression of noisy evolution dynamics. However, moving to the toggling frame defined by \emph{toggling frame Hamiltonian} $\Htog(t) \equiv\UcD(t)\Herr(t)\Uc(t)$, the error propagator satisfies the Schrodinger equation $i\dot{\Uerr}(t) = \Htog(t)\Uerr(t)$. Performing a Magnus expansion in this frame -- assuming convergence of the series~\cite{Magnus1954} -- we may write $\Uerr(\tau) = \exp\big[-i\sum_{\mu = 1}^\infty\EV_\mu(\tau)\cdot\sigvec\big]$ where the \emph{error vectors} $\EV_\mu(\tau)$ determine expansion coefficients of the Magnus series operators $\Phi_\mu(\tau)$ expressed in the basis of Pauli matrices (see \ref{App:DetailedFFDerivation}). We may then in principle approximate $\tilde{U}(t)$ to arbitrary accuracy by truncating the infinite series at an appropriate order.

%				-------------------------------------------------
%						subsub:SECTION: first-order Infidelity 
%				-------------------------------------------------
%\subsection{First-Order Infidelity $\langle a_1^2\rangle$}\label{SubSec:FirstOrderInfidelity} 
The operational fidelity $\mathcal{F}_{av}(\tau)=\frac{1}{4}\langle |\text{Tr}(\Uerr(\tau))|^2\rangle$ may now be fully expressed as an infinite power series over the ensemble-averaged magnitudes of the expansion vectors $\EV_\mu(\tau)$. In the limit of sufficiently weak noise\footnote{The first-order approximation has recently been experimentally tested and demonstrated to produce good agreement in the weak noise limit~\cite{SoareNatPhys2014}. For the noise field $\beta(t)$, this regime is sufficiently characterized by requiring $\xi^2\ll 1$, where the \emph{smallness parameter} is defined by $\xi^2\equiv\langle\beta^2(t)\rangle\tau^2\equiv \tau^2\int_{-\infty}^{+\infty}d\omega S_\beta(\omega)$~\cite{GreenPRL2012}. The condition $\xi^2<1$ is also required for the Magnus series to formally converge.}, however, it is appropriate to truncate the expansion to first-order yielding $\mathcal{F}_{av}(\tau) \approx 1 - \langle\normalfont{a}^2_1\rangle$ with $\langle\normalfont{a}^2_1\rangle\equiv\langle\EV_1(\tau)\EV_1^T(\tau)\rangle$ defining the \emph{first order infidelity}. Now, as set out in \ref{App:DetailedFFDerivation} the first-order error vector is related to the first-order Magnus term according to Eq. \ref{Magnus1}, yielding $\EV_1(\tau)\cdot\sigvec = \Phi_1(\tau) = \int_0^\tau dt \Htog(t)$. That is, the first-order infidelity $\langle\normalfont{a}^2_1\rangle$ is associated with the time-average of the toggling frame Hamiltonian over the total sequence duration. 

Expressing $\Htog(t)\equiv\vec{\mathcal{R}}(t)\cdot\sigvec$ in the Pauli basis, where the expansion vector $\vec{\mathcal{R}}(t)$ is some convolution of both control and noise fields, we obtain the computational expression $\EV_1(\tau) = \int_0^\tau dt\vec{\mathcal{R}}(t)$. Using the noise model assumptions outlined in Sec. \ref{Sec:PhysicalSetting}, and performing a number of Fourier-like transforms (see \ref{App:DetailedFFDerivation} for full details), we obtain a spectral representation of the form 
\begin{align}\label{FirstOrderInfidelityComputational}
\langle a_1^2\rangle &=
\frac{1}{2\pi}
\int_{-\infty}^{\infty}
\frac{d\omega}{\omega^2}
\Sd(\omega)
\Fd(\omega)+
\frac{1}{2\pi}
\int_{-\infty}^{\infty}
\frac{d\omega'}{\omega'^2}
\Sa(\omega') 
\Fa(\omega').
\end{align}
Here $\Sd(\omega)$ and $\Sa(\omega)$ denote the dephasing and amplitude noise power-spectral densities (PSDs). The dephasing $\Fd(\omega)$ and amplitude $\Fa(\omega)$ filter-transfer functions, on the other hand, capture the spectral response of the control sequence.   Moving forward, we will present the mathematical framework that permits calculation of these quantities for arbitrary control protocols.

\subsection{Defining the Control Space}\label{SubSec:ControlHamiltonian}
In order to realize specific noise filters, characterized by the filter-transfer functions introduced above, we require a simple framework to define the time-domain control operations that can be applied to the qubit. Representing the qubit state on the Bloch sphere, state manipulation maps to a rotation in $\mathbb{R}^3$ of the Bloch vector associated with a unitary transformation 
$\Unitary(\theta, \sig_{\nvec})\equiv \exp\left[-i\left(\sigvec\cdot\nvec\right)\theta/2\right]$, 
reflecting the homeomorphism between $SU(2)$ and $SO(3)$. The rotation generator $\sig_{\nvec} \equiv \nvec\cdot\sigvec\equiv n_x\sigx+n_y\sigy+n_z\sigz$ produces a rotation though an angle $\theta$ about the axis defined by the unit vector $\nvec\in\mathbb{R}^3$. 

\begin{figure}
\centering
\includegraphics[width=0.6\columnwidth]{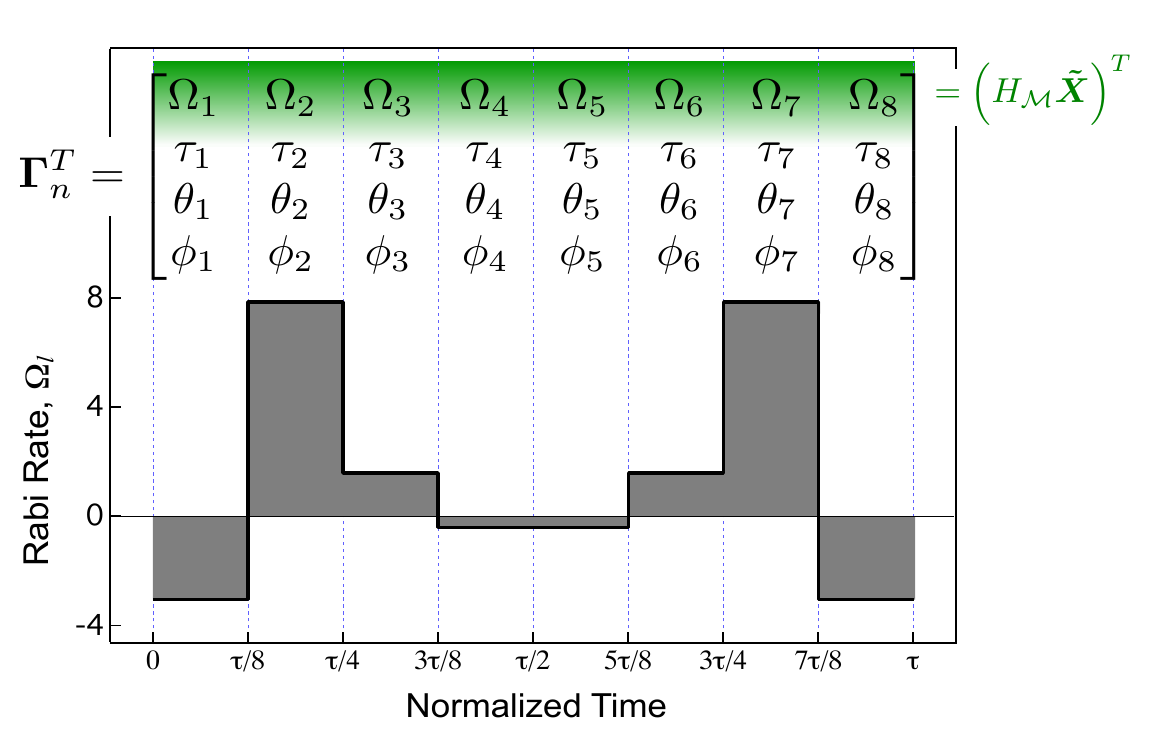}
\caption{Visualization of the available control space for an $n=8$ segment control sequence.  The filter is synthesized over the parameters presented in $\CPSMatElem_n$, whose transpose corresponds to the discrete time segments in the time-domain filter.  As an illustration, a time-varying Rabi rate (arbitrary units) is presented for each of the $l$ segments.  Synthesis of this waveform may be constructed in the Walsh basis using the Hadamard transform (notation upper right), as will be discussed in Sec.~\ref{Sec:FilterDesignByWalshSynthesis}.}  \label{Fig:FilterConstruction}
\end{figure}

We treat control protocols taking the form of an \emph{$n$-segment sequence} of such unitaries, executed over the time period $[0, \tau]$. This implies a natural partition of the total sequence duration  $\tau$ into $n$ subintervals $I_l = [t_{l-1},t_l]$, $l\in\{1,...,n\}$, such that the $l$th control unitary has duration $\tau_l = t_l-t_{l-1}$. Here $t_{l-1}$ and $t_l$ are the start and end times of the $l$th rotation respectively, and we define $t_0 \equiv 0$ and $t_n \equiv \tau$. In particular we consider control unitaries of the form 
\begin{align}
\label{PrimitivePulseForm}P_l &\equiv \Unitary(\Omega_l\tau_l,\sigphiL) = \exp\Big[-i\frac{\Omega_l}{2}\tau_l\sigphiL\Big]\\
\label{SpinOperatorDefinition}\sigphiL&\equiv \cos(\phi_l)\sigx+\sin(\phi_l)\sigy,
\end{align}
corresponding to the experimentally relevant case of a resonantly driven qubit. Here $\Omega_l$ is the Rabi rate during the $l$th time interval $[t_{l-1},t_l]$, and is assumed constant over the duration $\tau_l$ of the associated control interaction. During this interaction the rotation generator $\sigphiL$, parameterized by $\phi_l\in[0,2\pi]$, thus generates a rotation of the Bloch vector through an angle $\theta_l \equiv \Omega_l\tau_l$ about the axis $\nvec_l \equiv \left(\cos(\phi_l),\hspace{0.1cm}\sin(\phi_l),\hspace{0.1cm}0\right)$ in the $xy$ plane\footnote{For a resonantly driven qubit $\phi_l$ is the phase of the driving field and $\Omega_l$ is linearly proportional to the driving amplitude. }. The control Hamiltonian associated with this $n$-segment sequence takes the form  
\begin{align}\label{ControlHamiltonian}
H_c(t) = \sum_{l=1}^{n} G^{(l)}(t) 
\frac{\Omega_l}{2} 
\sigphiL
\end{align}
where the function $G^{(l)}(t)$ is 1 if $t\in I_l$ and zero otherwise. Controlled evolution \emph{during} the $l$th time interval is, under this Hamiltonian, consequently described by the unitary
\begin{align}\label{LocalUnitaryRotation}
\Uc(t,t_{l-1}) =\exp\Big[-i\frac{\Omega_l}{2}\sigphiL(t-t_{l-1})\Big].
\end{align}
That is, implementation of the $l$th completed rotation is equivalently denoted by the operator $P_l = \Uc(t_l,t_{l-1})$. For compactness we define the \emph{cumulative operator} \begin{align}
Q_l := P_lP_{l-1}...P_0,\hspace{1cm}P_0 := \Id
\end{align}
to capture the cumulative action of the first $l$ sequentially competed rotations. Hence the control propagator at any time $t$ may be written  
\begin{align}\label{ControlPropagatorComputational}
\Uc(t,0) = \sum_{l=1}^n G^{(l)}(t)U_c(t,t_{l-1})Q_{l-1}.
\end{align} 
$\Hc(t)$ is thus completely described by the sequence of $n$ triples $\{(\Omega_l,\tau_l,\phi_l)\}_{l=1}^n$, and each control operation is completely parameterized by the control variables according to $P_l = P_l(\theta_l,\Omega_l,\tau_l,\phi_l)$. Although not strictly an independent parameter it is useful to include $\theta_l = \Omega_l\tau_l$ in the argument to distinguish different \emph{realizations} of the same net rotation for different choices of $\Omega_l\tau_l$. We define the $\left(n\times4\right)$ $n$-segment matrix   
\begin{align}\label{TemplateCPSMatrix}
\CPSMatElem_n\hspace{0.25cm}=\hspace{0.25cm}
\begingroup
  \renewcommand*{\arraystretch}{1.15}%
  \kbordermatrix{
           & \Omega_l           & \tau_l           & \theta_l           & \phi_l     \cr
    P_1    & \Omega_1             & \tau_1       & \theta_1       & \phi_1 \cr
    P_2    & \Omega_2             & \tau_2       & \theta_2       & \phi_2 \cr
    \vdots & \vdots        & \vdots        & \vdots        &  \vdots  \cr
    P_n    & \Omega_n             & \tau_n       & \theta_n       & \phi_n \cr
  }%
\endgroup
\end{align}
to compactly describe any arbitrary $n$-segment unitary control sequence (see Fig.~\ref{Fig:FilterConstruction}). The entire space $\CtrlSpace_n$ of such control forms, referred to the \emph{$n$-segment control space}, and written formally
\begin{align}
\CtrlSpace_n:= \Big\{\CPSMatElem_n\big|\Omega_l,\theta_l,\tau_l>0,\hspace{0.2cm} \phi_l\in[0,2\pi],\hspace{0.2cm}l\in1,...,n,\hspace{0.2cm} \sum_l^n\tau_l = \tau\Big\}
\end{align}
thus corresponds to an infinite set of $\CPSMatElem_n$ matrices ranging continously over all possible values taken by the control variables. This general class of control, consisting of bounded-strength unitary sequences, includes familiar \emph{composite-pulse sequences} in NMR and DCGs in quantum information. We use the more general control space, however, to construct novel qubit gates specifically designed to filter non-Markovian noise.

%				-------------------------------------------------
%				     sub:SECTION: filter-transfer functions 
%				-------------------------------------------------
\subsection{Generalized filter-transfer Functions}\label{subSec:FilterFunctions:ComputationalDefinitions} 

We now present the computational forms of the filter-transfer functions $\Fd(\omega)$ and $\Fa(\omega)$ introduced in Eq. \ref{FirstOrderInfidelityComputational} for arbitrary $n$-segment control protocols implemented by Eq.~\ref{ControlHamiltonian}. As outlined above, the filter-transfer functions are completely parametrized by the control variables $\{(\Omega_l,\tau_l,\phi_l)\}_{l=1}^n\cong\CPSMatElem_n$. Here we only provide a summary of the relevant computational quantities, leaving the major derivations and full explanation to \ref{App:DetailedFFDerivation}. We start by writing 
\begin{align}
\label{DephasingFF}
\Fd(\omega) &:= \Big[
\Rd(\omega)
\Big]^*
\Big[
\Rd(\omega)
\Big]^T\hspace*{1cm}&&\text{(\emph{Dephasing Filter-Transfer Function})}\\
\label{AmplitudeFF}\Fa(\omega) &:=\Big[
\Ra(\omega)t
\Big]^*
\Big[
\Ra(\omega)
\Big]^T\hspace*{1.0cm}&&\text{(\emph{Amplitude Filter-Transfer Function})}
\end{align}
where the row vectors $\Rd(\omega)$, $\Ra(\omega)\in\mathbb{R}^3$ are obtained by Fourier transforming relevant time-domain functions associated with the control evolution dynamics. In \ref{App:CVsComputationalForms} we derive the explicit computational forms
\begin{align}\label{DephasingControlVector:ComputationalForm}
\Rd(\omega)&=\sum_{l=1}^{n}e^{i\omega t_{l-1}}\green\mathbf{R}^{P_l}_z\bla(\omega) \mathbf{\Lambda}^{(l-1)},\hspace{1cm}&\text{(\emph{Dephasing Control Vector})}\\
\label{AmplitudeControlVector:ComputationalForm}
\Ra(\omega) &=
\sum_{l=1}^n
\Big[
e^{i\omega t_{l-1}} - e^{i\omega t_l}
\Big]\ProjVec\HistMat\hspace{1cm}&\text{(\emph{Amplitude Control Vector})}
\end{align}
The row vector $\green\mathbf{R}^{P_l}_z\bla(\omega)\in\mathbb{R}^3$ captures the spectral response in the dephasing noise quadrature contributed during the $l$th unitary control segment. This takes the form (see \ref{LocalControlMatrixElements})

%\begin{align}\label{Eq:LocalDephasingControlVector}
%\green\mathbf{R}^{P_l}_z\bla(\omega) = \left(\begin{array}{ccc}\RPlzx, & \RPlzy, & \RPlzz\end{array}\right) 
%\end{align}
%where
%\begin{align}
%%EXPRESSION FOR R_zx(\omega)
%\label{RzPlx(omega)}
%&\RPlzx=\frac{\omega}{\omega^2-\Omega_l^2}
%\sin(\phi_l)
%\Big[\brown
%i\Omega_le^{i\omega\tau_l}\cos(\Omega_l\tau_l)
%+\omega e^{i\omega\tau_l}\sin(\Omega_l\tau_l) 
%-i\Omega_l
%\bla\Big]\\
%%EXPRESSION FOR R_zy(\omega)
%\label{RzPly(omega)}
%&\RPlzy =-\frac{\omega}{\omega^2-\Omega_l^2}
%\cos(\phi_l)
%\Big[\brown
%i\Omega_le^{i\omega\tau_l}\cos(\Omega_l\tau_l)
%+\omega e^{i\omega\tau_l}\sin(\Omega_l\tau_l) 
%-i\Omega_l
%\bla\Big]\\
%%EXPRESSION FOR R_zz(\omega)
%\label{RzPlz(omega)}
%&\RPlzz=\frac{\omega}{\omega^2-\Omega_l^2}
%\Big[\violet
%i\Omega_le^{i\omega\tau_l}
%\sin(\Omega_l\tau_l)
%-\omega e^{i\omega\tau_l}\cos(\Omega_l\tau_l)
%+\omega\bla
%\Big]
%\end{align}
\begin{align}\label{Eq:LocalDephasingControlVector}
\green\mathbf{R}^{P_l}_z\bla(\omega) = \frac{\omega}{\omega^2-\Omega_l^2}
\left(\begin{array}{c}
\sin(\phi_l)
\Big[\brown
i\Omega_le^{i\omega\tau_l}\cos(\Omega_l\tau_l)
+\omega e^{i\omega\tau_l}\sin(\Omega_l\tau_l) 
-i\Omega_l
\bla\Big]\\ 
-\cos(\phi_l)
\Big[\brown
i\Omega_le^{i\omega\tau_l}\cos(\Omega_l\tau_l)
+\omega e^{i\omega\tau_l}\sin(\Omega_l\tau_l) 
-i\Omega_l
\bla\Big]\\
\violet
i\Omega_le^{i\omega\tau_l}
\sin(\Omega_l\tau_l)
-\omega e^{i\omega\tau_l}\cos(\Omega_l\tau_l)
+\omega\bla
\end{array}\right)^T. 
\end{align}
%\begin{align}
%%EXPRESSION FOR R_zx(\omega)
%\label{RzPlx(omega)}
%&\RPlzx=\frac{\omega}{\omega^2-\Omega_l^2}\sin(\phi_l)
%\BofL\\
%%EXPRESSION FOR R_zy(\omega)
%\label{RzPly(omega)}
%&\RPlzy =-\frac{\omega}{\omega^2-\Omega_l^2}\cos(\phi_l)
%\BofL\\
%%EXPRESSION FOR R_zz(\omega)
%\label{RzPlz(omega)}
%&\RPlzz=\frac{\omega}{\omega^2-\Omega_l^2}
%\VofL
%\end{align}
%and we have defined
%\begin{align}
%\VofL &:= 
%\Big[\violet
%i\Omega_le^{i\omega\tau_l}
%\sin(\Omega_l\tau_l)
%-\omega e^{i\omega\tau_l}\cos(\Omega_l\tau_l)
%+\omega\bla
%\Big],\\
%\BofL&:=
%\Big[\brown
%i\Omega_le^{i\omega\tau_l}\cos(\Omega_l\tau_l)
%+\omega e^{i\omega\tau_l}\sin(\Omega_l\tau_l) 
%-i\Omega_l
%\bla\Big]
%\end{align}
We also define the \emph{lth-Segment Projection Vector} 
\begin{align}\label{ProjectionVector}
\ProjVec :=\frac{\Omega_l}{2}\big(\cos(\phi_l),\hspace{0.1cm}\sin(\phi_l),\hspace{0.1cm}0\big)
\end{align} 
to compactly express the control variables - namely Rabi rate and the rotation-axis vector, projected onto the $xy$ plane of the Bloch sphere - associated with evolution during the $l$th unitary. In fact, inspection of Eqs. \ref{DephasingControlVector:ComputationalForm} and \ref{AmplitudeControlVector:ComputationalForm} reveals that $\ProjVec$ is the computational analogue of $\mathbf{R}^{P_l}_z$ for the amplitude noise quadrature. The simpler dependence of $\ProjVec$ on the control variables, however, reflects the fact that amplitude noise in our model is always coaxial, and hence commutes with, the control. 

On the other hand, the $3\times3$ \emph{Control History Matrix} $\mathbf{\Lambda}^{(l-1)}$, defined by 
%\begin{eqnarray}
\begin{align}
\Lambda_{ij}^{(l-1)} = \frac{1}{2}\text{Tr}\Big[Q^\dagger_{l-1}\sig_iQ_{l-1}\sig_j\Big],%\hspace{1cm}\text{(\emph{Control History Matrix})}
\end{align}
%\end{eqnarray}
is the result of expanding the operator $Q^\dagger_{l-1}\sig_iQ_{l-1}\equiv\sum_j\Lambda_{ij}^{(l-1)}\sig_j$, with $i,j \in\{x,y,z\}$, in the Pauli basis, and identifying the coefficients $\Lambda_{ij}^{(l-1)}$  as the matrix elements. $\mathbf{\Lambda}^{(l-1)}$ thereby captures the accumulated effect of the previous $l-1$ \emph{completed} unitaries, implemented via the cumulative operator $Q_{l-1}$.

%In order to bound the complexity of this general task, however, we advance a method of filters construction in which the relevant control fields are synthesized using an existing set of well-characterized basis functions. Specifically, we employ the basis set of Walsh functions. In the next section we review the Walsh basis, providing the necessary background for Sec. \ref{Sec:FilterDesignByWalshSynthesis} where our method of filter synthesis is fully developed. 

%____________________%____________________%____________________%____________________%

%				SECTION 5: 	FILTER DESIGN BY WALSH SYNTHESIS

%General Filter Design 4   (tick)
%AMPMF Modulation 5.0  (tick)
%General Walsh Synthesis 5.1 (tick)
%WAM/WPM Synthesized Filters 5.2 (tick)
%Walsh Filter Design Rules 5.3

%____________________%____________________%____________________%____________________%

\section{Characteristics of Noise Filters}\label{Sec:CharacteristicsOfNoiseFilters}
The power of the noise filtering formalism lies in the simple interpretation of the filter-transfer functions $F_{i}(\omega)$, which may be characterized in a standard engineering approach, considering passbands, stopbands, and filter order~\cite{BiercukJPB2011,SuterPRA2011, GreenPRL2012, GreenNJP2013}. In particular, error suppression corresponds to minimizing $F_{i}(\omega)$, $i\in\{z,\Omega\}$ in the spectral region where the corresponding PSDs are non-negligible. This can, in principle, be achieved by judicious construction of the control sequence since the filter-transfer functions are completely parametrized in variables describing the time-domain control applied to the qubit.

We are now in a position to examine the characteristics of the filter-transfer functions for an arbitrary control sequence $\CPSMatElem_n$, formally indicating the functional dependence of the filter-transfer functions on the control variables by writing $F_i(\omega\tau) = F_i(\omega\tau;\CPSMatElem_n)$, $i\in\{z,\Omega\}$. Inversely, we may commence a study of filter design based on constructing control sequences satisfying some \emph{desired} filter property - our main goal.  We now advance the main mathematical framework used in this paper to study filter design, pulling together the ideas introduced in the previous sections.  

%							----------------------------------------------------------------
%							     	   sub:SECTION: General Filter Design
%							----------------------------------------------------------------

\subsection{The Filter Cost Function}
A definition of the cost function associated with filter performance - captured through the \emph{filter order} - leads us naturally to the imposition of constraints on the available space of controls.  This cost function therefore lies at the heart of our attempts to craft control protocols appropriate for a given noise environment.

From the spectral overlap in Eq. \ref{FirstOrderInfidelityComputational}, minimizing the infidelity contributed by the noise process $S_{i}(\omega)$ corresponds to minimizing the area under $ F_i(\omega\tau;\CPSMatElem_n)$ in the spectral region of interest.  We therefore define a cost function over a user-defined frequency band taking the form
\begin{align}\label{ErrorSuppressionFunctional}
A_i(\CPSMatElem_n)&:=\int_{\omega_{L}}^{\omega_c}d\omega F_i(\omega\tau;\CPSMatElem_n),\hspace{1cm}i\in\{z,\Omega\}
\end{align}
to diagnose the filtering effectiveness achieved by the control sequence $\CPSMatElem_n$. The smaller the integral $A_i(\CPSMatElem_n)$, the more effective the noise filtering over this band, in this noise quadrature. Since $\CPSMatElem_n$ is defined continuously over $\CtrlSpace_n$ for a given $n$, we may in principle construct a variational procedure over this control space to derive minimizing ``values'' of $\CPSMatElem_n$ satisfying a given cost function. In effect, the problem involves solving for paths in the control space over which the functional $A_i(\CPSMatElem_n)$ is minimized (up to some order). %Although in principle this can be achieved via a purely numerical search it is still a challenging procedure for unbounded $\CPSMatElem_n$. This motivates us in Section\ref{subSec:WalshSynthesizedFilters} to examine specific, constrained filter constructions facilitating analytic approaches to filter design.

Typically one would define the band $[\omega_L,\omega_c]$ over which the cost function is defined to fall within the \emph{stopband} of $F_i(\omega\tau)$, below which filtering generally takes place. In general the band $[\omega_L,\omega_c]$ may be tailored to target specific spectral regions in the noise PSD. Doing so may produce highly effective filtering over this narrow spectral region, though out-of-band behaviour can be quite poor if not specifically optimized\footnote{This effect is captured by the multiple slopes in Fig~\ref{Fig: WAMF_O2}h which clearly show the difference between the asymptotic zero-frequency \emph{roll-off} and the local slope over targeted regions $[\omega_L,\omega_c]$ in the stopband.}. %It is also possible to derive band-stop filters by minimizing the cost function in a narrow spectral region, even beyond the stobpand (such objects will always be high-pass filters for frequencies higher than the fastest control). 
%We now introduce two measures of the filter order, asymptotic and local.  

\subsection{The Filter Order} 
Again, following concepts from filtering in classical control engineering, we may define a \emph{filter order} which will play a central role in efficiently realizing effective noise filters.  We will mainly consider high-pass filters for low-frequency noise, setting $\omega_L = 0$ such that filtering takes place in the stopband up to the cutoff $\omega_c$. In this case it is useful to perform the Taylor expansion of the filter-transfer function about $\omega = 0$, written
\begin{align}\label{FFTaylorExpansionGeneral}
F_i(\omega\tau;\CPSMatElem_n) = \sum_{k=1}^{\infty}C^{(i)}_{2k}(\CPSMatElem_n)(\omega\tau)^{2k}
\end{align}
where the dependence of the expansion coefficients $C^{(i)}_{2k}$ on $\CPSMatElem_n$ has been made explicit, and we include only even powers of $\omega\tau$ due to the evenness of $F_i(\omega\tau)$. Assuming sufficiently low-frequency noise ($\omega_c<1/\tau$), the approximation $F(\omega\tau)\propto (\omega\tau)^{2p}$ holds for some $p$ associated with the most significant power law expansion term. This defines a high-pass filter with \emph{filter order} (determined by $p$) visualized as the slope in the stopband on a log-log plot\footnote{All stopbands ``turn on" with a finite response, the functional form of which determines the filter order and the effectiveness of noise suppression. In the stopband this is quantified by the slope, or \emph{roll off} in the language of filter design}.

Using this notation, and working in the low-frequency limit, we then say the control sequence $\CPSMatElem_n\in\CtrlSpace_n$ filters $\beta_i(t)$ noise to order $(p-1)$ over the band $[0,\omega_c]$ if $\CPSMatElem_n$ is a \emph{concurrent zero} of the first $(p-1)$ Taylor coefficients. That is, if $C^{(i)}_2(\CPSMatElem_n) = C^{(i)}_4(\CPSMatElem_n) = ... = C^{(i)}_{2(p-1)}(\CPSMatElem_n) = 0$. In this case we approximate $F_i(\omega\tau;\CPSMatElem_n)\approx C^{(i)}_{2p}(\CPSMatElem_n)(\omega\tau)^{2p}$ and consequently $A_i(\CPSMatElem_n)\approx C^{(i)}_{2p}(\CPSMatElem_n)\frac{(\omega_c\tau)^{2p+1}}{2p+1}$. Thus $\CPSMatElem_n$ is a $(p-1)$-order (high-pass) filter in the $i$th noise quadrature if the following equivalent conditions are satisfied
%\begin{align}\label{SuppressionConditions}
%p-1 \text{\emph{order order}}\iff\cases{F(\omega\tau;\CPSMatElem_n)\propto (\omega\tau)^{2p}\\
%C_2(\CPSMatElem_n) = C_4(\CPSMatElem_n) = ... = C_{2(p-1)}(\CPSMatElem_n) = 0\\
%\frac{A(\CPSMatElem_n)}{C_{2p}(\CPSMatElem_n)}=\mathcal{O}\Big(\frac{(\omega\tau_c)^{2p+1}}{2p+1}\Big)}%
%\end{align}
\begin{align}\label{pOrderFilterConditions}
\frac{A_i(\CPSMatElem_n)}{C^{(i)}_{2p}(\CPSMatElem_n)}=\mathcal{O}\Big(\frac{(\omega_c\tau)^{2p+1}}{2p+1}\Big)\hspace{0.2cm}\iff\hspace{0.2cm}
C^{(i)}_2(\CPSMatElem_n) = ... = C^{(i)}_{2(p-1)}(\CPSMatElem_n) = 0.
\end{align}
\noindent  This metric will play a central role in the analyses that follow.

It is important to disambiguate the asymptotic filter order $(p-1)$, introduced above for characterizing the behaviour near zero frequency, from a more general  metric capable of describing filter performance over an arbitrary spectral band. For this we introduce the \emph{local filter order} $(p^*-1)$ by the property that, over the band $[\omega_L,\omega_c]$ the filter-transfer function is well approximated by $F_i\propto(\omega\tau)^{2p^*}$. One may take the limit that $\omega_L\rightarrow\omega_c\rightarrow\omega^*$ and thereby obtain the \emph{instantaneous filter order}, effectively measuring the power-law behaviour at $\omega^*$. Both local and instantaneous filter order reduce to the asymptotic filter order over the stopband if over this region $F_i$ is well-characterized by its the zero-frequency behaviour.

%\begin{enumerate}[itemindent=4cm,labelsep=1cm]
%\item $\CPSMatElem_n$ is $p-1$ order filter\label{SuppressionCondition1}
%\item $F(\omega\tau;\CPSMatElem_n)\propto (\omega\tau)^{2p}$\hfill\refstepcounter{equation}(\theequation)\label{SuppressionCondition2}
%\item $C_2(\CPSMatElem_n) = C_4(\CPSMatElem_n) = ... = C_{2(p-1)}(\CPSMatElem_n) = 0$\hfill\refstepcounter{equation}(\theequation)\label{SuppressionCondition3}
%\item $\frac{A(\CPSMatElem_n)}{C_{2p}(\CPSMatElem_n)}=\mathcal{O}\Big(\frac{(\omega\tau_c)^{2p+1}}{2p+1}\Big)$\hfill\refstepcounter{equation}(\theequation)\label{SuppressionConditions}
%\end{enumerate}
%In particular, having defined $\CPSMatElem_n$ continuously over $\CtrlSpace_n$ for a given $n$, we may in principle construct a variational procedure over this control space to derive ``values'' of $\CPSMatElem_n$ satisfying a given cost function. In effect, the problem involves solving for paths in the control space over which the functional $A_i(\CPSMatElem_n)$ is minimized (up to order $p-1$). %Such paths describe \emph{classses} of control sequences filtering noise to order $p-1$. 

\subsection{Time-domain filter order vs. Magnus order}\label{AppSec:FiltervsMagnusOrder}
Both the asymptotic and instantaneous filter orders defined above for time-domain noise must be distinguished from the \emph{Magnus order} of error cancellation.  The latter is familiar from work in NMR in which quasi-static errors can be cancelled by suitable composite pulse sequence design.  The regime of quasistatic errors coincides with the DC limit for the time-dependent noise fields introduced in Sec. \ref{Sec:PhysicalSetting}. That is, the time-dependent noise fields reduce to scalar constants $\Bd\; (\Ba)$. The Magnus expansion terms in \ref{Magnus1}, now denoted $\Phi^\text{(DC)}_\mu$, are then evaluated strictly as time integrals over \emph{ideal} control operations, scaled by factors $\Bd^\mu\;(\Ba^\mu)$ specifying the power law dependence on the magnitude of these static offsets errors. A pulse sequence for which $\Phi^\text{(DC)}_1 = ...= \Phi^\text{(DC)}_{\mu-1} = 0$ is then said to compensate offset errors to Magnus order $(\mu-1)$. In this case the total error operator satisfies $\Phi^\text{(DC)}(\tau)=\mathcal{O}(\Phi^\text{(DC)}_\mu)$ and is dominated by the residual error proportional to the $\mu$th power in the magnitude of the error.

This is quite distinct from time-dependent noise where the error expansion used to calculate the fidelity contains terms of various Magnus order but equivalent time-dependent error norm in the ensemble average (see, \emph{e.g.} Eq. 1 in Ref.~\cite{GreenPRL2012}).  The net result is the observation that  \emph{high-order error suppression in the Magnus expansion does not imply high-order time-domain noise filtering}. This has been validated using experiments on trapped ions~\cite{SoareNatPhys2014}, and formalized rigorously in Ref.~\cite{ViolaFFFArXve2014}, where it has been shown that $p\leq\mu$, but $p^{*}$ over a user-defined band is unrelated to $\mu$.  Our focus throughout this work will be on crafting efficient noise filters rather than high-order error suppressing gates.

%The filter-transfer functions used to produce the definition for filter-order are the result of having made the first-order approximation in the Magnus expansion (Eq. \ref{Magnus1}) that $\Phi(\tau)\approx\Phi_1(\tau)$. Recent work has extended these concepts to a set of \emph{fundamental filter functions} serving as basic elements used to construct higher-order error terms as discussed in Ref.~\cite{Green2013}.  

%____________________%____________________%____________________%____________________%

%				SECTION 4 <--- (5): 	THE WALSH BASIS

%Motivate discussion about wtf Walsh is 5.0
%History of Walsh/Context. 5.0
%Paley Basis. 5.1
%Hadamard Basis
%____________________%____________________%____________________%____________________%

%\section{The Walsh Basis}\label{Sec:TheWalshBasis}

\section{Filter Design by Walsh Synthesis}\label{Sec:FilterDesignByWalshSynthesis}
Even with the general insights into the appropriate modulation protocols outlined above, it is desirable to bound the dimensionality of the control space, and hence the complexity of the filter-design task, by imposing physically motivated constraints on the form of $\CPSMatElem_n$. In practice the achievable filter order is typically limited by the number of unitary operations in the control sequence; one may increase $(p-1)$ at the cost of increasing $n$. From an experimental standpoint, faced with the physical limitation set by a maximum achievable Rabi rate, this cost manifests as a longer total sequence duration $\tau = \sig_l^n\tau_l$. This may offset the proposed benefit of the higher-order filter due to a longer noise interaction time. From a theoretical standpoint the cost is in the greater complexity of the variational search; the number of (free) variational parameters in $\CPSMatElem_n$ grow as $3n$ and the number of matrix products in Eqs. \ref{DephasingControlVector:ComputationalForm} and \ref{AmplitudeControlVector:ComputationalForm} grows as $n$. 

We are able to effectively bound the synthesis space while still achieving highly effective gates by \emph{synthesizing} relevant time-domain control fields in the basis set of Walsh functions - square wave analogues of the sines and cosines ~\cite{Beauchamp1975, HayesPRA2011} - using the concept of functional analysis.  Walsh functions are defined in a uniform piecewise-constant fashion  (Fig. \ref{Fig:WalshFunctions}), building intrinsic compatibility with discrete clocking~\cite{Hodgson2010} and classical digital logic. Since their formulation in the first half of the twentieth century~\cite{Walsh1923} they have played an important role in scientific and engineering applications. Their development and utilization has been strongly influenced by parallel developments in digital electronics and computer science since the 1960s, with Walsh-type transforms replacing Fourier transforms in a range of engineering applications such as communication, signal processing, pattern recognition, noise filtering and so forth~\cite{Tzafestas1985,Beauchamp1975}.

More recently the Walsh functions have been identified as an attractive resource in quantum information, with applications in time-resolved magnetometry using nitrogen-vacancy centres in diamond \cite{Cooper14} and in DD for digital-efficient pulse sequencing~\cite{HayesPRA2011}. Notably, in the latter scheme the decoupling performance was found to be determined by the distinct symmetry and spectral properties of the Walsh basis. These properties enable us to establish \emph{analytic design rules} (see Sec. \ref{subSec:WalshFilterDesignRues}) to further streamline Walsh-synthesized filter construction.

We begin by reviewing the relevant mathematical details of the Walsh basis. Two equivalent representations are introduced, \emph{Paley ordering} and the \emph{Hadamard representation}, which shall be used throughout this paper.
%. This so-called\emph{Walsh dynamical decoupling} (WDD) establishes a unified framework for describing both familiar and novel control protocols

%				--------------------------------------------------------------------------------------------------
%				   		sub-sub-SECTION: Paley Ordering  
%				--------------------------------------------------------------------------------------------------

\subsection{The Paley and Hadamard Representations}\label{SubSec:PaleyOrdering}
\begin{figure}
\centering
\includegraphics[width=0.6\columnwidth]{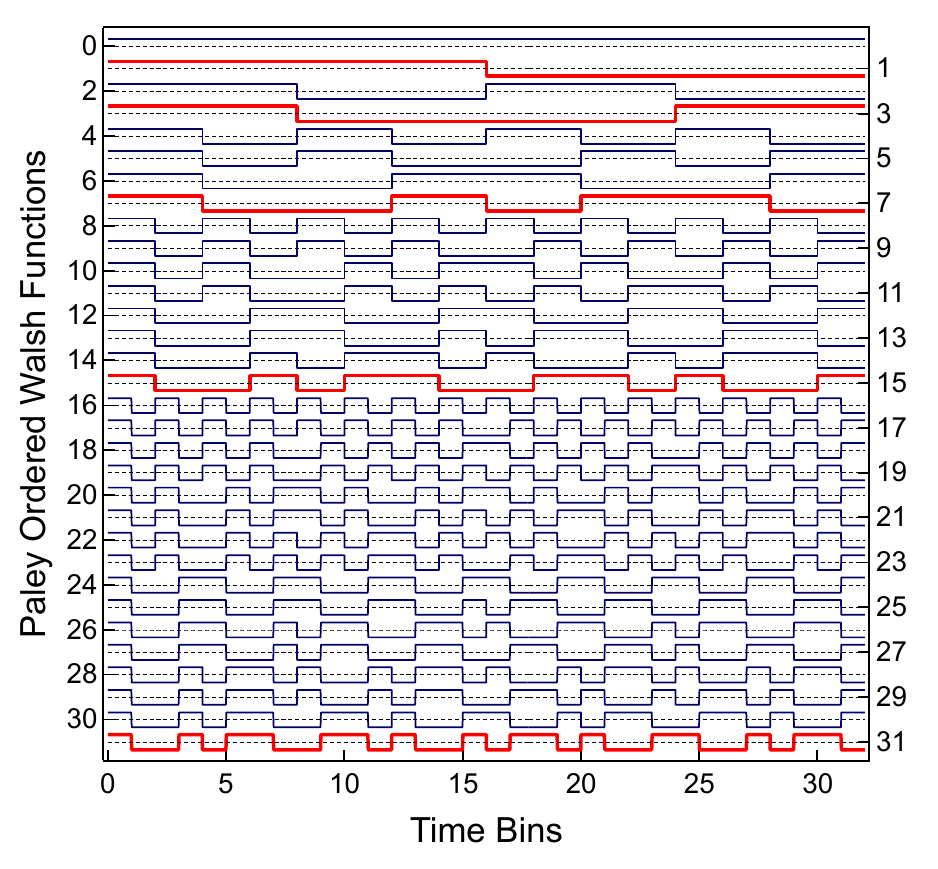}
\caption{The first 32 Paley-ordered Walsh functions $\PAL_{k}$, $k\in\{0,...,31\}$. Functions with maximum Hamming weight $r(k)$ (hence maximum number of Rademacher functions) for given $m(k)$, corresponding to Paley orders $2^{m(k)}-1$, are highlighted in red.}  \label{Fig:WalshFunctions}
\end{figure}

The set of Walsh functions $\Walsh_k:[0,1]\rightarrow\{\pm1\}$, $k\in\mathbb{N}$ form an orthonormal-complete family of binary-valued square waves defined on the unit interval. They are aperiodic and hence do not admit to a unique ordering, in contrast with the Fourier basis in which sinusoids are ordered by increasing \emph{frequency}.
%Thus, any square integrable function $f(x)$ on $[0,1]$ has a unique \emph{spectral decomposition} 
%\begin{align}
%&f(x) = \sum_{k=0}^\infty \WAMampP_k\Walsh_k(x)\hspace{0.25cm}\iff\hspace{0.25cm}\WAMampP_k := \int_0^1f(x)\Walsh_k(x)dx.
%\end{align}
%In this regard Walsh functions are the \emph{digital analogues} of the sines and cosines in Fourier analysis, and are naturally suited to synthesizing arbitrary (piecewise-constant) waveforms. 
A number of different orderings exist~\cite{Harmuth1969a,Harmuth1969b,Beauchamp1975} due to the different ways in which the basis elements may be defined. We employ the \emph{Paley ordering}~\cite{Paley1932} in which basis functions are generated from products of \emph{Rademacher functions}~\cite{Rademacher1922}, defined by %which lead to useful, recursive, Walsh-signal generation algorithms
\begin{align}\label{Rademacher}
R_j(x) := \text{sgn}\big[\sin(2^j\pi x)\big],\hspace{0.8cm} x\in[0,1],\hspace{0.8cm} j\ge0.
\end{align}
The $j$th Rademacher function $R_j(x)$ is thus a periodic square wave switching $2^{j-1}$ times between $\pm1$ over the interval $[0,1]$. The Walsh function of Paley order $k$,  here denoted $\PAL_k(x)$, is then defined by
\begin{align}\label{PALgeneration}
\PAL_k(x) = \prod_{j=1}^{m} R_j(x)^{b_j}
\end{align}
where $(b_m,b_{m-1},...,b_1)_2$ is the binary representation of $k$. That is, $k = b_m2^{m-1}+b_{m-1}2^{m-2}+...+b_{1}2^0$, where $m(k)$ indexes the most significant binary digit, having defined $b_m\equiv1$.  Consequently, $\PAL_k(x)$ has factors $R_j(x)$ whenever $b_j$ is a nonzero  binary digit of $k$; the total number of Rademacher functions in the construction of  $\PAL_k(x)$ is thus given by the number of nonzero $b_j$'s in $k$ - namely, the \emph{Hamming weight} denoted $r(k)$. For a given value of $m(k)$, the maximum number Rademacher functions therefore occur for $\PAL_{2^{m(k)}-1}(x)$. For example, setting $m(k) = 3$, a maximum of three Rademacher functions are used to construct $\PAL_7(x) = R_3(x)R_2(x)R_1(x)$, corresponding to the three nonzero digits $b_{1,2,3}$ in the binary expansion $k = 7 = (1,1,1)_2$.  For illustration, the first 32 Walsh functions in the Paley ordering are shown in Fig. \ref{Fig:WalshFunctions}. 

The discrete-timestep properties of these basis functions produce, under linear superposition, piecewise-constant waveforms with digitized segment lengths. In our framework these segments are used to specify the a modulation of the control field, ultimately defining a piecewise-constant sequence of unitaries. We therefore require a straightforward expression for the envelope of an arbitrary synthesis $\sum_{k = 0}^N \WAMampP_k\PAL_k(x)$. Due to the aperiodicity of the Walsh functions, however, a general expression in Paley ordering is difficult. To overcome this it is convenient to use the Hadamard representation. %This is particularly useful for efficiently constructing and \emph{deconstructing} Walsh-synthesized waveforms.

%Figure: Walsh Functions

%				--------------------------------------------------------------------------------------------------
%				   		sub-sub-SECTION: Hadamard Representation 
%				--------------------------------------------------------------------------------------------------

%\subsection{Hadamard Representation}\label{Sec:HadamardRepresentation}

%For our purposes we require an expression for the piecewise-constant structure of an arbitrary superposition of Walsh functions, and therefore desire \emph{a priori} knowledge of the locations of their various zero crossings. Due to the aperiodicity of the Walsh functions, however, a general expression is difficult. It is convenient instead to use the Hadamard representation. 

The unique sign-switching envelope of $\PAL_k(x)$ is determined by the sign-switching of the constituent Rademacher functions. Since any $R_j(x)$ switches sign uniformly $2^j$ times over the interval $[0,1]$, the fastest sign-modulation rate in $\PAL_k(x)$ derives from the highest order Rademacher function $R_{m(k)}(x)$, which switches sign $2^{m(k)}$ times over $[0,1]$. Provided $m(k)\le n$, we may therefore partition $[0,1]$ into $2^{n}$ equal time bins such that $\PAL_k(x)$ is constant valued on each bin. Any basis function $\PAL_k(x)$ then projects completely onto a digital vector in $\mathbb{R}^{2^n}$ with the $j$th element, $P^{(k)}_j\in\{\pm1\}$, defined by the value of $\PAL_k(x)$ in the $j$th bin. That is, $\PAL_k(x)$ is isomorphic to the discrete digital vector written
\begin{align}\label{VectorizedPAL}
\boldsymbol{P}^{(k)}_{2^n} \equiv \left[\begin{array}{cccc}P^{(k)}_1,&\hspace{0.2cm}P^{(k)}_2,&..., &P^{(k)}_{2^n}\end{array}\right].
\end{align}
This projection is possible for all $k\in\{0,1,...,2^n-1\}$ for which the condition $m(k)\le n$ is true, resulting in a set of $2^n$ vectors. Since these vectors inherit the orthogonality of the $\PAL_k(x)$, moreover, they form a \emph{discrete} Walsh basis spanning $\mathbb{R}^{2^n}$. 

In the Hadamard representation, these vectors occur as columns (rows) of the Hadamard matrix of dimension $2^n$. Using so-called \emph{Sylvester construction}~\cite{Horadam2007} the $2^n$-dimensional Hadamard matrix $H_{2^n}$ is generated recursively by 
\begin{align}
&H_{2^n} = \left[\begin{array}{cc}H_{2^{n-1}} & H_{2^{n-1}}\\H_{2^{n-1}}&-H_{2^{n-1}}\end{array}\right] = S^{\otimes n},\hspace{1cm}
S = \left[\begin{array}{cc}1 & 1\\1 &-1\end{array}\right],\hspace{1cm} H_1 = 1
\end{align} 
where $S$ is the Sylvester matrix, and $\otimes n$ denotes $n\ge1$ applications of the Kronecker product. In this construction $\boldsymbol{P}^{(k)}_{2^n}$ defines the $i(k)= 1+\sum_{j=1}^{m(k)}b_j2^{n-j}$  column (row) of $H_{2^n}$.  The orthogonality of the Walsh basis is thereby reflected in the familiar property that $H_{2^n}H_{2^n}^T = 2^nI$, implying the orthogonality of the Hadamard matrices.

%Now consider an arbitrary function $f(x) = \sum_{k = 0}^N \WAMampP_k\PAL_k(x)$ synthesized in the Walsh basis where $N$ sets the highest (Paley) ordered function in the construction. Then, from the above discussion, all these basis functions may be projected onto a Hadamard matrix with minimum dimension $\MinDim: = 2^{m(N)}$. A discrete representation of the function $f(x)$ therefore exists as a projection onto the column space of $H_\MinDim$ by writing
%\begin{align}\label{fHadRep}
%\boldsymbol{f} = H_{_\MinDim}\boldsymbol{\WAMampH},
%\hspace{0.75cm}\boldsymbol{\WAMampH}=  \big(\WAMampH_1,\hspace{0.1cm}  \WAMampH_2,...,\WAMampH_{_\MinDim}\big)^T,
%\hspace{0.75cm}\WAMampH_{i(k)} &=\cases{\WAMampP_k\hspace{0.4cm}\text{for}\hspace{0.4cm}0\le k\le N\\
%0\hspace{0.725cm}\text{for}\hspace{0.4cm}N<k<\MinDim}
%\end{align}
%where the column vector $\boldsymbol{\WAMampH}$ consists of the reordered Paley spectral amplitudes $\WAMampP_k$ according to the change of basis map specified by $i(k)$. Thus, in the Hadamard representation, the piecewise-constant structure of $f(x)$ is extracted from the vector $\boldsymbol{f}=\big(f_1,\hspace{0.1cm} f_2,...,f_{_\MinDim}\big)^T$, with $\boldsymbol{\WAMampH}$ representing the synthesis spectrum.

%							----------------------------------------------------------------
%							  sub:SECTION: (Walsh) Modulated Control Space
%							----------------------------------------------------------------

%				-------------------------------------------------
%				     sub:SECTION: WAPMFs CPS Representation 
%				-------------------------------------------------

The Hadamard representation of the Walsh functions has the distinct advantage of naturally specifying the piecewise-constant structure of time domain sequences constructed via linear combinations of Walsh functions.  Any function synthesized in the Paley-ordered Walsh basis, $f(x) = \sum_{k = 0}^{2^n-1} \fampP_k\PAL_k(x)$, has a vector representation in the column space of $H^{2^n}$. In this section we will use this observation to efficiently construct Walsh-synthesized filters, whose properties map compactly onto the Walsh spectrum.

\subsection{Walsh-Synthesized Filters}\label{subSec:WalshSynthesizedFilters}
%\subsection{Modulation Quadrature}
%It is desirable to impose structure on the form of $\CPSMatElem_n$ in order to constrain the dimensionality of the filter-design space and hence reduce the complexity of our task.  In practice the achievable filter order is typically limited by the number of unitaries operations in the control sequence; one may increase $(p-1)$ at the cost of increasing $n$. From an experimental standpoint, faced with the physical limitation set by a maximum achievable Rabi rate, this cost manifests as a longer total sequence duration $\tau = \sig_l^n\tau_l$. This may offset the proposed benefit of the higher-order filter due to a longer noise interaction time. From a theoretical standpoint the cost is in the greater complexity of the variational search; the number of (free) variational parameters in $\CPSMatElem_n$ grow as $3n$ and the number of matrix products in Eqs. \ref{DephasingControlVector:ComputationalForm} and \ref{AmplitudeControlVector:ComputationalForm} grows as $n$. 

The basic physics of commutation relations between Pauli operators immediately suggests an immediate constraint on the available modulation, broadly involving structuring of the Rabi rate or control phase in the time domain.
\begin{align}
\label{AmplitudeModulationDefinition}
&\CPSMatElem^\text{AMF}_n\cong\{(\tau_l,\Omega_l)\}_{l=1}^n, \hspace{0.25cm}\phi_l = \phi_0\hspace{0.25cm}\forall l\in\{1,...,n\}%&&\text{(\emph{Amplitude Modulated Filters})}\\
\\
\label{PhaseModulationDefinition}
&\CPSMatElem^\text{PMF}_n\cong\{(\tau_l,\phi_l)\}_{l=1}^n, \hspace{0.25cm}\Omega_l = \Omega_0\hspace{0.25cm}\forall l\in\{1,...,n\}%&&\text{(\emph{Amplitude Modulated Filters})}
\end{align}
referred to as \emph{single-axis amplitude-modulated filters} (AMFs) and \emph{constant-amplitude phase-modulated filters} (PMFs). These constrained forms may be used to design filters for dephasing and amplitude noise \emph{separately} using minimal control resources. For $\sigz$ (dephasing) noise it is sufficient to employ rotations about a single (orthogonal) axis in the $xy$ plane and therefore restrict attention to AMFs. On the other hand, unless implementing the trivial identity gate such that the total gate rotation angle $\Theta \equiv \sum_{l = 1}^\MinDim\theta_l=0$, strict amplitude modulation is insufficient for filtering amplitude noise. \footnote{This can be shown by Taylor expanding the amplitude noise filter fuction $\Fa(\omega\tau;\CPSMatElem^\text{AMF}_n)$ and deriving the result that $C^{(\Omega)}_2(\CPSMatElem^\text{AMF}_n) = \frac{1}{4}\big(\sum_{l = 1}^n\theta_l\big)^2$.} For nontrivial gates, amplitude noise filters generally require control over the rotation axis, and for this purpose we employ PMFs. 

In the Walsh synthesis framework, the modulated structures $\CPSMatElem^\text{AMF}_n$ and $\CPSMatElem^\text{PMF}_n$ are further constrained by synthesizing the time-domain Rabi rate $\Omega(t)$ or phase $\phi(t)$ as linear superpositions of Walsh functions
\begin{align}\label{PaleyOrderedWalshSynthesis}
\Omega(t) = \sum_{k = 0}^N \WAMampP_k\PAL_k(t/\tau),\hspace{1cm}\phi(t) = \sum_{k = 0}^N \WPMampP_k\PAL_k(t/\tau),\hspace{1cm}t\in[0,\tau].
\end{align} 
Here the synthesis spectra are denoted in terms of $\WAMampP_k$ ($\WPMampP_k$) to distinguish Walsh modulation in the amplitude (phase) quadratures. We refer to the resulting sequences as \emph{Walsh amplitude-} (WAMF) and \emph{phase-} (WPMF) \emph{modulated filters}. To compactly express these modulated control forms as sequences of unitaries we now employ the Hadamard representation. 

%				-------------------------------------------------
%				     sub:SECTION: Paley Synthesis
%				-------------------------------------------------
%\red 
%\subsubsection{Paley Synthesis}
%
%
%
%The set of Walsh functions $\Walsh_k:[0,1]\rightarrow\{\pm1\}$, $k\in\mathbb{N}$ form an orthonormal-complete family of binary-valued square waves defined on the unit interval. Thus, any square integrable function $f(x)$ on $[0,1]$ has a unique \emph{spectral decomposition} 
%\begin{align}
%&f(x) = \sum_{k=0}^\infty \WAMampP_k\Walsh_k(x)\hspace{0.25cm}\iff\hspace{0.25cm}\WAMampP_k := \int_0^1f(x)\Walsh_k(x)dx.
%\end{align}
%In this regard Walsh functions are the \emph{digital analogues} of the sines and cosines in Fourier analysis, and are naturally suited to synthesizing arbitrary (piecewise-constant) waveforms. The relevant mathematical details of the Walsh basis are presented below; we introduce two equivalent representations, \emph{Paley ordering} and the \emph{Hadamard representation}, which shall be used throughout this paper.

%This is particularly useful for \emph{efficiently constructing Walsh-synthesized waveforms}. 

Consider an arbitrary function $f(x) = \sum_{k = 0}^N \fampP_k\PAL_k(x)$ synthesized in the Walsh basis up to Paley order $N$. From Sec. \ref{SubSec:PaleyOrdering} all basis functions in this synthesis projected onto a Hadamard matrix of (minimum) dimension $\MinDim(N)\equiv 2^{m(N)}$. A discrete representation of the function $f(x)$ therefore exists as a projection onto the column space of $H_\MinDim$ by writing
\begin{align}\label{fHadRep}
\boldsymbol{f} = H_{_\MinDim}\boldsymbol{\fampH},
\hspace{0.75cm}\boldsymbol{\fampH}=  \big(\fampH_1,\hspace{0.1cm}  \fampH_2,...,\fampH_{_\MinDim}\big)^T,
\hspace{0.75cm}\fampH_{i(k)} &=\cases{\fampP_k\hspace{0.4cm}\text{for}\hspace{0.4cm}0\le k\le N\\
0\hspace{0.725cm}\text{for}\hspace{0.4cm}N<k<\MinDim}
\end{align}
where the column vector $\boldsymbol{\fampH}$ consists of the reordered Paley spectral amplitudes $\fampP_k$ according to the change of basis map specified by $i(k)= 1+\sum_{j=1}^{m(k)}b_j2^{n-j}$. Thus, in the Hadamard representation, the piecewise-constant structure of $f(x)$ is extracted from the vector $\boldsymbol{f}=\big(f_1,\hspace{0.1cm} f_2,...,f_{_\MinDim}\big)^T$, with $\boldsymbol{\fampH}$ representing the synthesis spectrum. In this case Eq. \ref{PaleyOrderedWalshSynthesis} implies the forms
%%%%%%%%%%%%%%%GENERAL FORM OF WAMF MATRIX
\begin{align}\label{Eq:WAMFGeneralForm}
 \CPSMatElem^\text{(WAMF)}_{\MinDim} =  \begin{blockarray}{ccccc}
    & \text{\scriptsize $\Omega_l$\normalsize}  &\text{\scriptsize $\tau_l$\normalsize}  & \text{\scriptsize $\theta_l$\normalsize}  &\text{\scriptsize $\phi_l$\normalsize} \\
    \begin{block}{c[c|ccc@{\hspace*{5pt}}]}
   \text{\scriptsize $P_1$\normalsize} & \BAmulticolumn{1}{c|}{\multirow{1}{*}{}}  &\tauMIN&\tauMIN\Omega_1& \phi_0\\
   % \cline{2-10}% don't use \hline
    \text{\scriptsize $P_2$\normalsize}
& \BAmulticolumn{1}{c|}{\multirow{2}{*}{$\Rabivec$}}
&\tauMIN
&\tauMIN\Omega_2
&\phi_0\\
    \vdots  & &\vdots&\vdots& \vdots &&&&&&&\\
    \text{\scriptsize $P_{\MinDim}$\normalsize}  & &\tauMIN&\tauMIN\Omega_\MinDim&\phi_0 &&&&&&&\\
    \end{block}
  \end{blockarray}\hspace{0.1cm},
\hspace{1cm}\Rabivec = H_\MinDim\boldsymbol{\WAMampH}%,\hspace{1cm}\MinDim(k)=2^{m(k)}
\end{align}

%%%%%%%%%%%%%%%GENERAL FORM OF WPMF MATRIX
\begin{align}\label{Eq:WPMFGeneralForm}
 \CPSMatElem^\text{(WPMF)}_{\MinDim} =  \begin{blockarray}{ccccc}
    & \text{\scriptsize $\Omega_l$\normalsize}  &\text{\scriptsize $\tau_l$\normalsize}  & \text{\scriptsize $\theta_l$\normalsize}  &\text{\scriptsize $\phi_l$\normalsize} \\
    \begin{block}{c[ccc|c@{\hspace*{5pt}}]}
   \text{\scriptsize $P_1$\normalsize} & \RabiWPMF &\tauMIN&\tauMIN\RabiWPMF& \BAmulticolumn{4}{c}{\multirow{1}{*}{}}\\
   % \cline{2-10}% don't use \hline
    \text{\scriptsize $P_2$\normalsize}
&\RabiWPMF  
&\tauMIN
&\tauMIN\RabiWPMF
& \BAmulticolumn{4}{c}{\multirow{2}{*}{$\vec{\boldsymbol{\phi}}$}}\\
    \vdots  & \vdots &\vdots&\vdots& &&&&&&&\\
    \text{\scriptsize $P_{\MinDim}$\normalsize}  & \RabiWPMF &\tauMIN&\tauMIN\RabiWPMF& &&&&&&&\\
    \end{block}
  \end{blockarray}\hspace{0.1cm},
\hspace{1cm}\vec{\boldsymbol{\phi}} = H_\MinDim\boldsymbol{\WPMampH}%,\hspace{1cm}\MinDim(k)=2^{m(k)}
\end{align}
\bla
with $\MinDim \equiv 2^{m(N)}$ and $\tau_\MinDim \equiv \tau/\MinDim$. The Rabi rate- or phase-modulation is thus defined by the vectors $\Rabivec = \big(\Omega_1,\hspace{0.1cm},\Omega_2,...,\Omega_\MinDim\big)^T$ and $\vec{\boldsymbol{\phi}} =  \big(\phi_1,\hspace{0.1cm},\phi_2,...,\phi_\MinDim\big)^T$ whose components 
\begin{align}
\Omega_l \equiv \Omega_l(\WAMampP_0, \WAMampP_1,...,\WAMampP_N), 
\hspace{1cm}
\phi_l\equiv\phi_l(\WPMampP_0, \WPMampP_1,...,\WPMampP_N), \hspace{1cm}
l\in\{1,...,\MinDim\}
\end{align}
specify the control variables during $l$th timestep. %are paramerized by the Walsh spectra.  %$\boldsymbol{\WAMampP}$($\boldsymbol{\WPMampP}$)
In this case $\tau_{l}$ takes a fixed discrete value, though consecutive segments with the same values of $\Omega_l$ and $\phi_l$ may be combined sequentially to form effective operations of longer duration.
%The degree of freedom associated with $\tau_l$ has thus apparently been removed. In fact the choice of $\tau_l$ has been transformed into the choice of Walsh basis functions in the synthesis, each with its characteristic temporal profile. 
The remaining degrees of freedom reside in the functional dependencies of $\Omega_l(\boldsymbol{\WAMampP})$ and $\phi_l(\boldsymbol{\WPMampP})$ on the Walsh spectra,
\footnote{We use the vectors $\boldsymbol{\WAMampP} \equiv(\WAMampP_0,\hspace{0.1cm} \WAMampP_1,...,\WAMampP_N)$ and $\boldsymbol{\WPMampP} \equiv (\WPMampP_0,\hspace{0.1cm} \WPMampP_1,...,\WPMampP_N)$ to compactly write the Paley ordered Walsh spectra implied by Eq. \ref{PaleyOrderedWalshSynthesis} in synthesizing $\Omega(t)$ and $\phi(t)$.}
%We ignore the $\theta_l$ colum of $\CPSMatElem_\MinDim$ in accounting for the remaining degrees of freedom since these are dependent variables.
the explicit forms of which are determined by the Hadamard matrix equations above. 

The reduced control space, now compactly parameterized by the Walsh spectra, thus consists of bounded-strength unitary sequences inheriting the discrete timing properties of the Walsh basis. This contrasts with similar composite pulse methods in NMR and quantum information~\cite{Khodjasteh2009, Khodjasteh2010, MerrillArXv2012} which generally rely on structures defined in continuous time\footnote{Pulse periods taking non-integer multiples values of $\tau_\text{min}$ then have intrinsic conflict with implementation in discretized time via digital control, giving rise to residual errors.}. In the next section we identify useful properties of the Walsh basis which capture filter performance and hence inform effective filter design. %Moreover our focus is essentially different from these approaches: we aim to filter non-Markovian \emph{time-dependent noise}, not compensate for \emph{static offset} detuning or pulse-length errors~\cite{MerrillArXv2012}. 
%In our work these, and other, properties are exploited to establish \emph{analytic design rules} to streamline Walsh-synthesized filter construction and constrain our search to a managable subspace of $\CtrlSpace_n$. 
%Thus we are able to efficiently produce piecewise-constant amplitude- or phase-modulated waveforms, with digitized segment lengths parameterized in the Walsh spectra. 

%				-------------------------------------------------
%				     sub:SECTION: Walsh Filter Design Rules
%				-------------------------------------------------
\subsection{Analytic Filter-Design Rules}\label{subSec:WalshFilterDesignRues}

%Not italics.  also these are hard to follow and aren't necessarily design rules.  I'd suggest:
%(i) Alternate modulation quadratures for dephasing or amplitude noise
%(ii) Impact of gate angle on allowable modulation
%(iii) Restricting the control space by symmetry considerations and target rotation
%(iv) Achievable filtering characteristics determined by n and r(k)

%The WAMF (WPMF) constructs  - and hence their filter characteristics - consequently map onto the Walsh spectra $\boldsymbol{\WAMampP} (\boldsymbol{\WPMampP})$ in terms of which they are parameterized. This enables us to produce a range of useful analytic filter-design rules based on symmetry and spectral properties of the Walsh basis, detailed below. 

%Walsh-synthesis restricts the search space for filter design to the subspace of  $\CtrlSpace_\MinDim$ effectively spanned by the Walsh spectra. That is, 
From Eqs.  \ref{Eq:WAMFGeneralForm} and \ref{Eq:WPMFGeneralForm} the WAMF (WPMF) constructs are completely parameterized by the Walsh spectra $\boldsymbol{\WAMampP}^{(i)}$, $i\in\{z,\Omega\}$. Here, for compactness, we denote $\boldsymbol{\WAMampP}^{(z)}(\boldsymbol{\WAMampP}^{(\Omega)}) = \boldsymbol{\WAMampP}( \boldsymbol{\WPMampP})$. Filter properties and gate characteristics consequently map onto the basis functions in the synthesis. 

To target these properties it is convenient to partition the Walsh spectrum $\boldsymbol{\WAMampP}^{(i)}\equiv(\boldsymbol{\WAMampP}^{(i)}_\nu,\boldsymbol{\WAMampP}^{(i)}_\rho)$ into spectral-amplitude classes to be treated as \emph{variational} ($\boldsymbol{\WAMampP}^{(i)}_\nu$) and  \emph{fixed parameters} ($\boldsymbol{\WAMampP}^{(i)}_\rho$). Making the formal substitution $\CPSMatElem_\MinDim\rightarrow\boldsymbol{\WAMampP}^{(i)}$,  the cost function in Sec. \ref{Sec:CharacteristicsOfNoiseFilters} is consequently re-expressed 
%It is convenient to partition the spectrum into  denoted further denote the spectral partition $\boldsymbol{\WAMampP}^{(i)}\equiv(\boldsymbol{\WAMampP}^{(i)}_\nu,\boldsymbol{\WAMampP}^{(i)}_\rho)$ to separate Walsh spectral amplitudes  to treat as  \emph{variational} ($\boldsymbol{\WAMampP}^{(i)}_\nu$) and  \emph{fixed parameters} ($\boldsymbol{\WAMampP}^{(i)}_\rho$). 
\begin{align}\label{WalshCostFunction}
A_i(\boldsymbol{\WAMampP}^{(i)}_\nu;\boldsymbol{\WAMampP}^{(i)}_\rho)&:=\int_{\omega_L}^{\omega_c}d\omega F_i(\omega\tau;\boldsymbol{\WAMampP}^{(i)}_\nu;
\boldsymbol{\WAMampP}^{(i)}_\rho),\hspace{1.25cm}i\in\{z,\Omega\}
\end{align}
where it is understood that $A_i$ is minimized over the space spanned by $\boldsymbol{\WAMampP}^{(i)}_\nu$ with $\boldsymbol{\WAMampP}^{(i)}_\rho$ held constant. Similarly, high-pass $(p-1)$-order filters satisfy the conditions
\begin{align}\label{pFilterWalsh}
\frac{A_i(\boldsymbol{\WAMampP}^{(i)})}{C_{2p}(\boldsymbol{\WAMampP}^{(i)})}=\mathcal{O}\Big(\frac{(\omega\tau_c)^{2p+1}}{2p+1}\Big)\hspace{0.2cm}\iff\hspace{0.2cm}
C^{(i)}_2(\boldsymbol{\WAMampP}^{(i)}) = ... = C^{(i)}_{2(p-1)}(\boldsymbol{\WAMampP}^{(i)}) = 0.
\end{align}

We are now in a position to establish a range of analytic filter-design rules to refine our search space and streamline Walsh synthesis leveraging approaches similar to electrical or digital signal filter construction. In particular, the well defined spectral properties and symmetries of the Walsh functions may be used to inform effective filter construction with improved performance. These include 
\begin{enumerate}
\item Alternate modulation quadratures for dephasing or amplitude noise
\item Restricting Walsh synthesis by symmetry considerations
\item Constraining Walsh spectra for target gate angle
\item Achievable filtering characteristics determined by $m(k)$ and $r(k)$
\end{enumerate}  
We address each of these in turn.

%--------------------%--------------------%--------------------%--------------------%--------------------

(i) \emph{Alternate modulation quadratures for dephasing or amplitude noise - } As the most basic element of design, we first reiterate the statements made above establishing  WAMFs (WPMFs) as useful for filtering dephasing (amplitude) noise \emph{separately}. In Sec. \ref{Sec:UWMFs}, however, we derive universal noise filters by concatenating these two filter constructs.
%--------------------%--------------------%--------------------%--------------------%--------------------

(ii) \emph{Restricting Walsh synthesis by symmetry considerations - } As with the cosines (sines) constituting the Fourier basis, the Walsh basis separates into so-called CAL (SAL) functions with even (odd) parity. Restricting the synthesis to the CAL subset ensures the modulated waveform has time-reversal symmetry about the sequence midpoint $\tau/2$. This can be a convenient and effective method in filter design, in line with the observation in dynamic decoupling literature~\cite{QECLidar2013,Souza2012} that sequence performance is often improved using time-symmetric over -asymmetric building blocks\footnote{Our studies have not produced proof that this symmetry is strictly necessary. In fact for WPMFs it is not required. However WAMF constructions possessing time-reversal symmetry do appear to yield results more readily, and all WAMFs we have discovered have this property.}. 

%--------------------%--------------------%--------------------%--------------------%--------------------

(iii) \emph{Constraining Walsh spectra for target gate angle - }
%In filter construction we may further constrain the form of a candidate pulse sequence by imposing required phys- ical properties on the sequence, such as fixing the total rotation angle of the Bloch vector in order to implement a target logic operation. 
Imposing desired physical properties on a candidate control sequence may generally be achieved by holding some subset $\boldsymbol{\WAMampP}^{(i)}_\rho$ of the Walsh-spectral amplitudes constant. For example, we may fix the total rotation angle of the Bloch vector in order to implement a target logic operation. For WAMFs this involves a very straightforward constraint on the Walsh spectrum: the total rotation angle depends only on $\WAMampP_0$. This can be seen as follows. First observe for Paley orders $k\ge1$ the Walsh functions are balanced in the sense that $\int_0^1\PAL_k(x)dx = \delta_{0k}$, where $\delta_{ij}$ denotes the Kronecker delta. For WAMFs the total gate angle $\Theta \equiv \int_0^\tau dt \Omega(t)$ therefore takes the form
\begin{align*}
\Theta  = \int_0^\tau\sum_{k=1}^N\WAMampP_k\PAL_k(t/\tau)dt= \tau\sum_{k=1}^N\WAMampP_k\int_0^1\PAL_k(x)dx = \tau\sum_{k=1}^N\WAMampP_k\delta_{0k} = \WAMampP_0\tau.
\end{align*}
The effective gate rotation, $\theta = \Theta\hspace{0.1cm}\text{mod}\hspace{0.1cm}2\pi$, is consequently given by 
\begin{align}\label{NetBlochRotationConstraint}
\theta = \WAMampP_0\tau\hspace{0.1cm}\text{mod}\hspace{0.1cm}2\pi
\end{align} 
implying the necessary constraint on $\WAMampP_0$ for a desired $\theta$. 

%--------------------%--------------------%--------------------%--------------------%--------------------

%(iv) \emph{Time Reversal Symmetry (Restricted Synthesis) - }
%We may also exploit the distinct parity of the Walsh functions. As with the cosines (sines) constituting the Fourier basis, the Walsh basis separates into so-called CAL (SAL) functions with even (odd) parity. Restricting the synthesis to the CAL subset ensures the modulated waveform has time-reversal symmetry about the sequence midpoint $\tau/2$. This can be a convenient and effective method in filter design, in line with the observation in dynamic decoupling literature~\cite{QECLidar2013,Souza2012} that sequence performance is often improved using time-symmetric over -asymmetric building blocks\footnote{Our studies have not produced proof that this symmetry is strictly necessary. In fact for WPMFs it is not required. However WAMF constructions possessing time-reversal symmetry do appear to yield results more readily, and all WAMFs we have discovered have this property.}. 

%--------------------%--------------------%--------------------%--------------------%--------------------

(iv) \emph{Achievable filtering characteristics determined by $m(k)$ and $r(k)$ - } The achievable filter order over the entire stopband is essentially limited by the number of constituent control operations: one may achieve higher $p$ at the cost of higher $n$. For the Walsh-synthesized filters in Eqs.  \ref{Eq:WAMFGeneralForm} and \ref{Eq:WPMFGeneralForm}, with $N$ the highest-order basis function, $n \equiv 2^{m(N)}$. Hence higher-order Walsh functions generally produce higher-order filters.

For high-pass filters further insight is gained by examining the low-frequency spectral properties of the $\PAL_k(t/\tau)$. This reflects the fact that the filter-transfer functions are closely related to Fourier transforms of relevant time-domain control functions. In particular, the Fourier transform of $\PAL_k(t/\tau)$, near zero frequency, has a power-law expansion~\cite{HayesPRA2011}
\begin{align}\label{Eq:HayesFourier-PALIdentity}
\FT_x\big[\PAL_k(x)\big]\propto (\omega\tau)^{r(k)}
\end{align}
where $r(k)$ is the Hamming weight. Here $x \equiv t/\tau$ is a non-dimensional time-domain variable and $\FT_x\big[\PAL_k(x)\big]$ denotes the forward Fourier transform of $\PAL_k(x)$ from $x$ to the (nondimensional) angular frequency variable $\omega\tau$ in Fourier space. Increasing the low-frequency \emph{roll-off} is therefore associated with maximizing $r(k)$ for a given number of control operations $n = 2^{m(k)}$. This corresponds to maximizing the number of Rademacher functions in the construction\footnote{Maximizing the number of Rademacher functions does \emph{not} correspond to maximizing the switching rate of $\PAL_k(x)$. In fact, for a given $m(k)$ the maximum switching rate for $\PAL_k(x)$ corresponds to $k = 2^{m(k)-1}$, which consists of the single Rademacher function $R_{m(k)-1}(x)$.} and immediately identifies Paley orders $k = 2^\alpha -1$, $\alpha\in\mathbb{N}$, (see Fig. \ref{Fig:WalshFunctions}) as key design resources. 

\section{Walsh Amplitude Modulated Filters (WAMFs)}\label{Sec:WAMFs}

Having introduced the basic physical picture and mathematical basis for Walsh filter synthesis, we move on to demonstrate explicit realizations of WAMFs for dephasing noise. Both first and second-order filters with high-pass filter characteristics are constructed.

%				-------------------------------------------------
%				sub:SECTION:first-order WAMFs
%				-------------------------------------------------
\subsection{First-Order WAMFs}\label{SubSec:FirstOrderWAMFs}
We begin by considering first-order filters for dephasing noise implementing target single-qubit rotations. Construction begins by considering the design rules (i)-(iv) outlined in Sec. \ref{subSec:WalshFilterDesignRues}. For filtering noise in this quadrature (i) implies we should employ the WAMF construction (Eq. \ref{Eq:WAMFGeneralForm}). In this case, invoking (iii), the requirement of implementing nontrivial gates dictates we include Paley order $k = 0$ in the synthesis. The average Rabi rate (and hence rotation angle) is then determined by $\WAMampP_0$, the spectral amplitude of $\PAL_0(t/\tau)$. The remaining synthesis choices include basis functions of Paley order $k>0$ and are in principle unbounded.  

As a first application, we pursue the construction minimizing the number $n = 2^{m(N)}$ of unitary operations in the synthesized sequence such that error suppression is still attainable.  In line with design rules (ii) and (iv), time-reversal symmetry is ensured and the number of Rademacher functions is maximized by reducing the remaining synthesis choices to the single basis function $\PAL_3(t/\tau)$ (Fig. \ref{Fig:WalshFunctions}). Hence, in this simple example, $N = 3$, and $\MinDim(N) = 2^{m(N)} = 2^2$, yielding 4-segment gates with segment lengths $\tau_\MinDim = \tau/4$. These represent the lowest-order constructions with error suppression capabilites. 

\begin{figure}[!ht]
\centering
\includegraphics[width=0.98\columnwidth]{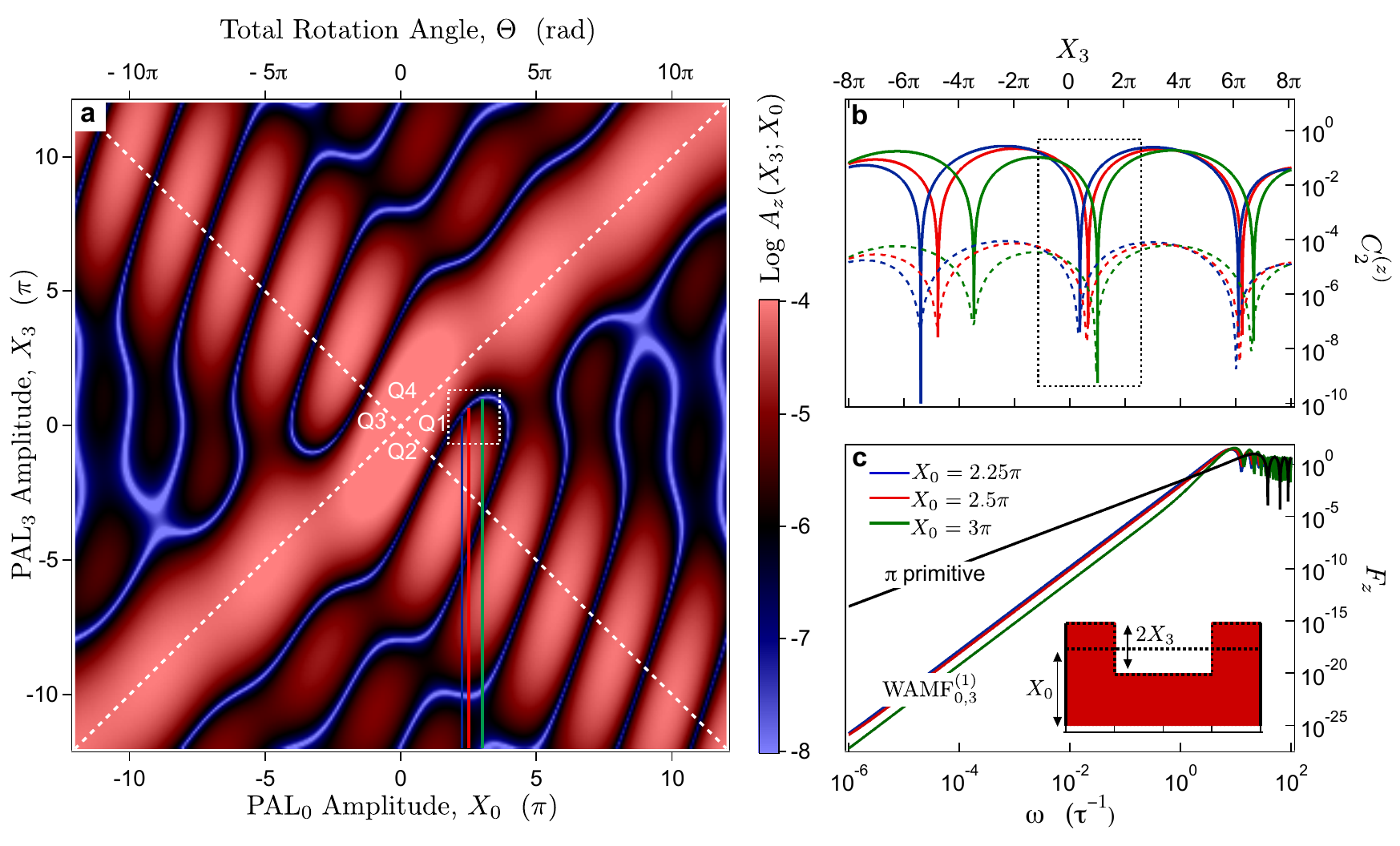}
\caption{Construction of WAMF$_{0,3}^{(1)}$ for dephasing noise filtering. \textbf{a}) Log-scale color plot of the cost function $A_z(\WAMampP_3;\WAMampP_0)$ integrated over $\omega\in [10^{-9}, 10^{-1}]\tau^{-1}$. Total gate angle $\Theta = \WAMampP_0$ $(\tau\equiv1)$. Blue regions indicate minima in $A_z(\WAMampP_3;\WAMampP_0)$, implying optimized filter synthesis. Coloured lines (\blu blue\bla, \red red\bla, \green green\bla) at $\WAMampP_0 = \big(\blu2\frac{1}{4}\bla,\hspace{0.1cm} \red 2\frac{1}{2}\bla,\hspace{0.1cm} \green3\bla\big)\pi$ correspond to rotation angles $\theta = \big(\blu \frac{\pi}{4}\bla,\hspace{0.1cm} \red \frac{\pi}{2}\bla,\hspace{0.1cm} \green\pi\bla\big)$. These lines terminate at values $\WAMampP_3 = \big(\blu0.36...\bla, \red 0.65...\bla, \green1\bla\big)\pi$ on a blue contour (boxed) and indicate representative points in the $X_0X_3$ plane for which first-order filtering is achieved. In this plot, for $|\WAMampP_3|>|\WAMampP_0|$ the Rabi rates $\WAMampP_{\pm}$ have opposite sign, implying a $\pi$-phase shift in addition to amplitude modulation (e.g. see Eq. \ref{Eq:PiPhaseShift}). We therefore distinguish quadrants Q1 \& Q3 in the $\WAMampP_0\WAMampP_3$ plane in which $|\WAMampP_3|\le|\WAMampP_0|$ (strict amplitude modulation) and Q2 \& Q4 in which $|\WAMampP_3|>|\WAMampP_0|$ (sign-switching amplitude modulation).
%The \blu blue\bla, \red red\bla, and \green green \bla lines correspond to the choices $\WAMampP_0 = \big(\blu2\frac{1}{4}\bla, \red 2\frac{1}{2}\bla, \green3\bla\big)\pi$, implying effective gate angles $\theta = \big(\blu \frac{\pi}{4}\bla, \red \frac{\pi}{2}\bla, \green\pi\bla\big)$. These lines, terminating at values $\WAMampP_3 = \big(\blu0.36...\bla, \red 0.65...\bla, \green1\bla\big)\pi$ indicate values in the $X_0X_3$ plane optimized for filter construction. 
%The \blu blue\bla, \red red\bla, and \green green \bla lines correspond to the choices $\WAMampP_0 = \big(\blu2\frac{1}{4}\bla, \red 2\frac{1}{2}\bla, \green3\bla\big)\pi$, implying effective gate angles $\theta = \big(\blu \frac{\pi}{4}\bla, \red \frac{\pi}{2}\bla, \green\pi\bla\big)$. 
\textbf{b}) \emph{Solid lines}: first order Taylor coefficient $C_2^{(z)}(\WAMampP_3;\WAMampP_0)$ as a function of $\WAMampP_3$ with $\WAMampP_0 = \big(\blu2\frac{1}{4}\bla,\hspace{0.1cm} \red 2\frac{1}{2}\bla, \hspace{0.1cm} \green3\bla\big)\pi$; zeros appear as dips on log-scale. \emph{Dotted lines}: one-dimensional slices of $A_z(X_3;X_0)$ for same fixed values of $X_0$. Boxed dips correspond to boxed points in a) where the colored lines intersect with the blue contour. \textbf{c}) filter-transfer functions for the spectrally optimized WAMF$_{0,3}^{(1)}$ gates identified by the boxed features in a) and b).}  \label{Fig: WAMF_O1}
\end{figure}

The Rabi rate is consequently written $\Omega(t) = \WAMampP_0\PAL_0(t/\tau)+\WAMampP_3\PAL_3(t/\tau)$. Physically,  $\WAMampP_3$ specifies the modulation depth of the resulting Rabi rate envelope (see inset to Fig. \ref{Fig: WAMF_O1}c) while  $\WAMampP_0$ determines the average value as described above.  Accordingly, for a particular target rotation, we treat $\WAMampP_0$ as a \emph{fixed} parameter (see Eq. \ref{NetBlochRotationConstraint}) while $\WAMampP_3$ is treated as a \emph{variational} parameter by which to optimize the (dephasing) cost function (Eq. \ref{WalshCostFunction}). Thus, values of $\WAMampP_3$ for which $A_z(\WAMampP_3;\WAMampP_0)$ is minimized specify the optimum modulation depths for an effective filter. 

Using Eq.~\ref{fHadRep} the Walsh synthesis spectrum is represented $\boldsymbol{\WAMampH}\equiv\big(\WAMampH_1,\hspace{0.1cm}\WAMampH_2,\hspace{0.1cm}\WAMampH_3,\WAMampH_4\big)^T = \big(\WAMampP_0,\hspace{0.1cm}0,\hspace{0.1cm}0,\hspace{0.1cm}\WAMampP_3\big)^T$, yielding the Hadamard representation of the modulated Rabi rate 
\begin{align}\label{WAMFO1HadamardRepresentation}
&\Rabivec= \left[\begin{array}{cccc}1&1&1&1\\1&-1&1&-1\\1&1&-1&-1\\1&-1&-1&1\end{array}\right]\left[\begin{array}{cccc}  \WAMampP_0\\ 0\\ 0\\ \WAMampP_3\end{array} \right] = \left[\begin{array}{cccc}  \WAMampP_0+\WAMampP_3\\ \WAMampP_0-\WAMampP_3\\ \WAMampP_0-\WAMampP_3\\\WAMampP_0+\WAMampP_3\end{array} \right].
\end{align}
The resulting WAMF construction, denoted WAMF$_{0,3}^{(1)}$, is thus represented 
\begin{align}\label{WAMFOrder1CPSMatrix}
\text{WAMF}^{(1)}_{0,3}\hspace{0.25cm}=\hspace{0.25cm}
\begingroup
  \renewcommand*{\arraystretch}{1.5}%
  \kbordermatrix{
           & \Omega_l           & \tau_l           & \theta_l           & \phi_l     \cr
    P_1    &\frac{\WAMFRabiPlus}{\tau}             & \frac{\tau}{4}       & \frac{\WAMFRabiPlus}{4}        & 0 \cr
    P_2    & \frac{\WAMFRabiMin}{\tau}             &\frac{\tau}{4}        & \frac{\WAMFRabiMin}{4}    & 0 \cr
    P_3    & \frac{\WAMFRabiMin}{\tau}             &\frac{\tau}{4}        & \frac{\WAMFRabiMin}{4}    & 0 \cr
    P_4   &\frac{\WAMFRabiPlus}{\tau}             & \frac{\tau}{4}       & \frac{\WAMFRabiPlus}{4}         & 0 \cr
  }%
\endgroup
\end{align}
where $\WAMampP_{\pm} := \WAMampP_0\pm\WAMampP_3$, and the superscript in this notation denotes first-order filter capabilities. Hence these gates inhabit the two-dimensional control space spanned by the $\WAMampP_0\WAMampP_3$ plane (see Fig. \ref{Fig: WAMF_O1}a).
%This choice is motivated by the fact that Paley order $k = 3$ corresponds to the lowest-order non-constant Walsh function with even symmetry about $\tau/2$ (Fig. \ref{Fig:WalshFunctions}). Hence WAMF$_{0,3}^{(1)}$ is the simplest Walsh amplituded-modulated construct both possessing time-reversal symmetry about the sequence midpoint and capable of implementing nontrivial gate angles, due to the inclusion of $\WAMampP_0$ (Eq. \ref{NetBlochRotationConstraint}).
The dephasing filter-transfer function $\Fd(\omega;\WAMampP_3;\WAMampP_0)$ for an arbitrary WAMF$_{0,3}^{(1)}$ gate is derived by substituting Eq. \ref{WAMFOrder1CPSMatrix} into Eq. \ref{DephasingFF}. The cost function $A_z(\WAMampP_3;\WAMampP_0)$ may then be numerically integrated. 

Fig.  \ref{Fig: WAMF_O1}a shows a two-dimensional representation of $A_z(\WAMampP_3;\WAMampP_0)$ integrated over the stopband $\omega\in [10^{-9}, 10^{-1}]\tau^{-1}$. The value of $\text{Log}_{10}\big[A_z(\WAMampP_3;\WAMampP_0)\big]$ is indicated by the color scale. Total sequence length is normalized to $\tau=1$ in this data, so the total gate rotation angle $\Theta \equiv \WAMampP_0$ is given directly by the $X_0$-axis. %Minima in $A(\WAMampP_3;\WAMampP_0)$ are shown as blue regions, representing points in the control space for which these gates also perform as first-order high-pass filters. 
As can be seen, for any fixed $\WAMampP_0$ there exist quasi-periodic tunings of $\WAMampP_3$ which minimize the cost function. In other words, we have a prescription for synthesizing \emph{spectrally-optimized dephasing filters which implement arbitrary rotation angles}. 
Interestingly, the point $(\WAMampP_0,\hspace{0.1cm}\WAMampP_3)\equiv( \green 3\pi\bla,\hspace{0.1cm}\green \pi\bla)$ reproduces the previously derived first-order DCG NOT construction~\cite{khodjasteh2009dcg}.

Blue regions, where the cost function has been minimized, represent \emph{first-order} filters for low-frequency noise due to the restrictions placed on the synthesis space~\footnote{These points may also be derived using Nelder-Mead optimization of $A_z(\WAMampP_0;\WAMampP_3)$ over the two-dimensional domain. This method is useful for more complex constructions (see Sec. \ref{subSec:SecondOrderWAMFs}) where spectral optimization becomes a more multi-dimensional task. }.  To demonstrate that these optimized $\text{WAMF}^{(1)}_{0,3}$ gates perform as first-order filters we Taylor expand $\Fd(\omega;\WAMampP_3;\WAMampP_0)$ as in Eq. \ref{pFilterWalsh}, and derive an easy analytic expression for the first order coefficient
\begin{align}
C^{(z)}_2(\WAMampP_3;\WAMampP_0) = 4\Bigg[\frac{(\WAMampP_0-\WAMampP_3)\sin(\frac{\WAMampP_0}{2})+2\WAMampP_3\sin(\frac{\WAMampP_0-\WAMampP_3}{4})}{\big(\WAMampP_0^2-\WAMampP_3^2\big)}\Bigg]^2.
\end{align}
In principle we may now solve $C^{(z)}_2(\WAMampP_0;\WAMampP_3)=0$ to find values of $\WAMampP_3$ giving first-order filters for a given $\WAMampP_0$. 
%These roots may be found numerically, but for our purposes it is sufficient to observe such roots do indeed exist. 
In Fig. \ref{Fig: WAMF_O1}b we plot $C^{(z)}_2(\WAMampP_0;\WAMampP_3)$ (solid lines) as a function of $\WAMampP_3$ for the choices $\WAMampP_0 = \big(\blu2\frac{1}{4}\bla, \hspace{0.1cm}\red 2\frac{1}{2}\bla,\hspace{0.1cm} \green3\bla\big)\pi$ as above. Zeros of  $C^{(z)}_2(\WAMampP_0,\WAMampP_3)$, appearing as dips on the log scale, occur quasi-periodically in $\WAMampP_3$ and match with points in Fig. \ref{Fig: WAMF_O1}a where corresponding lines of constant $X_0$ intersect with the blue contours. To demonstrate this we plot one-dimensional slices of $A_z(X_3;X_0)$ for fixed values $\WAMampP_0 = \big(\blu2\frac{1}{4}\bla,\hspace{0.1cm} \red 2\frac{1}{2}\bla,\hspace{0.1cm} \green3\bla\big)\pi$ (dotted lines). We find the minima in $A_z(X_3;X_0)$ align with zeros of $C^{(z)}_2(\WAMampP_0;\WAMampP_3)$, implying the blue contours in the $X_0X_3$ plane do indeed produce first-order filters, with $(p-1) = 1$. The boxed zeros near $\WAMampP_3 = \big(\blu0.36...\bla, \red 0.65...\bla, \green1\bla\big)\pi$ correspond to the termination points of the colored lines in Fig \ref{Fig: WAMF_O1}a) (also boxed). These indicate representative points in the $X_0X_3$ plane producing first-order filters with nontrivial rotations. In particualr, these filters implement $\theta = \big(\blu \frac{\pi}{4}\bla, \hspace{0.1cm}\red \frac{\pi}{2}\bla, \hspace{0.1cm}\green\pi\bla\big)$ rotations.

The corresponding dephasing filter-transfer functions for these three optimized gates are shown in Fig. \ref{Fig: WAMF_O1}c. As expected from Eq. \ref{FFTaylorExpansionGeneral}, with $C_2^{(z)} = 0$, these approximately satisfy $\Fd\propto(\omega\tau)^4$ in the stopband, producing first-order filters with $(p-1) = 1$. For comparison we include the dephasing filter-transfer function for a primitive $\pi$ rotation where $\Fd\propto(\omega\tau)^2$, implying $(p-1)=0$. The steeper slopes, or \emph{roll-offs}, for the optimized WAMF$_{0,3}^{(1)}$ gates captures this difference. This filter design method, and the performance of the $\text{WAMF}^{(1)}_{0,3}$ filters, has recently been experimentally validated by our group~\cite{SoareNatPhys2014}.

\subsection{Second-Order WAMFs}\label{subSec:SecondOrderWAMFs}

We now consider higher-order dephasing filters by increasing the number $n$ of segments in the sequence. In particular we consider 8-segment gates. Construction again begins by considering the design rules (i)-(iv) outlined in Sec. \ref{subSec:WalshFilterDesignRues}. 

Using (i) and (iii) we employ the WAMF construction and include Paley order $k=0$ to ensure nontrivial rotation angles. Extending to 8-segments, however, increases the acessible range of Walsh functions in the synthesis as identified in design rule (iv). Specifically we extend the synthesis to Paley orders $k\le 7$ corresponding to the complexity class $m(k)\le 3$, implying a $2^3 = 8$ segments construction in the Hadamard representation. We denote these constructions by WAMF$_{0:7}^{(2)}$ where the superscript indicates second-order filtering capabilities, as will be shown. Imposing time-reversal symmetry about $\tau/2$ further restricts the synthesis space to $k\in\{3,5,6\}$, corresponding to CAL functions referenced in design rule (ii). We therefore study synthesized filters with spectral amplitudes partitioned into fixed  $\boldsymbol{\WAMampP}_\rho = \WAMampP_0$ and variational $\boldsymbol{\WAMampP}_\nu = (\WAMampP_3, \WAMampP_5, \WAMampP_6)$ classes. 

%Referring to design rule (ii), we therefore set $\boldsymbol{\WAMampP}_\rho \equiv \WAMampP_0$ and treat $\WAMampP_0$ as a \emph{fixed} parameter. On the other hand we set $\boldsymbol{\WAMampP}_\nu  \equiv \WAMampP_3$ and treat $\WAMampP_3$ as a \emph{variationonal} parameter with respect to which the (dephasing) cost function (Eq. \ref{WalshCostFunction}) is minimized. Values of $\WAMampP_3$ for which $A_z(\WAMampP_3;\WAMampP_0)$ is minimized thus specify the optimum modulation depth for an effective filter. 
As a representative example we set $\WAMampP_0 = 3\pi$ and restrict attention to filters implementing a net $\pi$ rotation ($\tau\equiv1$). Our cost function consequently takes the form $A_z(\WAMampP_3,\WAMampP_5,\WAMampP_6;3\pi)=\int_{\omega_L}^{\omega_c}\Fd(\omega\tau;\WAMampP_3,\WAMampP_5,\WAMampP_6;3\pi)d\omega$, implying a three-dimensional \emph{variational} control space over which to derive spectrally-optimized filters. We accomplish this using a Nelder-Mead search to minimize $A_z(\boldsymbol{\WAMampP}_\nu;3\pi)$ over the $\boldsymbol{\WAMampP}_\nu$-domain.
 
\begin{figure}[!ht]
\hspace{-0.7cm}%\centering
\includegraphics[width=1.1\columnwidth]{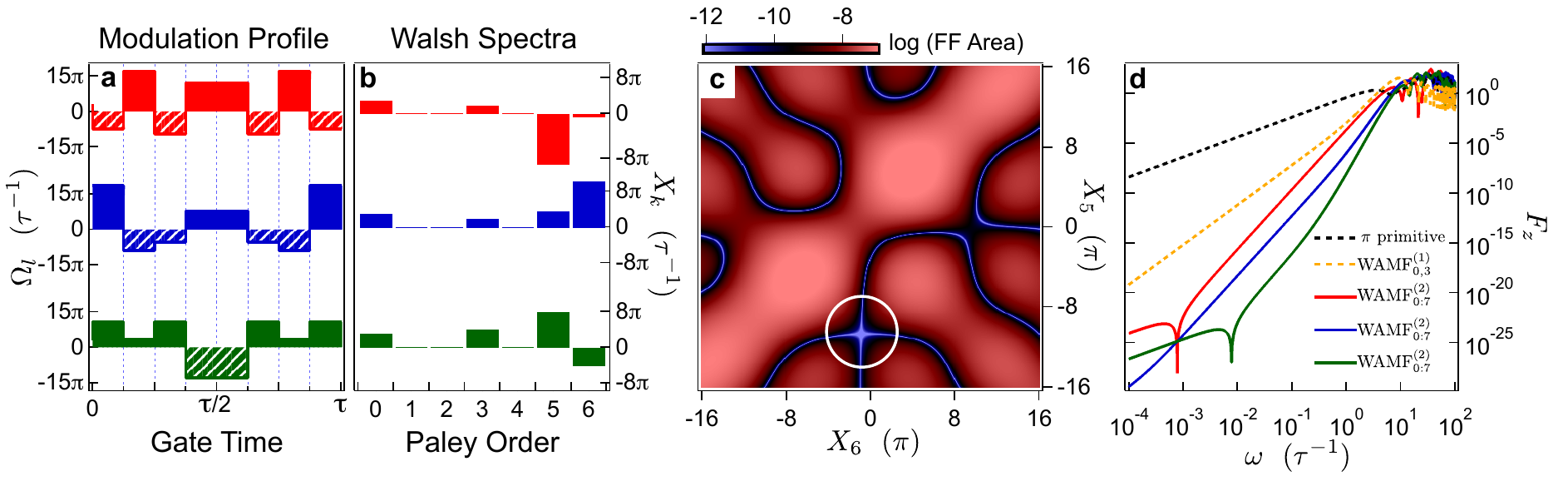}
\caption{Construction of WAMF$_{0:7}^{(2)}$ for dephasing noise filtering. \textbf{a}) Representative amplitude-modulated profiles for spectrally-optimized 8-segment WAMF$_{0:7}^{(2)}$ gates. Vertical axes indicates Rabi rate values $\Omega_l$ in units of $1/\tau$ for the 8-segments. \textbf{b}) Corresponding (Paley ordered) Walsh spectra. Vertical axes indicate values of the Walsh spectral amplitudes $X_k$ in units of $1/\tau$. Optimized spectra obtained via Nelder-Mead search. \textbf{c}) Log-scale color plot of the cost function $A_z(\WAMampP_5,\WAMampP_6)$ (integrated over $\omega\in [10^{-9}, 10^{-1}]\tau^{-1}$) defined on representative two-dimensional cross section of $\boldsymbol{\WAMampP}_\nu$-domain. Blue regions indicate minima in $A_z(\WAMampP_5,\WAMampP_6)$, implying  \emph{second-order} optimized filter synthesis. ``Cross-region'' (circled) indicates robustness region with respect to errors in $\WAMampP_{5,6}$. \textbf{d}) Dephasing filter-transfer functions for the optimized WAMF$_{0:7}^{(2)}$ gates in a), compared against primitive $\pi_x$ rotation and optimized $\pi_x$ WAMF$_{0,3}^{(1)}$ gate. For the blue, red and green traces the cost function $A_z(X_3,X_5,X_6)$ was defined over the band $[\omega_L,\omega_c]$ with $\omega_c = \tau^{-1}$ and $\omega_L = (10^{-4},10^{-3},10^{-2})\tau^{-1}$.} 
\label{Fig: WAMF_O2}
\end{figure}
%\caption{Construction of WAMF$_{0:7}^{(2)}$ for dephasing noise filtering. a) Log-scale color plot of the cost function $A_z(\pi;\WAMampP_5, \WAMampP_6;3\pi)$ integrated over $\omega\in [10^{-9}, 10^{-1}]\tau^{-1}$ and computed over a two-dimensional cross-section in the $\WAMampP_5\WAMampP_6$ plane. Blue regions indicate paths over which the cost function may be minimized to \emph{second-order}. ``Cross-region'' (circled) indicates regimes in which local minima are robust to errors some subset of Walsh spectra. b-d) Representative amplitude-modulated profiles for spectrally-optimized 8-segment WAMF$_{0:7}^{(2)}$ gates. Vertical axis represents $\Omega$, the Rabi rate per time step. e-g) Corresponding optimized Walsh spectra, obtained using Nelder-Mead seach over the $\boldsymbol{\WAMampP_\nu}$ control space. h) Corresponding dephasing filter-transfer functions, compared against primitive $\pi_x$ pulse and and effective $\pi_x$ WAMF$_{0,3}^{(1)}$ gate. }  
\label{Fig: WAMF_O2}

Representative examples of spectrally-optimized WAMF$_{0:7}^{(2)}$ constructions are shown in Fig. \ref{Fig: WAMF_O2}.  The 8-segment time-domain amplitude-modulated profiles are represented in Fig. \ref{Fig: WAMF_O2}a, with corresponding Walsh spectra shown in Fig. \ref{Fig: WAMF_O2}b. The blue, red and green spectra were obtained using a Nelder-Mead optimization of the cost function $A_z(X_3,X_5,X_6)$ defined over $[\omega_L,\omega_c]$ with $\omega_c = \tau^{-1}$ and $\omega_L = (10^{-4},10^{-3},10^{-2})\tau^{-1}$ respectively. The corresponding dephasing filter-transfer functions $\Fd(\omega)$ are plotted as solid  blue, red and green traces in Fig. \ref{Fig: WAMF_O2}d. Within the respective cost function bands these satisfy $\Fd\propto(\omega\tau)^{2p^*}$, with the \emph{instantaneous  filter order}  ranging between $2<(p^*-1)<3.8$ at various points.\footnote{Since the  WAMF$_{0:7}^{(2)}$ gates in Fig.  \ref{Fig: WAMF_O2} were derived by optimizing the cost function over local regions in the stopband, the asymptotic filter order $(p-1)$ associated with Taylor expanding $\Fd(\omega)$ about $\omega = 0$ is not a meaningful descriptor of these filters. Hence we do not expect $C^{(z)}_{2,4} = 0$ and do not pursue such a calculation. Instead the instantaneous filter order is used.}
%implying a \emph{local filter order} between second- and third-order. 
For comparison we also plot the dephasing filter-transfer function for a primitive $\pi$ rotation (black dashed trace) and an optimized WAMF$_{0,3}^{(1)}$ gate (yellow dashed trace). These respectively satisfy  $\Fd\propto(\omega\tau)^{2,4}$ over the whole stopband and are well characterized by the (asymptotic) filter orders $(p-1)=0,1$.
%For the WAMF$_{0:7}^{(2)}$ gates we also see very steep notches in the viscinity of discontinuities in the filter-transfer functions. These are due to having defined (and minimized) the cost function over a narrow spectral band $[\omega_L,\omega_c]$ around these notches. 

Fig \ref{Fig: WAMF_O2}a shows a two-dimensional representation of $A_z(\WAMampP_5,\WAMampP_6)$ defined on a two-dimensional cross-section of the $\boldsymbol{\WAMampP}_\nu$-domain, holding $\WAMampP_3$ fixed, and integrated over the stopband $\omega\in [10^{-9}, 10^{-1}]\tau^{-1}$. The value of $\text{Log}_{10}\big[A_z(\WAMampP_5,\WAMampP_6)\big]$ is indicated by the color scale. Areas in blue indicate minima in  $A_z(\WAMampP_5,\WAMampP_6)$, indicating optimized paths in $\WAMampP_5\WAMampP_6$ plane the over which effective filters may be found. Notably, it is possible to find ``cross-regions'' (circled) in which the spectral amplitudes $\WAMampP_5$ and $\WAMampP_6$ may independently be varied substantially without the cost function moving off a local minima. This potentially indicates the existence of classes of WAMFs which may be robust to errors in the Walsh spectrum itself.

%__________________________________________________________________________________________

%				SECTION: WALSH PHASE MODULATED FILTERS
%__________________________________________________________________________________________

\section{Walsh Phase Modulated Filters (WPMFs)}\label{Sec:WPMFs}

We now turn to filters for amplitude-damping noise constructed via phase modulation using the WPMF construction set out in Eq. \ref{Eq:WPMFGeneralForm}.  Following the same procedure described above for WAMFs, one can implement a Nelder-Mead search to derive spectrally-optimized WPMFs which implement nontrivial rotations. For these constructions, however, the target rotation angle is dependent on both the Rabi rate and the sequence of phase modulations. Consequently it is less straightforward to impose a constraint during the optimization procedure to ensure a particular target rotation. Although we do not pursue the general problem in detail in this paper, we demonstrate the approach in this and the following sections, deriving a family of WPMFs in which the synthesis space is limited to a variety of simple combinations of Walsh function. 

In the remainder of this section we study a variation on the strict WPMF structure which resolves the difficulty of imposing a target rotation and enables us to make some useful connections with existing composite pulse sequences in NMR. This variation involves partitioning the control modulation into a target rotation $P(\theta,0)$ followed by a sequence of phase-modulated \emph{identity} operations $\prod_{l = 1}^{\MinDim} P(2\pi,\phi_l)$ with the $\phi_l$ chosen such that amplitude noise is filtered to some order. Here the operator $P(\theta,\phi)$ denotes the rotation through angle $\theta$ about $\sig_\phi$. By insisting these $\MinDim$ ``correction'' segments are all identity operations, the phase modulations do not produce complicated rotation paths and the net rotation is determined simply by the initial target pulse. 

We assume a constant Rabi rate $\RabiWSK$ so that each correction segment has equal duration $\tau_{2\pi} = 2\pi/\RabiWSK$. Provided $\MinDim$ is a power of 2, the phase modulation describing the correction sequence may therefore be constructed as a Walsh-synthesized waveform consisting of Paley orders $k\le\MinDim-1$. The simplest such ``synthesis'' derives from a single Walsh function $\PAL_k(x)$ with spectral weight $Y_k$, yielding the sequence
\begin{align}\label{Eq:WAMF(SK1)GeneralForm}
\text{WPMF}^{(c)}_{k}(\theta) \equiv \hspace{0.09cm}
  \begin{blockarray}{ccccc}
    & \text{\scriptsize $\Omega_l$\normalsize}  &\text{\scriptsize $\tau_l$\normalsize}  & \text{\scriptsize $\theta_l$\normalsize}  &\text{\scriptsize $\phi_l$\normalsize} \\
    \begin{block}{c@{\hspace*{4pt}}[ccc|c@{\hspace*{5pt}}]}
   \text{\scriptsize $P_1$\normalsize} & {\hspace*{6pt}}\RabiWSK  &\tau_\theta&\theta& \BAmulticolumn{4}{c}{\multirow{1}{*}{$0$}}\\
    \cline{2-10}% don't use \hline
    \text{\scriptsize $P_2$\normalsize}&{\hspace*{6pt}}\RabiWSK  &\tau_{2\pi}&2\pi& \BAmulticolumn{4}{c}{\multirow{3}{*}{$\vec{\boldsymbol{\phi}}_c$}}\\
    \vdots  & \vdots &\vdots&\vdots& &&&&&&&\\
    \text{\scriptsize $P_{\MinDim+1}$\normalsize}  & {\hspace*{6pt}}\RabiWSK &\tau_{2\pi}&2\pi& &&&&&&&\\
    \end{block}
  \end{blockarray},\hspace{1cm}\vec{\boldsymbol{\phi}}_c = \WPMampP_k\boldsymbol{P}^{(k)}_{\MinDim},\hspace{0.9cm}\MinDim(k)=2^{m(k)}
\end{align}
where, as in Eq. \ref{VectorizedPAL}, the column vector $\boldsymbol{P}^{(k)}_{\MinDim}$ specifies the sequence of values taken by $\PAL_k(\frac{t-\tau_\theta}{\tau-\tau_\theta})$ over the interval $[\tau_\theta,\tau]$ partitioned into the minimum $\MinDim(k)$ equal time bins. We include the superscript $(c)$ and write the vectorized phase $\boldsymbol{\vec{\phi}}_c$ to indicate Walsh modulation during the ``correction stage'' of the sequence,  disambiguating this from the strict WPMF structure. For a given $\theta$ we may now treat $\WPMampP_k$ as a tuning parameter which may be optimized by minimizing the cost function $A_\Omega(Y_k;\theta)\equiv\int_0^{\omega_c}d\omega \Fa(\omega\tau;Y_k;\theta)$. The optimized $\WPMampP_k$ are thereby defined as an implicit function of $\theta$. 

In fact we may analytically show this construction yields first-order filters for amplitude noise by Taylor expanding $\Fa(\omega)$ and solving the first-order filter condition $C_2^{(\Omega)}(\WPMampP_k;\theta) = 0$. 
%Numerical search over the $\WPMampP_k\theta$ plane indicates the cost function can be minimized to first-order, implying first-order amplitde noise filters. 
We compute $C_2^{(\Omega)}{(\WPMampP_k;\Omega)} = \big[\theta+2\pi\MinDim(k)\cos(\WPMampP_k)\big]^2/4$, implying the optimized Walsh spectral amplitude
\begin{align}\label{Eq: WPMF(SK1)_OmptimumAmp}
\WPMampP_k(\theta) = \cos^{-1}\Big(-\frac{\theta}{2\pi\MinDim(k)}\Big).
\end{align}
%\begin{align}
%&k=1\hspace{1cm}&&C_2^{(\Omega)} = \frac{1}{4}(\theta+4\pi\cos(\WPMampP_k))^2\\
%&k=2,3\hspace{1cm}&&C_2^{(\Omega)} = \frac{1}{4}(\theta+8\pi\cos(\WPMampP_k))^2\\
%&k=4,5,6,7\hspace{1cm}&&C_2^{(\Omega)} = \frac{1}{4}(\theta+16\pi\cos(\WPMampP_k))^2\\
%&...&&...\\
%&k=.\hspace{1cm}&&C_2^{(\Omega)} = \frac{1}{4}(\theta+2^{m(k)+1}\pi\cos(\WPMampP_k))^2\\
%\end{align}
%\begin{align}
%&k=1\hspace{1cm}&&\Phi^\text{(DC)}_1 = \frac{1}{2}(\theta+4\pi\cos(Y_k))\\
%&k=2,3\hspace{1cm}&&\Phi^\text{(DC)}_1 = \frac{1}{2}(\theta+8\pi\cos(Y_k))\\
%&k=4,5,6,7\hspace{1cm}&&\Phi^\text{(DC)}_1 = \frac{1}{2}(\theta+16\pi\cos(Y_k))\\
%&...&&...\\
%&k=.\hspace{1cm}&&\Phi^\text{(DC)}_1 = \frac{1}{2}(\theta+2\MinDim(k)\pi\cos(Y_k))\\
%\end{align}

\begin{table}[bp]
  \centering
\renewcommand{\arraystretch}{1.3}
\begin{tabular}{c|cccc||cc|}
\cline{2-7}
 & \multicolumn{4}{c||}{WPMF$^{(c)}_k$  Construction}  &\multicolumn{2}{c|}{Amplitude Errors}\\
\cline{2-7}
 &{$\hspace{0.1cm}k\hspace{0.1cm}$}  & $\MinDim(k)$ &$\vec{\boldsymbol{\phi}}$ &$\WPMampP_k(\theta)$  &$(\mu-1)$ &  $(p-1)$  \\
\hline
\multicolumn{1}{|c|}{SK1$(\theta)$}& $1$ &$2$  & $\WPMampP_1(\theta)\PAL_1$ &$\cos^{-1}\big(-\theta/4\pi\big)$  &$1$ &$1$ \\
\multicolumn{1}{|c|}{P2$(\theta)$}& $3$ &$4$  & $\WPMampP_3(\theta)\PAL_3$ &$\cos^{-1}\big(-\theta/8\pi\big)$   &$2$ &$1$ \\
\hline
\end{tabular}
\caption{Filter characteristics of WPMF$^{(c)}_k$ constructions corresponding to well-known NMR sequences, SK1 and P2, originally designed to compensate for \emph{static} amplitude errors to first and second \emph{Magnus order} respectively (see \ref{AppSec:FiltervsMagnusOrder}).}\label{Table:WalshNMRSequences}
\end{table}
\noindent On the other hand, computing the second-order Taylor coefficient and substituting in the first-order-optimized spectral value in Eq. \ref{Eq: WPMF(SK1)_OmptimumAmp}, one finds $C_4^{(\Omega)}\ne0$. Hence the WPMF$^{(c)}_k$ sequence has $(p-1) = 1$ filtering properties. 

We make the interesting observation that Eqs. \ref{Eq:WAMF(SK1)GeneralForm} and \ref{Eq: WPMF(SK1)_OmptimumAmp} produce the first-order Solovay-Kitaev SK1($\theta$) sequence~\cite{Brown2004,Brown2005}  and the second-order Wimperis passband P2($\theta$) sequence~\cite{MerrillArXv2012} by setting $k = 1,3$ respectively. Hence these well-known NMR sequences, originally designed to compensate for \emph{static} amplitude errors to first and second \emph{Magnus order} respectively (see \ref{AppSec:FiltervsMagnusOrder}), appear in the Walsh filter space as phase-modulated filters for non-Markovian amplitude noise. Table \ref{Table:WalshNMRSequences} summarizes this. 

Another remarkable result is found using the synthesis $\vec{\boldsymbol{\phi}}_c = H_\MinDim\boldsymbol{\WPMampH}$ over the two Walsh functions $\PAL_0$ \& $\PAL_3$, setting $\boldsymbol{\WPMampH} \equiv \big(\WPMampP_0,\hspace{0.1cm}0,\hspace{0.1cm}0,\hspace{0.1cm}\WPMampP_3\big)^T$ in analogy with the Walsh spectrum defining amplitude modulation in the WAMF$_{0,3}^{(1)}$ construction. The first-order filtering condition $C_2^{(\Omega)}(\WPMampP_0,\WPMampP_3,\Omega_0;\theta) = 0$ then implies solutions
\begin{align}
\boldsymbol{\WPMampH} = \big(2\phi_\text{BB1},\hspace{0.1cm}0,\hspace{0.1cm}0,\hspace{0.1cm}-\phi_\text{BB1}\big)^T,\hspace{1cm}\phi_\text{BB1}(\theta) \equiv \cos^{-1}\big(-\frac{\theta}{4\pi}\big),\hspace{1cm}\Omega_0 = \frac{4\pi+\theta}{\tau}%\Omega_0 = 4\pi+\theta$, $(\WPMampP_0,\WPMampP_3) = (2,-1)\phi_\text{BB1}$, where 
\end{align}
yielding the Wimperis broadband BB1 sequence ~\cite{WimperisBB11994}
\begin{align}
\text{BB1}(\theta)\hspace{0.25cm}=\hspace{0.25cm}
\begingroup
  \renewcommand*{\arraystretch}{1.5}%
  \kbordermatrix{
           & \Omega_l           & \tau_l           & \theta_l           & \phi_l     \cr
    P_1   &\Omega_0             & \tau_\theta       & \theta         & 0 \cr
    P_2    &\Omega_0             & \frac{\tau-\tau_\theta}{4}       &\pi        & \WPMFPhasePlus \cr
    P_3    & \Omega_0             &\frac{\tau-\tau_\theta}{4}        & \pi    & \WPMFPhaseMin \cr
    P_4    & \Omega_0             &\frac{\tau-\tau_\theta}{4}        & \pi    & \WPMFPhaseMin \cr
    P_5   &\Omega_0            & \frac{\tau-\tau_\theta}{4}       & \pi       & \WPMFPhasePlus \cr
  }%
\endgroup\hspace{0.25cm}\cong\hspace{0.25cm}
\begingroup
  \renewcommand*{\arraystretch}{1.5}%
  \kbordermatrix{
           & \Omega_l           & \tau_l           & \theta_l           & \phi_l     \cr
    P_1   &\Omega_0             & \tau_\theta       & \theta         & 0 \cr
    P_2    &\Omega_0             & \frac{\tau-\tau_\theta}{4}       &\pi        & \phi_\text{BB1} \cr
    P_3    & \Omega_0             &\frac{\tau-\tau_\theta}{2}        & 2\pi    & 3\phi_\text{BB1} \cr
    P_4   &\Omega_0            & \frac{\tau-\tau_\theta}{4}       & \pi       & \phi_\text{BB1} \cr
  }%
\endgroup.
\end{align}
Here $\WPMampP_{\pm} := \WPMampP_0\pm\WPMampP_3$ and in the last equality we have collapsed the array to show the BB1 construction explicitly. Computing the second-order Taylor coefficient and in substituting the first-order-optimized spectral values, however, we find $C_4^{(\Omega)}\ne0$. Thus, although BB1 was originally derived to compensate \emph{static} amplitude errors to second \emph{Magnus order}, it only provides \emph{first-order} noise filtering. 

%\begin{table}[bp]
%  \centering
%\renewcommand{\arraystretch}{1.3}
%\begin{tabular}{c|ccccc||cc|}
%\cline{2-8}
% & \multicolumn{5}{c||}{Walsh Construction}  &\multicolumn{2}{c|}{Amplitude Errors}\\
%\cline{2-8}
% &{$\hspace{0.1cm}k\hspace{0.1cm}$}  & $\MinDim(k)$ &  $\text{WCPM}_{k}(\theta)$ &$\vec{\boldsymbol{\phi}}$ &$\WPMampP_k(\theta)$  &$\mu-1$ &  $p-1$  \\
%\hline
%\multicolumn{1}{|c|}{SK1$(\theta)$}& $1$ &$2$ &$\text{WCPM}_{1}(\theta)$ & $\WPMampP_1(\theta)\PAL_1$ &$\cos^{-1}\big(-\theta/4\pi\big)$  &$1$ &$1$ \\
%\multicolumn{1}{|c|}{P2$(\theta)$}& $3$ &$4$ &$\text{WCPM}_{3}(\theta)$  & $\WPMampP_3(\theta)\PAL_3$ &$\cos^{-1}\big(-\theta/8\pi\big)$   &$2$ &$1$ \\
%\hline
%\end{tabular}
%\caption{Well-known NMR sequences, SK1 and P2, originally designed to compensate for amplitude errors to orders $\mu-1 = 1,2$ respectively. These sequences turn up in Walsh basis as WAMFs performing time-domain amplitude noise filtering to order $p-1 = 1$.}\label{Table:WalshNMRSequences}
%\end{table}

%__________________________________________________________________________________________

%						SECTION: WALSH ROTARY SPIN ECHO
%__________________________________________________________________________________________

\section{Walsh Rotary Spin Echo (WRSE)}\label{Sec:WRSE}

In this section we treat a sub-class of Walsh modulated filters which may be described either in terms of phase- or amplitude-modulation. The phase-modulation consists of applying a sequence of $\pi$-phase shifts, relative to some offset $\phi_0$, on the driving field with a constant amplitude $\RabiWPMF$.  This construction generalizes the rotary spin echo (RSE) sequence from NMR, analogous to the Hahn-echo sequence for driven systems, consisting of a single $\pi$-phase shift applied at the sequence midpoint $\tau/2$.  In quantum information RSE has been employed, for example, in relaxation noise spectroscopy~\cite{YanNature2013} and in mitigating low-frequency off-resonant noise~\cite{Gustavsson2012} in driven superconducting flux qubits.  In contrast with previous approaches, our generalization permits higher-order filter performance in both amplitude and dephasing quadratures and may prove of significant use. 

\begin{figure}[h]
\centering
\includegraphics[width=0.9\columnwidth]{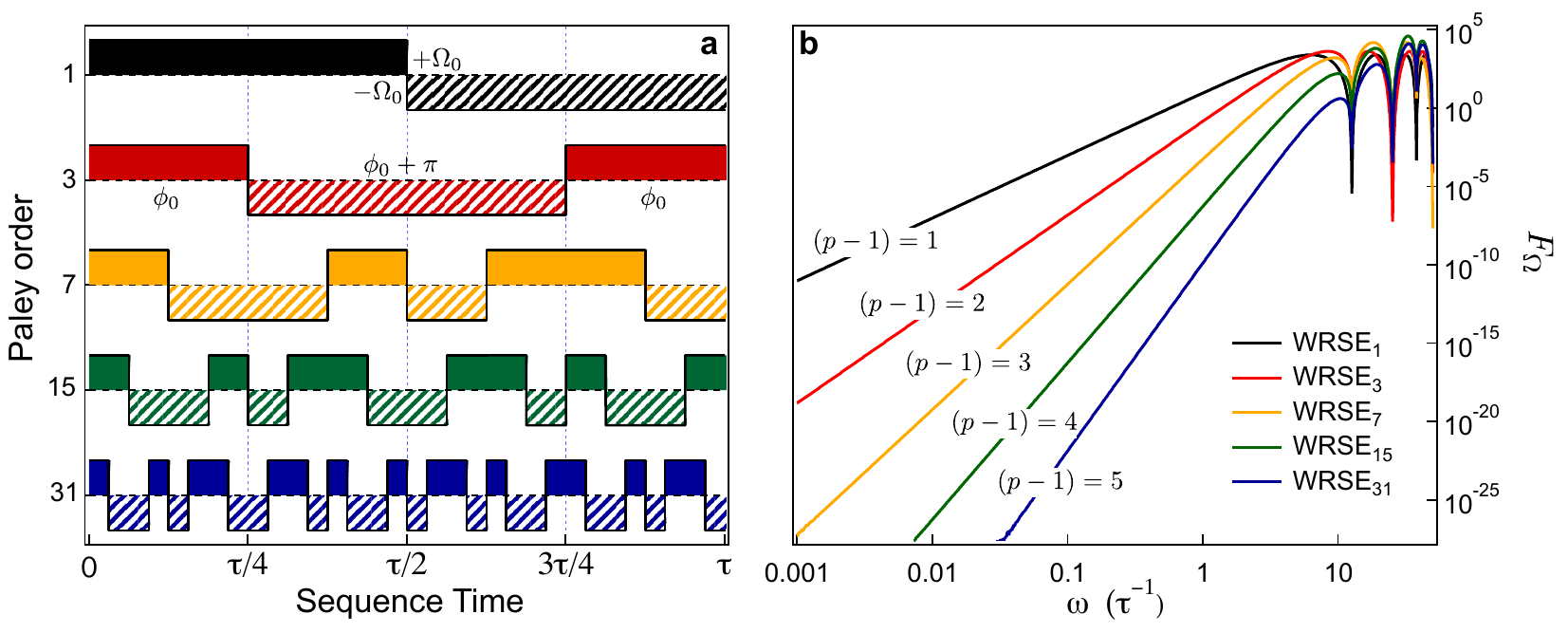}
\caption{WRSE$_k$ amplitude noise filter characteristics, $k\in\{1,3,7,15,31\}$. \textbf{a}) Modulation profiles of WRSE$_k$ sequences. Amplitude modulation involves switching the sign $\pm\Omega_0$; this is equivalent to holding $\Omega_0$ constant and instead shifting the phase $\phi_0$ by $\pi$, indicated by the hatching. \textbf{b}) Corresponding amplitude filter-transfer functions $\Fa(\omega)$, showing filter order increases with Hamming weights $r(k)\in\{1,2,3,4,5\}$.} \label{Fig:WRSEAmplitude}
\end{figure}

In our construction the temporal profile of the phase is expressed $\phi(t) = \phi_0+\frac{\pi}{2}(1-y(t))$, where  $y(t)\in\{\pm1\}$ is a binary-valued \emph{switching function} defined to change sign at the application of each $\pi$-shift. Specifically, we consider sequences based on Walsh functions by defining the switching function 
\begin{align}
y(t):= \PAL_k(t/\tau),\hspace{1cm}t\in\{0,\tau\}
\end{align}
Consequently the phase has Walsh synthesis $\phi(t) = \WPMampP_0\PAL_0(t/\tau)+\WPMampP_k\PAL_k(t/\tau)$   where $\WPMampP_0 = \phi_0+\frac{\pi}{2}$ and $\WPMampP_k = -\frac{\pi}{2}$. These gates only perform the identity operation with $\Theta \equiv 0$ owing to the property that $\sig_{\phi_0+\pi} = -\sig_{\phi_0}$ (see Eq. \ref{SpinOperatorDefinition}). This implies the direction of unitary rotation is reversed by the application of each $\pi$ shift, and since any Walsh function of Paley order $k>0$ is equally distributed between values $\pm 1$ over the domain these rotations perfectly cancel, yielding zero total rotation. This is formally equivalent to modulating the \emph{sign} of the Rabi rate and holding the phase $\phi_0$ constant (see \ref{AppSec:WRSEDerivations}), as schematically illustrated in Fig. \ref{Fig:WRSEAmplitude}a. These sequences, referred to as \emph{Walsh rotary spin echo} order $k$ (WRSE$_k$), thus take the form 
\begin{align}\label{Eq:WRSEPulseForm}
\text{WRSE}_k \equiv \prod_{l=1}^{\MinDim(k)}\exp\Big(i\frac{\RabiWPMF}{2}\tauMIN\sig_{\phi_l}\Big),\hspace{1cm}\vec{\boldsymbol{\phi}} = \big(\phi_1,\hspace{0.1cm}\phi_2,...,\phi_\MinDim\big)^T=H_\MinDim\boldsymbol{\WPMampH}
\end{align}
where, referring to Eqs. \ref{Eq:WPMFGeneralForm} and \ref{fHadRep}, $\MinDim(k)\equiv2^{m(k)}$ and the Walsh spectrum $\boldsymbol{\WPMampH}$ has only two nonzero elements: $\WPMampH_1 =  \phi_0+\frac{\pi}{2}$ and $\WPMampH_{i(k)} = -\frac{\pi}{2}$. This naming convention reflects the fact that the WRSE family generalize the traditional RSE sequence: in particular, WRSE$_1\equiv$ RSE. We now proceed to derive the filtering properties of the WRSE family, starting with the amplitude quadrature then moving to the dephasing quadrature.

%				-------------------------------------------------
%				sub:SECTION: WRSE as Amplitude Filters
%				-------------------------------------------------

\subsection{WRSE as Amplitude Filters}

The amplitude filter-transfer function, in the stopband, is determined by the Hamming weight of the chosen Walsh function by
\begin{align}\label{AmplitudeFilterFunctionRollOff}
\Fa(\omega\tau) \propto (\omega\tau)^{2(r(k)+1)}.
\end{align}
Comparing this with the low-frequency approximation $\Fa(\omega\tau;k) \propto (\omega\tau)^{2p}$ from Eq. \ref{pOrderFilterConditions}, we conclude the time-domain filter order is given by $(p-1) = r(k)$. That is, high-pass filter performance is completely determined by the Hamming wieght. Fig. \ref{Fig:WRSEAmplitude}c demonstrates this by plotting $\Fa(\omega)$ for the WRSE$_k$ sequences with $k\in\{1,3,7,15,31\}$, corresponding to the Hamming weights $r(k)\in\{1,2,3,4,5\}$. The corresponding filter order increase is clear from the steepening roll-off. 

\noindent
This result follows from deriving $\Fa(\omega)$ for Eq. \ref{Eq:WRSEPulseForm} which, owing to the fact that the noise Hamiltonian in this quadrature always commutes with control operations, takes the relatively simple analytical form (see \ref{AppSubSec:WRSEAmpFF})
\begin{align}\label{Eq:WRSEAmpFF}
\Fa(\omega) =\frac{\RabiWPMF^2}{4}\Big|
\sum_{l=1}^\MinDim\Pk_l
\Big[
e^{i\omega t_{l-1}} - e^{i\omega t_l}
\Big]
\Big|^2.
\end{align}
Our key insight now is to observe the sum inside the modulus square above satisfies $\sum_{l=1}^\MinDim\Pk_l\big[e^{i\omega t_l}-e^{i\omega t_{l-1}}\big] = i\omega\tau\FT_x\big[\PAL_k(x)\big]$, where $x\equiv t/\tau$ is a non-dimensional time-domain variable. We now invoking Eq. \ref{Eq:HayesFourier-PALIdentity} in design rule (iii), namely $\FT_x\big[\PAL_k(x)\big]\propto (\omega\tau)^{r(k)}$, to map the low-frequency spectral properties onto the Hamming weight $r(k)$ of the chosen Walsh function. Substituting this into Eq. \ref{Eq:WRSEAmpFF} then yields Eq. \ref{AmplitudeFilterFunctionRollOff}. 

%i\omega\tau\FT_x\big[\PAL_k(x)\big] = \sum_{l=1}^\MinDim\Pk_l\big[e^{i\omega t_l}-e^{i\omega t_{l-1}}\big] %\int_0^1 \PAL_k(x)e^{i\omega\tau x}dx 
%\end{align}
%where $\FT_x[\cdot]$ denotes the foreward fourier transform from the $x = t/\tau$ domain, to Fourier domain associated with the nondimensional angular frequency $\omega\tau$. We now invoke a useful results proved by \emph{Hayes et al.} ~\cite{HayesPRA2011} that 
%\begin{align}\label{Eq:HayesFourier-PALIdentity}
%\FT_x\big[\PAL_k(x)\big]\propto (\omega\tau)^{r(k)},
%\end{align}
%This was originally proved by \emph{Hayes et al.} ~\cite{HayesPRA2011} to establish the relationship between the order of (dephasing) error suppression achieved by so-called Walsh Dynamic Decoupling (WDD$_k$) sequences, and the Hamming weight $r(k)$ of the Walsh function $\PAL_k$ there used to sequence a train of instantaneous $\pi$-pulses. 

%				-------------------------------------------------
%				sub:SECTION: WRSE as Dephasing Filters
%				-------------------------------------------------

\subsection{WRSE as Dephasing Filters}

%Fig: Taylor Expansion Coefficients 2,4,6 for RSE PAL(3) 

The dephasing filter performance for WRSE$_k$ is more complicated to study as  noise terms in this quadrature do not commute with our control, obfuscating a compact expression for $\Fd(\omega)$. It is convenient instead to study the zeros of the Taylor coefficients $C^{(z)}_{2j}$ as in Eq. \ref{pOrderFilterConditions}. Since the Rabi rate is the only free variable in the WRSE$_k$, for a given $k$, it follows the Taylor coefficients are functions only of $\RabiWPMF$. Filtering to order $(p-1)$ then corresponds to the condition
\begin{figure}[]
\centering
\includegraphics[width=0.92\columnwidth]{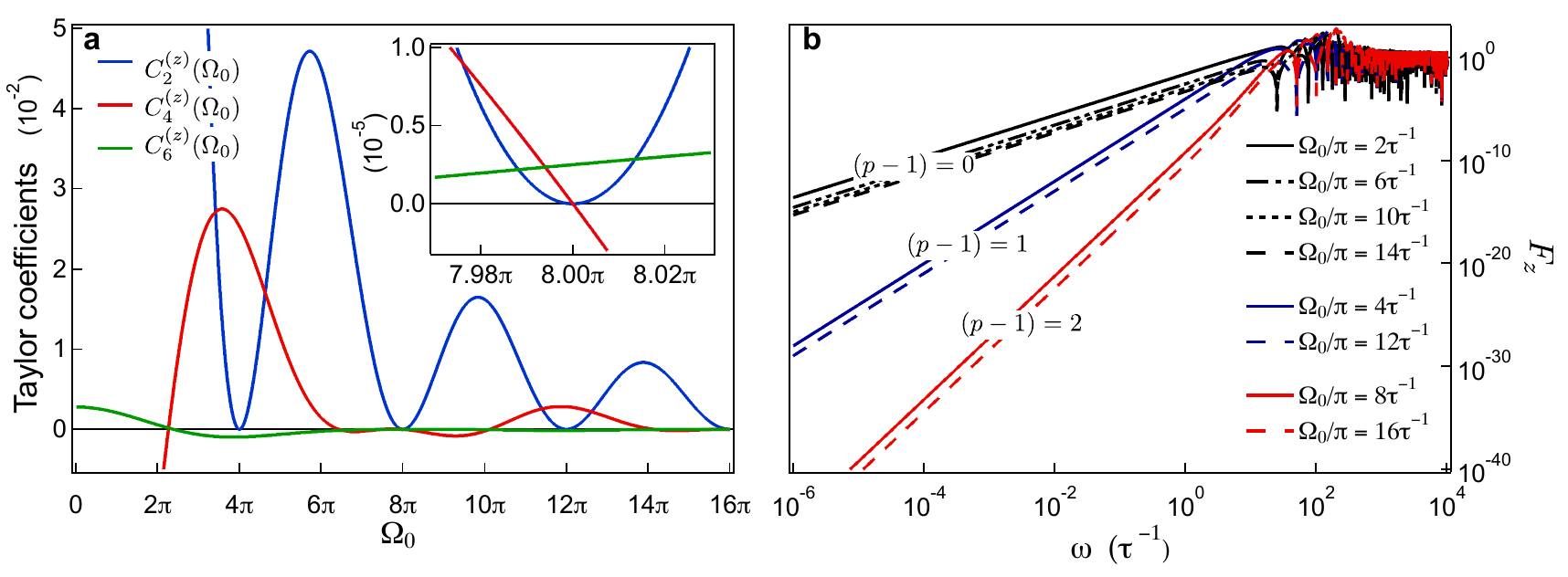}\\
\caption{WRSE$_3$ dephasing noise filter characteristics. \textbf{a}) Taylor expansion coefficients $C^{(z)}_{2,4,6}(\Omega_0;3)$ for  $\Fd(\omega)$. The inset shows typical behaviour: $\Omega_0=8\pi$ is a concurrent zero of $C^{(z)}_{2,4}(\Omega_0;3)$ , but not of $C^{(z)}_{6}(\Omega_0;3)$. Hence WRSE$_3$ can only filter up to second-order. \textbf{b}) Dephasing filter-transfer functions for WRSE$_3$ corresponding to $\Omega_0 = 2\pi q$, $q\in\{1,..,8\}$. When $\Omega_0$ is a multiple of $8\pi$ we achieve second-order filters, that is $(p-1) = 2$.} \label{WRSE_C246_PAL(3)}
\end{figure}
\begin{align}\label{Eq:WRSEDephFitlerCondition}
C^{(z)}_2(\eta;k) = C^{(k)}_4(\eta;k) = ... = C^{(z)}_{2(p-1)}(\eta;k) = 0
\end{align}
where $\eta$ denotes some value of $\Omega_0$ for which the above coefficients are concurrently zero. Here we have included the Paley order as a parameter of the coefficients. Analysis shows, however, concurrent zeros exist only for $j\in\{1,2\}$. We effectively obtain the following ``no-go theorem'': WRSE$_k$ \emph{sequences perform as (high-pass) dephasing noise filters up to (but not beyond) second order}. This result may be of use to characterize the relevant quadrature of an unkown noise source, by probing with higher-order WRSE$_k$ sequences and determining the resulting filtering properties. 

The general insight supporting this statement is developed in \ref{AppSubSec:WRSEforDephasing}, by explicitly studying the representative case for WRSE$_3$. The Taylor coefficients $C^{(z)}_{2,4,6}$ for this representative case are plotted in Fig. \ref{WRSE_C246_PAL(3)}a as functions of $\Omega_0$. Zeros of $C^{(z)}_{2}$ occur at multiples of $4\pi$, and concurrent zeros of $C^{(z)}_{2,4}$ occur at multiples of $8\pi$. However, since $C^{(z)}_6$ is never zero at multiples of $8\pi$ (see inset to Fig. \ref{WRSE_C246_PAL(3)}a), it follows we can never achieve higher than second-order filtering. We verify this by examining the slope of $\Fd(\omega;\Omega_0)$ for the values $\Omega_0 = 2\pi q$, $q\in\{1,...,8\}$ (see Fig. \ref{WRSE_C246_PAL(3)}b). Similar considerations for other values of $k$ generalize the result.

\section{Universal Walsh Modulated Filters (UWMFs)}\label{Sec:UWMFs}

In the previous sections we have considered WAMF and WPMF gates which implement target qubit rotations while filtering, to some order, \emph{either} dephasing or amplitude noise respectively. In this section we derive filters for universal noise by concatenating both filter types into a composite structure that filters both noise quadratures simultaneously, while still implementing a target qubit rotation. We refer to such constructions as \emph{universal Walsh modulated filters} (UWMFs). 

The general approach is conveniently illustrated through a particular concatenated structure obtained by embedding the WPMF$^{(c)}_{1}(\theta)\equiv \text{SK1}(\theta)$ gate (Sec. \ref{Sec:WPMFs}) within the various segments of the WAMF$_{0,3}^{(1)}$ gate (Sec. \ref{SubSec:FirstOrderWAMFs}). The former is explicitly written 
\begin{align}\label{SK1CPSMatrix}
\text{WPMF}^{(c)}_1(\theta)\hspace{0.1cm}&\equiv\hspace{0.1cm}
\begingroup
  \renewcommand*{\arraystretch}{1.35}%
  \kbordermatrix{
           & \Omega_l           & \tau_l           & \theta_l           & \phi_l     \cr
    P_1    &  \Omega_0            &\tau_\theta       & \theta        & 0 \cr
    P_2    & \Omega_0             &\tau_{2\pi}        &2\pi     & \phiSK \cr
    P_3    & \Omega_0             &\tau_{2\pi}        &2\pi        & -\phiSK \cr
  }%
\endgroup
%\hspace{0.5cm}&\phiSK(\theta): = \cos^{-1}\Big(-\frac{\theta}{4\pi}\Big)
\end{align}
\begin{align}
\Omega_0 = \frac{\theta+4\pi}{\tauSK},\hspace{0.7cm}\tau_\theta = \frac{\theta}{\Omega_0},\hspace{0.7cm}\phiSK(\theta)\equiv Y_1(\theta) = \cos^{-1}\Big(-\frac{\theta}{4\pi}\Big).
\end{align}

\begin{figure}[tp]
\centering
\includegraphics[width=0.7\columnwidth]{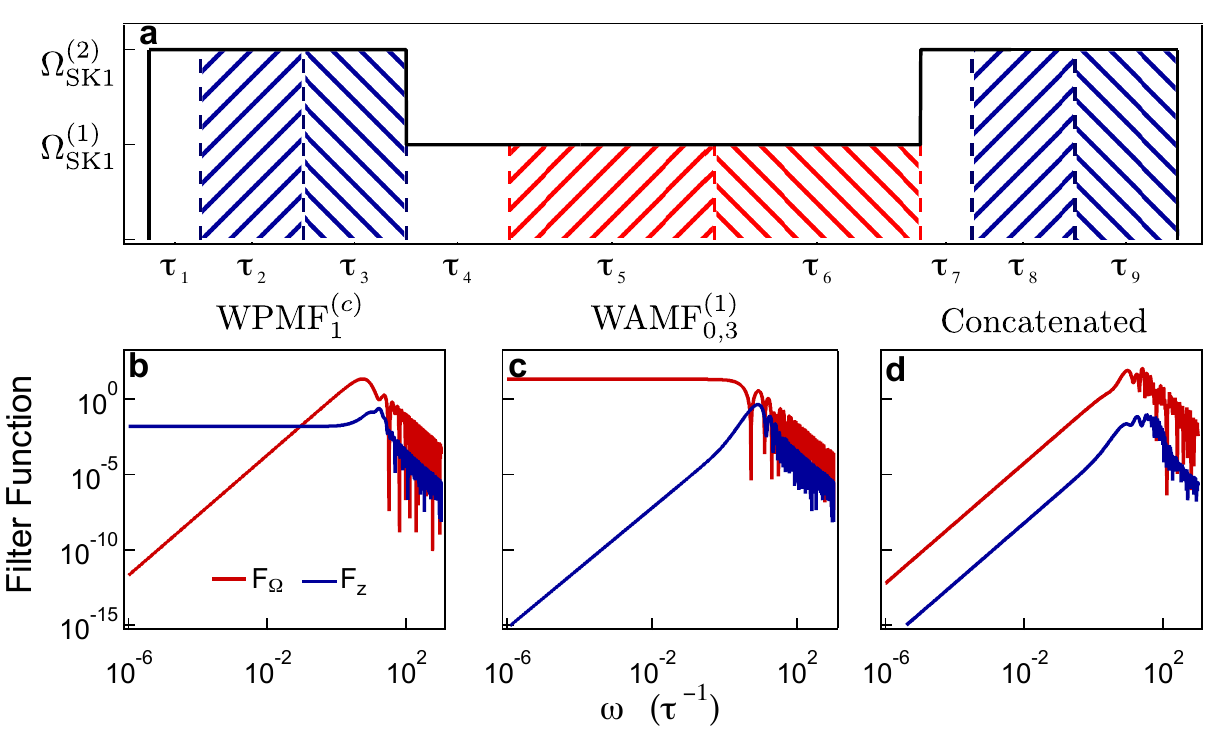}
\caption{\label{Fig:Concatenation}Concatenation scheme for universal noise suppression. \textbf{a}) Concatenation of WPMF$^{(c)}_1\equiv$SK1 within WAMF$_{0,3}^{(1)}$ sequence yielding UWMF$_{1,SK1}$. White fill indicates rotations enacted with $\phi = 0$; orientation of hatching denotes SK1 phase flips $\phi = \pm \phiSK$. \textbf{b}) filter-transfer functions for WPMF$^{(c)}\equiv$SK1 sequence. \textbf{c}) filter-transfer functions for four-segment WAMF$_{0,3}^{(1)}$ sequence. \textbf{d}) filter-transfer functions for concatenated sequence.}
\end{figure}
\noindent where $\tauSK$ denotes the total implementation time and the subscripts refer to to the equivalence, as in Table \ref{Table:WalshNMRSequences}, of this filter to the phase-modulated 3-pulse NMR sequence SK1$(\theta) := P_3(2\pi,-\phiSK)P_2(2\pi,\phiSK)P_1(\theta,0)$. Again, $P(\theta,\phi)$ denotes the rotation through angle $\theta$ about $\sig_\phi$. The dephasing- and amplitude-quadrature filter-transfer functions for Eq. \ref{SK1CPSMatrix} are shown in Fig. ~\ref{Fig:Concatenation}b, demonstrating first-order filtering of amplitude but not dephasing noise. Referring to Eq. \ref{WAMFOrder1CPSMatrix},  the amplitude modulated WAMF$_{0,3}^{(1)}$ filter is similarly written $P_3(\frac{\WAMFRabiPlus}{4},0)P_2(\frac{\WAMFRabiMin}{2},0)P_1(\frac{\WAMFRabiPlus}{4},0)$. As a reminder, the noise-filtering performance of WAMF$_{0,3}^{(1)}$ is shown in Fig. ~\ref{Fig:Concatenation}c, demonstrating first-order filtering of dephasing but not  amplitude noise.

The basic concatenation procedure is now to replace each constant-amplitude pulse in WAMF$_{0,3}^{(1)}$ with the constant-amplitude phase-modulated sequence implementing the equivalent rotation. Doing so effectively distributes the dephasing filter across the composite sequence, each subsequence of which filters amplitude noise. In our example, this takes place via the operator substitutions
\begin{align}
\defcol P_1\bla (\WAMFRabiPlus/4,0) \ampcol&\rightarrow \defcol \text{SK1}^{(1)}\bla(\WAMFRabiPlus/4),\hspace{1.5cm}\tauSK^{(1)} = \tau/4\\
\ampcol P_2\bla(\WAMFRabiMin/2,0) &\rightarrow  \ampcol \text{SK1}^{(2)}\bla(\WAMFRabiMin/2),\hspace{1.5cm}\tauSK^{(2)} = \tau/2\\
\defcol P_3\bla (\WAMFRabiPlus/4,0) &\rightarrow \defcol \text{SK1}^{(3)}\bla(\WAMFRabiPlus/4),\hspace{1.5cm}\tauSK^{(3)} = \tau/4
\end{align} 
where, as indicated, each SK1 is equal in duration to the original WAMF$_{0,3}^{(1)}$ pulse being replaced. We denote this concatenated structure by UWMF$_{1,\text{SK1}}$. The temporal profile is shown in Fig. ~\ref{Fig:Concatenation}a. The SK1 phase flips $\phi = \pm\phiSK$  are indicated by the oppositely oriented hatching within each constant-amplitude segment of the WAMF$_{0,3}^{(1)}$ envelope; $\phi = 0$ is indicated by white fill. Once again the total gate rotation is is determined by Eq. \ref{NetBlochRotationConstraint} and, as in Fig. \ref{Fig: WAMF_O1}, $\WAMampP_3$ may be treated as an independent tuning parameter to achieve first-order filtering against dephasing noise. The dephasing and amplitude filter-transfer functions for the concatenated and tuned sequence (in this case for a net $\pi$ rotation) are shown in Fig. ~\ref{Fig:Concatenation}d, indicating effective filtering of both amplitude and dephasing noise.  Below we detail two alternative constructions for realizing the UWMF structure.  %Further details for the concatenation method are presented in \ref{AppSec:UWMFs}, where we detail two alternative constructions. 

\subsection{Concatenation Method 1: Constrain Sequencing of WAMF Envelope}
The amplitude-modulated envelope of the WAMF$_{0,3}^{(1)}$ construction, as defined by Eq. \ref{WAMFOrder1CPSMatrix}, may be viewed as a sequence of 3 piecewise-constant rotations $\vec{\boldsymbol{\theta}} =\big(\frac{\WAMFRabiPlus}{4},\hspace{0.15cm}\frac{\WAMFRabiMin}{2},\hspace{0.15cm}\frac{\WAMFRabiPlus}{4}\big)$, implemented over durations $\big(\frac{\tau}{4},\hspace{0.15cm}\frac{\tau}{2},\hspace{0.15cm}\frac{\tau}{4}\big)$ with Rabi rates $\Rabivec =\big(\frac{\WAMFRabiPlus}{\tau},\hspace{0.15cm}\frac{\WAMFRabiMin}{\tau},\hspace{0.15cm}\frac{\WAMFRabiPlus}{\tau}\big)$. In our first concatenation method each rotation $\theta_l$, $l\in\{1,2,3\}$, is replaced by a constant-amplitude phase-modulated operation SK1$(\theta_l)$, implemented over a duration $\tauSK^{(l)}$ equal to that of the original rotation. That is, we constrain $\big(\tauSK^{(1)},\hspace{0.15cm}\tauSK^{(2)},\hspace{0.15cm}\tauSK^{(3)}\big) = \big(\frac{\tau}{4},\hspace{0.15cm}\frac{\tau}{2},\hspace{0.15cm}\frac{\tau}{4}\big)$. The constant Rabi rate $\OmegaSK^{(l)} = \theta_l/\tauSK^{(l)}$ driving each SK1 sequence is thus given by 
\begin{align}
\defcol \OmegaSK^{(1)} \bla = \defcol\OmegaSK^{(3)} \bla= \frac{1}{\tau}(\WAMFRabiPlus+16\pi),\hspace{1cm} \ampcol\OmegaSK^{(2)}\bla &= \frac{1}{\tau}(\WAMFRabiMin+8\pi).
\end{align}
The composite structure may then be written
\begin{align}\label{UWMFMethod1GammaMatrix}
\text{UWMF}_{\text{SK1},1}\hspace{0.1cm}&\equiv\hspace{0.1cm}
\begingroup
  \renewcommand*{\arraystretch}{1.35}%
  \kbordermatrix{
           & \Omega_l           & \tau_l           & \theta_l           & \phi_l     \cr
     P_1    &  \defcol\OmegaSK^{(1)}            &\defcol\tau_1       & \defcol\frac{\WAMFRabiPlus}{4}       &\defcol 0 \cr
    P_2    & \defcol\OmegaSK^{(1)}             &\defcol\tau_2        &\defcol2\pi     & \defcol\phiSK^{(1)} \cr
    P_3    & \defcol\OmegaSK^{(1)}             &\defcol\tau_3        &\defcol2\pi        & \defcol-\phiSK^{(1)} \cr
  P_4    &  \ampcol\OmegaSK^{(2)}            &\ampcol\tau_4      & \ampcol\frac{\WAMFRabiMin}{2}       & \ampcol0 \cr
    P_5    & \ampcol\OmegaSK^{(2)}             &\ampcol\tau_5      &\ampcol2\pi     & \ampcol\phiSK^{(2)} \cr
    P_6    & \ampcol\OmegaSK^{(2)}             &\ampcol\tau_6        &\ampcol2\pi        & \ampcol-\phiSK^{(2)} \cr
  P_7    &  \defcol\OmegaSK^{(3)}           &\defcol\tau_7       & \defcol\frac{\WAMFRabiPlus}{4}        &\defcol 0 \cr
    P_8    & \defcol\OmegaSK^{(3)}             &\defcol\tau_8        &\defcol2\pi     &\defcol \phiSK^{(3)} \cr
    P_9    & \defcol\OmegaSK^{(3)}             &\defcol\tau_9        &\defcol2\pi        &\defcol -\phiSK^{(3)} \cr
  }%
\endgroup
%\hspace{0.5cm}&\phiSK(\theta): = \cos^{-1}\Big(-\frac{\theta}{4\pi}\Big)
\end{align}
%where the Rabi rates are given by  
 
%and the pulse durations $\tau_l = \theta_l/\Omega_l$,$\hspace{0.1cm}l\in\{1,...9\}$ are given by 
\begin{align}
\defcol\tau_{1,7} &= \frac{\tau}{4}\Bigg(\frac{\WAMFRabiPlus}{\WAMFRabiPlus+16\pi}\Bigg),
\hspace{1cm}\defcol\tau_{2,3,8,9}\bla= \tau\Bigg(\frac{2\pi}{\WAMFRabiPlus+16\pi}\Bigg)\\
\ampcol\tau_4&= \frac{\tau}{2}\Bigg(\frac{\WAMFRabiMin}{\WAMFRabiMin+8\pi}\Bigg),
\hspace{1.67cm}\ampcol\tau_{5,6}\bla= \tau\Bigg(\frac{2\pi}{\WAMFRabiMin+8\pi}\Bigg)
\end{align}
%
%\begin{align}
%&\defcol\tau_1= \frac{T}{4}\Bigg(\frac{a_+}{a_++16\pi}\Bigg),\hspace{1cm}\tau_2= T\Bigg(\frac{2\pi}{a_++16\pi}\Bigg),\hspace{1cm}\tau_3= T\Bigg(\frac{2\pi}{a_++16\pi}\Bigg)\\
%&\ampcol\tau_4= \frac{T}{2}\Bigg(\frac{a_-}{a_-+8\pi}\Bigg),\hspace{1.25cm}\tau_5= T\Bigg(\frac{2\pi}{a_-+8\pi}\Bigg),\hspace{1.25cm}\tau_6=T\Bigg(\frac{2\pi}{a_-+8\pi}\Bigg)\\
%&\defcol\tau_7= \frac{T}{4}\Bigg(\frac{a_+}{a_++16\pi}\Bigg),\hspace{1cm}\tau_8= T\Bigg(\frac{2\pi}{a_++16\pi}\Bigg),\hspace{1cm}\tau_9= T\Bigg(\frac{2\pi}{a_++16\pi}\Bigg)
%\end{align}
\begin{align}
\defcol\phiSK^{(1)}\bla = \defcol\phiSK^{(3)}\bla = \cos^{-1}\Big(-\frac{\WAMFRabiPlus}{16\pi}\Big),\hspace{1cm}
\ampcol\phiSK^{(2)} \bla= \cos^{-1}\Big(-\frac{\WAMFRabiMin}{8\pi}\Big).
\end{align}
%and the rotaion axis angles are given by 
Using this method one finds tunings of $\WAMampP_0$ and $\WAMampP_3$ such that both dephasing and amplitude noise are filtered to first-order, as in Fig. \ref{Fig:Concatenation}.  These do \emph{not}, however, correspond directly to the optimum Walsh coefficients found for simple WAMF$_{0,3}^{(1)}$ construction shown in Fig. \ref{Fig: WAMF_O1}a. Rather, an equivalent tuning plot may be generated over the $\WAMampP_0\WAMampP_3$ domain, essentially identical to Fig. \ref{Fig: WAMF_O1}a but with minima shifted by a constant factor. The second method, detailed below, involves a slightly different construction in which we recover the original WAMF$_{0,3}^{(1)}$ tuning plot. 

%				-------------------------------------------------
%				subsub:SECTION: SK1-WAMF Method 2
%				-------------------------------------------------

\subsection{Concatenation Method 2: Constrain Sequencing of Target SK1 Rotations}
In the second construction we impose the constraint that $\tau_1:\tau_4:\tau_7 = 1:2:1$. That is, the \emph{target rotations} within the three successive SK1 sequences follow the same timing sequence as the three constant-amplitude pulses being replaced, previously constituting the amplitude-modulated WAMG$_1$ envelope. Thus we write 
$\big(\tau_1,\hspace{0.15cm}\tau_4,\hspace{0.15cm}\tau_7\big) = \nu\big(1,\hspace{0.15cm}2,\hspace{0.15cm}1\hspace{0.15cm})$ where $\nu$ is some fraction of the total duration $\tau$ of the composite structure to be determined. 
%
%In the previous method, the \emph{non-identity} pulses in the SK1-WAMF construction occur at pulse numbers $(1,4,7)$. Note the durations of these pulses only satisfy the ratio
%if $\tau_4 = 2\tau_1 = 2\tau_7$, that is if
%\begin{align}
%\frac{T}{2}\Bigg(\frac{a_-}{a_-+8\pi}\Bigg) = 2\times\frac{T}{4}\Bigg(\frac{a_+}{a_++16\pi}\Bigg) 
%\end{align}
%One easily finds this is only true if 
%\begin{align}
%a_0 = 3a_3
%\end{align}
%As far as WAMFs are concerned, this is only true when we recover the original NOT DCG, for which $(a_0,a_3) = (3\pi,\pi)$. For arbitrary WAMFs \ref{NonIdentityRotationRatios} is not satisfied. This fact is at the heart of why method 1 fails to produce tunings of $a_0$ and $a_3$ which agree with those found produced in the contour plot in Fig. \ref{Fig: WAMF_O1}. We can overcome by approaching the construction from the opposite direction, starting with the constraing that

Now, we know pulses $(1,4,7)$ execute the rotations $\big(\frac{\WAMFRabiPlus}{4},\hspace{0.15cm}\frac{\WAMFRabiMin}{2},\hspace{0.15cm}\frac{\WAMFRabiPlus}{4}\big)$, given by the third row of Eq. \ref{WAMFOrder1CPSMatrix}. Hence the Rabi rates for these pulses in the composite structure must take the form $\big(\Omega_1,\hspace{0.15cm}\Omega_4,\hspace{0.15cm}\Omega_7\big) =\big(\WAMFRabiPlus,\hspace{0.15cm}\WAMFRabiMin,\hspace{0.15cm}\WAMFRabiPlus\big)/4\nu$. However each SK1 sequence has constant Rabi rate and we therefore conclude
\begin{align}\label{Method2SK1RabiRates}
\defcol\OmegaSK^{(1)}\bla = \defcol\OmegaSK^{(3)}\bla= \frac{1}{4\nu}\WAMFRabiPlus,\hspace{1cm}\ampcol\OmegaSK^{(2)}\bla = \frac{1}{4\nu}\WAMFRabiMin.
\end{align} 
Now, the duration of the composte sequence must satisfy 
\begin{align}
\tau &= \sum_{l = 1}^9\frac{\theta_l}{\Omega_1}\\
&= 2\Bigg(\frac{2\pi}{\OmegaSK^{(1)}}\Bigg)+2\Bigg(\frac{2\pi}{\OmegaSK^{(2)}}\Bigg)+2\Bigg(\frac{2\pi}{\OmegaSK^{(3)}}\Bigg) + 4\nu\\
&= 4\pi\Bigg(\frac{1}{\OmegaSK^{(1)}}+\frac{1}{\OmegaSK^{(2)}}+\frac{1}{\OmegaSK^{(3)}}\Bigg) + 4\nu.
\end{align}
Substituting in the results from Eq. \ref{Method2SK1RabiRates} we therefore find $\tau = 16\pi\nu\big(2/\WAMFRabiPlus+1/\WAMFRabiMin\big) + 4\nu$. Or, solving for $\nu$, 
\begin{align}
\nu = \frac{1}{4}\tau\big(1+\pi\kappa\big)^{-1},\hspace{1cm}\kappa:= 4\Bigg(\frac{2}{\WAMFRabiPlus}+\frac{1}{\WAMFRabiMin}\Bigg), 
\end{align}
concluding the construction. The composite structure may then be written

\begin{align}\label{UWMFMethod2GammaMatrix}
\text{UWMF}_{\text{SK1},1}\hspace{0.1cm}&\equiv\hspace{0.1cm}
\begingroup
  \renewcommand*{\arraystretch}{1.35}%
  \kbordermatrix{
           & \Omega_l           & \tau_l           & \theta_l           & \phi_l     \cr
     P_1    &  \defcol\frac{\WAMFRabiPlus}{4\nu}            &\defcol\nu       & \defcol\frac{\WAMFRabiPlus}{4}       &\defcol 0 \cr
    P_2    & \defcol\frac{\WAMFRabiPlus}{4\nu}            &\defcol\frac{8\pi\nu}{\WAMFRabiPlus}        &\defcol2\pi     & \defcol\phiSK^{(1)} \cr
    P_3    & \defcol\frac{\WAMFRabiPlus}{4\nu}            &\defcol\frac{8\pi\nu}{\WAMFRabiPlus}       &\defcol2\pi        & \defcol-\phiSK^{(1)} \cr
  P_4    &  \ampcol\frac{\WAMFRabiMin}{4\nu}            &\ampcol2\nu      & \ampcol\frac{\WAMFRabiMin}{2}       & \ampcol0 \cr
    P_5    & \ampcol\frac{\WAMFRabiMin}{4\nu}            &\ampcol\frac{8\pi\nu}{\WAMFRabiMin}     &\ampcol2\pi     & \ampcol\phiSK^{(2)} \cr
    P_6    & \ampcol\frac{\WAMFRabiMin}{4\nu}           &\ampcol\frac{8\pi\nu}{\WAMFRabiMin}       &\ampcol2\pi        & \ampcol-\phiSK^{(2)} \cr
  P_7    &  \defcol\frac{\WAMFRabiPlus}{4\nu}          &\defcol\nu     & \defcol\frac{\WAMFRabiPlus}{4}        &\defcol 0 \cr
    P_8    & \defcol\frac{\WAMFRabiPlus}{4\nu}            &\defcol\frac{8\pi\nu}{\WAMFRabiPlus}       &\defcol2\pi     &\defcol \phiSK^{(3)} \cr
    P_9    & \defcol\frac{\WAMFRabiPlus}{4\nu}         &\defcol\frac{8\pi\nu}{\WAMFRabiPlus}      &\defcol2\pi        &\defcol -\phiSK^{(3)} \cr
  }%
\endgroup
%\hspace{0.5cm}&\phiSK(\theta): = \cos^{-1}\Big(-\frac{\theta}{4\pi}\Big)
\end{align}
\begin{align}
\defcol\phiSK^{(1)}\bla = \defcol\phiSK^{(3)}\bla = \cos^{-1}\Big(-\frac{\WAMFRabiPlus}{16\pi}\Big),\hspace{1cm}
\ampcol\phiSK^{(2)} \bla= \cos^{-1}\Big(-\frac{\WAMFRabiMin}{8\pi}\Big).
\end{align}
Minima of the cost function in the $\WAMampP_0\WAMampP_3$ domain for this construction are found in identical regions to those shown in Fig. \ref{Fig: WAMF_O1}a for the simple WAMF$_{0,3}^{(1)}$ construction. Substituting these minimizing values of $\WAMampP_{0,3}$ into Eq. \ref{UWMFMethod2GammaMatrix} thus optimizes the \emph{concatenated} structure, yielding a desired net rotation (dictated by $\WAMampP_0$)which filters both amplitude and dephasing noise to first order simultaneously.

%__________________________________________________________________________________________

%					SECTION: BANDLIMITED PULSE EFFECTS
%__________________________________________________________________________________________

\section{Effect of Bandwidth Limits on Walsh Filters}\label{Sec:EffectOfBandwidthLimitsOnWalshFilters}

In the preceding sections of this paper filter design is based on optimizing the Walsh spectrum from which the relevant control structures are synthesized. This necessarily assumes perfectly square waveforms. Real control hardware, however, may suffer from bandwidth limitations which 'smooth out' the squareness of the pulse on the timescale of the application, leading to reduced filter performance. Here we show the assumption of perfect square pulses may be readily relaxed, with useful filter construction a simple matter of re-optimization in the Walsh-synthesis framework. %\red Namely, analytic design rules, discretized segment durations, and a minimum-dimensional control space over which to perform filter optimization.\bla

To illustrate the general procedure we consider the $\MinDim$-segment WAMF. Each segment implements a rotation through angle $\theta_l = \Omega_l\tau_l$ over duration $\tau_l=\tau/\MinDim$ and with constant Rabi rate $\Omega_l$, $l\in\{1,...,\MinDim\}$. The squareness of the resulting amplitude-modulated waveform may be relaxed by replacing the constant value $\Omega_l$ with an arbitrarily varying function of time in each segment. 

In order to achieve this we consider Walsh synthesis over the \emph{rotation angle} implemented in a single segment rather than the Rabi rate.  That is, we write $\theta_l =  \theta_l(\WAMampP_0, \WAMampP_1,...,\WAMampP_N)$ with the dependence on the Walsh spectra defined by the Hadamard-matrix equation  
\begin{align}
\vec{\boldsymbol{\theta}} = \big(\theta_1,\hspace{0.1cm},\theta_2,...,\theta_\MinDim\big)^T = (\tau/\MinDim)H_\MinDim\boldsymbol{\WAMampH}.
\end{align}
Defined in this way, the $\MinDim$-segment arbitrary-pulse sequence shares with the WAMF construction the property that the total gate rotation angle $\Theta = \sum_{l=1}^n\theta_l = \WAMampP_0\tau$ is completely determined by the spectral amplitude of $\PAL_0$. The symmetry-based design rules similarly carry over and filter optimization proceeds in the same manner as for ordinary Walsh-modulated control by minimizing the filter cost function with respect to the Walsh spectrum.

As a natural example of this approach we assume a Gaussian profile $G_l(t;\mu_l,\sigma_l)$ defined on $t\in[t_{l-1},t_l]$ with mean $\mu_l$ and standard deviation $\sigma_l$. Specifically, we construct
%\begin{align}
%\mu_l = (t_{l-1}-t_l)/2,\hspace{1cm} \sig_l = g\tau/n,\hspace{1cm}
%\end{align}
%
\begin{align}
&G_l(t;\mu_l,\sigma_l):=\frac{\theta_l}{C_l\sigma_l\sqrt{2\pi}}\exp\Big[{-\frac{(t-\mu_l)^2}{2\sigma_l^2}}\Big],\hspace{0.75cm}\mu_l = \frac{t_{l-1}+t_l}{2},\hspace{0.75cm}\sigma_l =  g\tau/\MinDim
\end{align}
with $\mu_l$ the segment midpoint and $\sigma_l$ expressed as a multiple $g$ of the segment duration. The normalizing factor
\begin{align}
C_l:=\int_{t_{l-1}}^{t_l}\frac{1}{\sigma_l\sqrt{2\pi}}\exp\Big[{-\frac{(t-\mu_l)^2}{2\sigma_l^2}}\Big]dt
\end{align}
is included to ensure the total rotation implemented by the Gaussian pulse in the $l$th segment is given by $\int_{t_l}^{t_{l-1}} G_l(t;\mu_l,\sigma_l)dt = \theta_l$. We now impose the same structure on the segment rotations $\theta_l$ as done for WAMFs in the Walsh-synthesis framework, such that the smooth-pulse sequence remains strictly parametrized in the Walsh spectrum $\boldsymbol{\WAMampP}$.

 \begin{figure}[bp]
 \centering
\includegraphics[width=0.8\columnwidth]{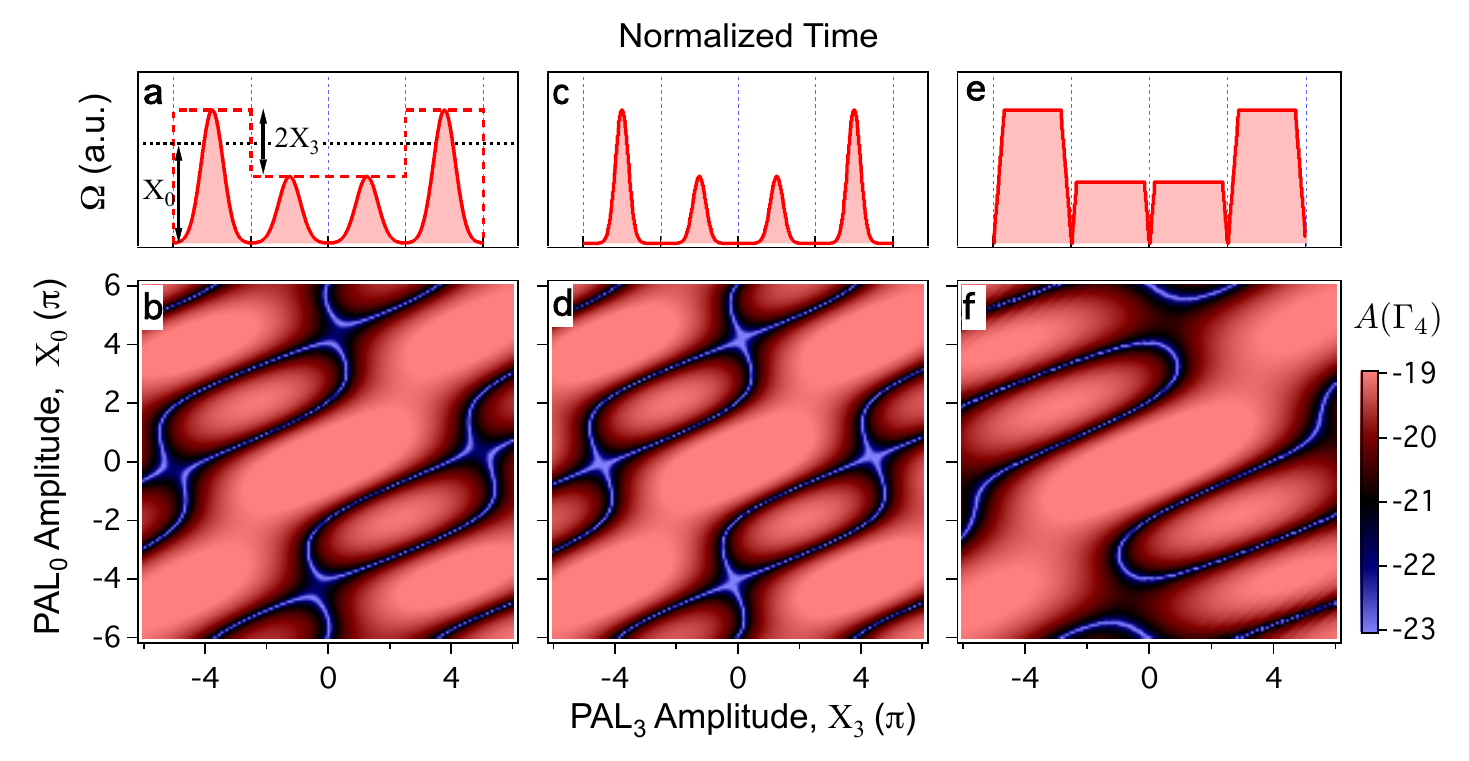}
\caption{\label{Fig:2DCostFunctionGaussianPulses}Construction of the first-order Walsh amplitude modulated dephasing-suppressing filter using shaped pulse segments.   a, c, e) Schematic representation of Walsh synthesis for a four segment gate of discrete Gaussian or trapezoidal segments.  Walsh synthesis determines the overall amplitude of individual pulses with fixed duration and standard deviation, setting the effective pulse area in each segment.   The metric $g$ takes value $1/6$ in panel (a), and $1/12$ in panel (c).  e) Trapezoidal pulses are characterized by a constant slope such that all angles are a fraction of a square waveform defined as $F\frac{\pi}{2}$.  Here $F=0.992$. In all pulse constructions the pulse profile is computed over 100 discrete time steps, permitting calculation of relevant filter-transfer functions.   b, d, f) Two-dimensional representation of the integral metric defining our target cost function, $A(\CPSMatElem_{4})$ integrated over the stopband $\omega\in [10^{-9}, 10^{-6}]\tau^{-1}$ for the corresponding pulse forms above. Areas in blue minimize $A(\CPSMatElem_{4})$, representing effective filter constructions. The $X_{0}$ determines the net rotation enacted in a gate while $X_{3}$ determines the modulation depth, as represented in a). }
\end{figure}

 For concreteness, we examine the Gaussian-pulse variation on the 4-segment WAMF$_{0,3}^{(1)}$ filter described by Eq. \ref{WAMFOrder1CPSMatrix} for two different Gaussian profiles, illustrated in Figs.~\ref{Fig:2DCostFunctionGaussianPulses}a,c.   The cost function $A_z(\WAMampP_3;\WAMampP_0) = \int_{\omega_L}^{\omega_c}d\omega \Fd(\omega\tau;\WAMampP_3;\WAMampP_0)$ may be computed by partitioning the time domain into a large number $N_s$ of subintervals on which the continuous Gaussian envelope is treated as approximately constant. Figs. \ref{Fig:2DCostFunctionGaussianPulses}b,d show a two-dimensional representation of $A_z(\WAMampP_3;\WAMampP_0)$ integrated over the interval $\omega\in [10^{-9}, 10^{-6}]\tau^{-1}$. The value of $\text{Log}_{10}\big[A_z(\WAMampP_3;\WAMampP_0)\big]$ is indicated by the color scale. Total sequence length is normalized to $\tau=1$ in this data, so the total gate rotation angle $\Theta \equiv \WAMampP_0$ is given directly by the $X_0$-axis. Regions in blue represent effective (first-order) filter constructions, where the cost function is minimized. 

Comparing with Fig. \ref{Fig: WAMF_O1} we conclude useful filter construction using Gaussian pulses is a simple matter of re-optimization in the Walsh-synthesis framework. This is readily achieved using a Nelder-Mead optimization of $A_z(\WAMampP_3;\WAMampP_0)$ for any particular choice of $g,\omega_L,\omega_c,\WAMampP_0$ or $N_s$.   Minor changes in the filter performance and optimal constructions arise with changes in Gaussian pulse parameters such as $g$.  Comparison with pulses constructed using a trapezoidal form (Fig.~\ref{Fig:2DCostFunctionGaussianPulses}e) we find a different optimization outcome that more closely approximates standard square pulses.  Nonetheless, these results show that, irrespective the specific pulse form, re-optimization over the Walsh coefficients remains a direct method to construct useful filters.  In cases where unknown waveform distortion is likely in hardware, it is possible to implement automated feedback mechanisms, as has previously been demonstrated in dynamical decoupling experiments~\cite{BiercukNature2009}.  

We may also explore the impact of finite modulation bandwidth on the application of square pulses if re-optimization of the waveform is, for some reason, not possible.  In order to explore these effects we systematically relax the infinite-modulation-bandwidth assumption underlying any square-pulse approximation by processing the ideal time-domain profile through a bandlimited digital filter with a user-defined cutoff. This results in an imperfect (bandlimited) profile envelope, effectively due to a reconstruction based on a truncated Fourier series. These profile distortions, manifesting as implementation errors, reduce filter performance, quantified by an increase in the area under the corresponding filter-transfer function.  

\begin{figure}[h]
\centering
\includegraphics[width=1\columnwidth]{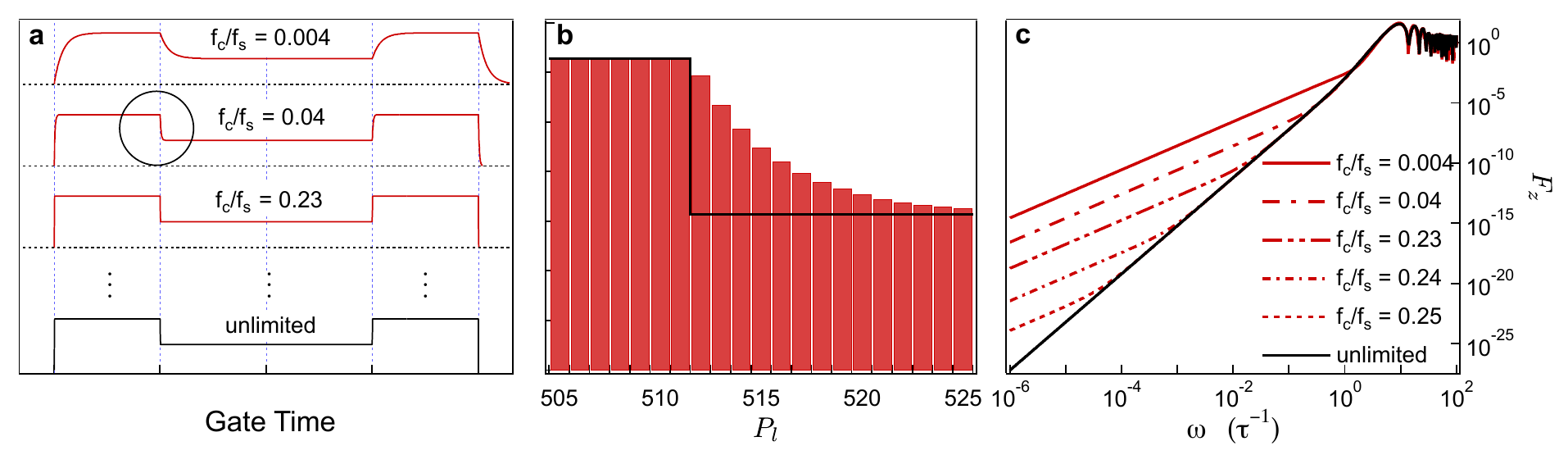}
\caption{Effect of bandwidth limits on Walsh filters. \textbf{a}) Bandlimited reconstructions of WAMF$_{0,3}^{(1)}$ modulation envelope as a function of first-order Butterworth filter parameter $f_c/f_s$. \textbf{b}) \emph{Red}: Closeup of circled region in a); \emph{Black} ideal square profile. \textbf{c}) filter-transfer functions for profiles in a).}\label{Fig:BandLimited}
\end{figure}

In this example we again consider the amplitude-modulated profile associated with the WAMF$_{0,3}^{(1)}$ gate ($X_0 = 3\pi, X_3=\pi$), and impose band limitations using a digital first-order Butterworth filter. Here the time domain is partitioned into $1/f_s$ subintervals, where $f_s = 1/2^{11}$ is the sampling rate of the digital filter. The bandlimited envelope of reconstructed waveform is then a function of the Butterworth cutoff frequency $f_c$. As the cutoff approaches the Nyquist frequency $f_s/2$ (the maximum possible value), the bandlimited effects are reduced and the reconstructed waveform approaches the ideal square-pulse envelope. Fig. \ref{Fig:BandLimited}a illustrates this as we increase $f_c/f_s$. The corresponding filter characteristics are shown in Fig. \ref{Fig:BandLimited}c where we plot $\Fd(\omega)$ as a function of $f_c/f_s$. To compute these filter-transfer functions we treat the reconstructed waveforms as amplitude-modulated sequences consisting of $n = 1/f_s$ segments whose Rabi-rates are determined by the bandlimited envelope (see Fig. \ref{Fig:BandLimited}b). The solid black traces in Figs. \ref{Fig:BandLimited}a,b,c respectively show the ideal profile and corresponding filter-transfer function, against which the filter performance of the bandlimited gates are benchmarked. As we decrease the cutoff the integrated area under $\Fd(\omega)$ gradually increases. However as this manifests first in the low-frequency region, even these bandlimited gates can provide useful filtering, given sufficient hardware precision. Similarly motivated studies in dynamic decoupling show analogous low-frequency filter performance decay with pulse-timing errors.

%__________________________________________________________________________________________

%				SECTION: CONCLUSION
%__________________________________________________________________________________________

\section{Conclusion}\label{Sec:Conclusion}

As the size and complexity of quantum information processing technologies increase, resource-efficiency will play a vital role in selecting methods designed to reduce errors in quantum coherent hardware systems. The pressure to minimize quantum-hardware overhead is likely to make \emph{open-loop control} protocols a key element in the design of error-robust quantum information systems~\cite{JonesPRX2012}. 
For these to be practically useful, however, it is important to move toward realistic control and noise models.   

Decoherence in real driven systems is predominantly due to low-frequency \emph{correlated} noise environments. This strongly motivates our study of bounded-strength control as a noise filtering problem using time-dependent, non-Markovian error models.   Moreover, in contrast with traditional DD schemes, the added complications of treating bounded-strength control - due to the continual presence of noise interactions during control operations and the resulting nonlinear dynamics - necessitates a streamlined approach to the design of noise-filtering control.   The generalized filter-transfer function framework we employ takes as input experimentally measurable characteristics of the environment - namely noise power spectra - and provides a simple framework for both control construction and the calculation of predicted operational fidelities.  It also efficiently captures the control nonlinearities implicit in situations where control and noise Hamiltonians do not commute.  We have exploited these strengths to pursue a simple variational procedure for filter design by minimizing a cost function over the relevant control space.

A key strength of the method we have presented is derived from our use of functional analysis for the crafting of effective noise-filtering control protocols.  In particular, employing the Walsh basis brings an intuitive set of analytic design rules for filter construction that further constrain the possible filter-construction space~\cite{HayesPRA2011}.  For instance, a user-imposed limit on the acceptable number of pulse segments in a filter construction impose additional constraints due to Walsh-function symmetry, spectral properties, and the level of recursiveness of the Walsh functions (measured by the Hamming weight of the Walsh Paley-ordered index).

In addition to efficiency of synthesis is the intrinsic compatibility with hardware controllers that comes with the selection of Walsh functions as our basis set.  This is particularly important in the layered architecture for quantum information systems mentioned throughout this paper.  In such a setting, it rapidly becomes undesirable to mandate a significant amount of communication between the physical qubit layer and hardware at the highest levels of system abstraction.  This suggests that controllers implementing dynamic error suppression protocols (here producing noise filters for arbitrary driven operations) should be reasonably simple to implement in standard digital hardware and should require only limited communication bandwidth to higher levels of the system.

These considerations are explicitly met in crafting control solutions from the Walsh basis.  First, the Walsh functions are defined using integer multiples of a fundamental clock period, meaning that limitations of finite timing precision in the definition of a control protocol are automatically inbuilt.  Further, given a particular Walsh-modulated control protocol is entirely defined by its Walsh spectrum, programming of the controller can in principle be reduced to a simple vector of numbers representing the Walsh spectrum and minimum timestep.  All other information \emph{e.g.} the total time, total number of timesteps, etc., is carried implicitly in the spectrum.  Moreover, the actual Walsh-function \emph{generation} is compatible with simple hardware systems (adding of various harmonics of a fundamental square-wave clock) and when Walsh synthesis is performed at the level of the controller hardware, this may provide a path to on-the-fly synthesis of the required modulation waveform.  Such capabilities also reduce the complexity of running automated hardware-driven optimization procedures for finding relevant control waveforms~\cite{BiercukNature2009}, by allowing efficient generation of many trial waveforms without the need for large memory stores at the local controller.
 
Synthesizing all of these considerations the Walsh-modulated noise filters we have developed in this work provide one of the first solutions for error suppression at the physical-qubit level simultaneously meeting the physical and engineering requirements we outline above for scalable control solutions.  Using this framework we have derived a range of novel filters, chiefly WAMFs for dephasing noise and WPMFs for amplitude noise. Both are capable of spectral optimization subject to physically motivated constraints such as implementing target qubit rotations. These design forms are also compatible with concatenation for filtering universal noise. Interestingly, our approach unifies a number of existing composite pulse sequencing schemes; we have revealed how Walsh-modulated filter construction naturally incorporates familiar sequences (e.g., DCG, SK1, P2, BB1) in a non-Markovian time-dependent noise context. This potential to incorporate other approaches may prove useful in building a consistent picture of the scope and applicability of the many and varied schemes that continue to be developed by the quantum control community. These considerations make the Walsh basis an attractive design platform and we believe this simple framework will provide a straightforward path for the development of improved quantum control techniques.

\emph{Acknowledgements:}  We thank K. Brown, J.T. Merrill and L. Viola for useful discussions.  This work partially supported by the US Army Research Office under Contract Number  W911NF-11-1-0068, and the Australian Research Council Centre of Excellence for Engineered Quantum Systems CE110001013.

%\begin{acknowledgments}
%This work partially supported by the US Army Research Office under Contract Number  W911NF-11-1-0068, and the Australian Research Council Centre of Excellence for Engineered Quantum Systems CE110001013, and the Office of the Director of National Intelligence (ODNI), Intelligence Advanced Research Projects Activity (IARPA), through the Army Research Office. All statements of fact, opinion or conclusions contained herein are those of the authors and should not be construed as representing the official views or policies of IARPA, the ODNI, or the U.S. Government.
%\end{acknowledgments}

%__________________________________________________________________________________________

%							     BIBLIOGRAPHY
%__________________________________________________________________________________________

\nocite{*}

\section*{References}
\bibliography{WMFBibliographyNJP}

\providecommand{\newblock}{}
\begin{thebibliography}{10}
\expandafter\ifx\csname url\endcsname\relax
  \def\url#1{{\tt #1}}\fi
\expandafter\ifx\csname urlprefix\endcsname\relax\def\urlprefix{URL }\fi
\providecommand{\eprint}[2][]{\url{#2}}

\bibitem{Clarke2004}
Harlingen D~J~V, Plourde B~L~T, Robertson T~L, Reichardt P~A and Clarke J 2004
  Decoherence in flux qubits due to $1/f$ noise in josephson junctions {\em
  Quantum Computing and Quantum Bits in Mesoscopic Systems\/} ed Leggett A,
  Ruggiero B and Silvestini P (New York: Kluwer Academic Press) pp 171--184

\bibitem{Faoro2004}
Faoro L and Viola L 2004 {\em Phys. Rev. Lett.\/} {\bf 92} 117905

\bibitem{Bylander2011}
Bylander J, Gustavsson S, Yan F, Yoshihara F, Harrabi K, Fitch G, Cory D~G,
  Nakamura Y, Tsai J~S and Oliver W~D 2011 {\em Nat. Phys.\/} {\bf 7} 565--570

\bibitem{Harmon2007}
Zhang W, Dobrovitski V~V, Santos L~F, Viola L and Harmon B~N 2007 {\em Phys.
  Rev. B\/} {\bf 75} 201302

\bibitem{BiercukQIC2009}
Biercuk M~J, Uys H, VanDevender A~P, Shiga N, Itano W~M and Bollinger J~J 2009
  {\em Quantum Information and Computation\/} {\bf 9} 920--949

\bibitem{rutman1978}
Rutman J 1978 {\em Proceedings of the IEEE\/} {\bf 66} 1048

\bibitem{Tarn80}
Tarn T~J, Huang G and Clark J~W 1980 {\em Mathematical Modelling\/} {\bf 1} 109

\bibitem{Tarn2003}
Clark J~W, Lucarelli D~G and Tarn T~J 2003 {\em Int. J. Mod. Phys. B\/} {\bf
  17} 5397

\bibitem{James2007}
Bouten L, Handel R~V and James M~R 2007 {\em SIAM J. Control Optim.\/} {\bf 46}
  2199

\bibitem{James2009}
Nurdin H~I, James M~R and Petersen I~R 2009 {\em Automatica\/} {\bf 45} 1837

\bibitem{Viola1999}
Viola L, Lloyd S and Knill E 1999 {\em Phys. Rev. Lett.\/} {\bf 83} 4888

\bibitem{Altafini2013}
Ticozzi F, Nishio K and Altafini C 2013 {\em IEEE Trans. Auto. Control\/} {\bf
  58} 74--85

\bibitem{QECLidar2013}
Lidar D~A and Brun T~A 2013 {\em Quantum Error Correction\/} (New York:
  Cambridge University Press)

\bibitem{JonesPRX2012}
Jones N~C, Van~Meter R, Fowler A~G, McMahon P~L, Kim J, Ladd T~D and Yamamoto Y
  2012 {\em Phys. Rev. X\/} {\bf 2} 031007

\bibitem{BiercukNature2009}
Biercuk M~J, Uys H, VanDevender A~P, Shiga N, Itano W~M and Bollinger J~J 2009
  {\em Nature\/} {\bf 458} 996

\bibitem{LongStorage}
Khodjasteh K, Sastrawan J, Hayes D, Green T~J, Biercuk M~J and Viola L 2013
  {\em Nat. Commun.\/} {\bf 4} 2045

\bibitem{RBDD}
Du J, Rong X, Zhao N, Wang Y, Yang J and Liu R~B 2009 {\em Nature\/} {\bf 461}
  1265

\bibitem{DavidsonDD}
Sagi Y, Almog I and Davidson N 2010 {\em Phys. Rev. Lett.\/} {\bf 105} 053201

\bibitem{khodjasteh2009dcg}
Khodjasteh K and Viola L 2009 {\em Phys. Rev. Lett.\/} {\bf 102} 080501

\bibitem{Khodjasteh2010}
Khodjasteh K, Lidar D~A and Viola L 2010 {\em Phys. Rev. Lett.\/} {\bf 104}(9)
  090501

\bibitem{DasSarmaGate}
Wang X, Bishop L~S, Kestner J~P, Barnes E, Sun K and Das~Sarma S 2012 {\em Nat.
  Commun.\/} {\bf 3} 997

\bibitem{Suter_Interleaved}
Souza A~M, \'Alvarez G~A and Suter D 2012 {\em Phys. Rev. A\/} {\bf 86} 050301

\bibitem{HansonInterleaved}
van~der Sar T, Wang Z, Blok M, Bernien H, Taminiau T, Toyli D, Lidar D,
  Awschalom D, Hanson R and Dobrovitski V 2012 {\em Nature\/} {\bf 484} 82--86

\bibitem{RBInterleaved}
Liu G~Q, Po H~C, Du J, Liu R~B and Pan X~Y 2013 {\em Nat. Comms.\/} {\bf 4}
  2254

\bibitem{UhrigPRA2012}
Fauseweh B, Pasini S and Uhrig G 2012 {\em Phys. Rev. A\/} {\bf 85} 022310

\bibitem{Vandersypen2004}
Vandersypen L~M~K and Chuang I~L 2005 {\em Rev. Mod. Phys.\/} {\bf 76} 1037

\bibitem{MerrillArXv2012}
Merrill J~T and Brown K~R 2012 {\em arXiv:1203.6392\/}

\bibitem{Kabytayev2014}
Kabytayev C, Green T~J, Khodjasteh K, Biercuk M~J, Viola L and Brown K~R 2014
  {\em Phys. Rev. A\/} {\bf 90} 012316

\bibitem{ViolaEDD}
Viola L and Knill E 2003 {\em Phys. Rev. Lett.\/} {\bf 90} 037901

\bibitem{JonesNJP2012}
Jones N~C, Ladd T~D and Fong B~H 2012 {\em New J. Phys.\/} {\bf 14} 093045

\bibitem{CaiNJP2012}
Cai J~M, Naydenov B, Pfeiffer R, McGuinness L~P, Jahnke K~D, Jelezko F, Plenio
  M~B and Retzker A 2012 {\em New J. Phys.\/} {\bf 14} 113023

\bibitem{FanchiniPRA2007}
Fanchini F~F, Hornos J~E~M and d~J~Napolitano R 2007 {\em Phys. Rev. A\/} {\bf
  75} 022329

\bibitem{XuPRL2012}
Xu X, Wang Z, Duan C, Huang P, Wang P, Wang Y, Xu N, Kong X, Shi F, Rong X and
  Du J 2012 {\em Phys. Rev. Lett.\/} {\bf 109} 070502

\bibitem{BermudezPRA2012}
Bermudez A, Schmidt P~O, Plenio M~B and Retzker A 2012 {\em Phys. Rev. A\/}
  {\bf 85} 040302

\bibitem{ChaudhryPRA2012}
Chaudhry A~Z and Gong J 2012 {\em Phys. Rev. A\/} {\bf 85} 012315

\bibitem{LemmerNJP2013}
Lemmer A, Bermudez A and Plenio M~B 2013 {\em New J. Phys.\/} {\bf 15} 083001

\bibitem{SoareNatPhys2014}
Soare A, Ball H, Hayes D, Sastrawan J, Jarratt M~C, McLoughlin J~J, Zhen X,
  Green T~J and Biercuk M~J 2014 {\em Nat. Phys.\/} {\bf 10}

\bibitem{ViolaFFFArXve2014}
Silva G~P and Viola L 2014 {\em arXiv:1408.3836\/}

\bibitem{KurizkiPRL2001}
Kofman A and Kurizki G 2001 {\em Phys. Rev. Lett.\/} {\bf 87} 270405

\bibitem{KurizkiPRL2004}
Kofman A~G and Kurizki G 2004 {\em Phys. Rev. Lett.\/} {\bf 93} 130406

\bibitem{TransferFunction}
Girod B, Rabenstein R and Stenger A 2001 {\em Signals and Systems\/} (New York:
  Wiley)

\bibitem{BiercukJPB2011}
Biercuk M~J, Doherty A~C and Uys H 2011 {\em J. Phys. B\/} {\bf 44} 154002

\bibitem{Martinis2003}
Martinis J~M, Nam S, Aumentado J, Lang K~M and Urbina C 2003 {\em Phys. Rev.
  B\/} {\bf 67} 094510

\bibitem{Kuopanportti2008}
Kuopanportti P, Mottonen M, Bergholm V, Saira O~P, Zhang J and Whaley K~B 2008
  {\em Phys. Rev. A\/} {\bf 77} 032334

\bibitem{UysPRL2009}
Uys H, Biercuk M~J and Bollinger J 2009 {\em Phys. Rev. Lett.\/} {\bf 103}
  040501

\bibitem{UhrigPRL2007}
Uhrig G 2007 {\em Phys. Rev. Lett.\/} {\bf 98} 100504

\bibitem{CywinskiPRB2008}
Cywinski L, Lutchyn R~M, Nave C~P and Sarma S~D 2008 {\em Phys. Rev. B\/} {\bf
  77} 174509

\bibitem{GreenPRL2012}
Green T~J, Uys H and Biercuk M~J 2012 {\em Phys. Rev. Lett.\/} {\bf 109} 020501

\bibitem{GreenNJP2013}
Green T~J, Sastrawan J, Uys H and Biercuk M~J 2013 {\em New J. Phys.\/} {\bf
  15} 095004

\bibitem{HayesPRL2014}
Hayes D, Clark S~M, Debnath S, Hucul D, Inlek I~V, Lee K~W, Quraishi Q and
  Monroe C 2012 {\em Phys. Rev. Lett.\/} {\bf 109} 020503

\bibitem{HayesPRA2011}
Hayes D, Khodjasteh K, Viola L and Biercuk M~J 2011 {\em Phys. Rev. A\/} {\bf
  84} 062323

\bibitem{GreenArXve2014}
Green T~J and Biercuk M~J 2014 {\em arXiv:1408.2749\/}

\bibitem{OwrutskyPRA2012}
Owrutsky P and Khaneja N 2012 {\em Phys. Rev. A\/} {\bf 86} 022315

\bibitem{Hodgson2010}
Hodgson T~E, Viola L and D'Amico I 2010 {\em Phys. Rev. A\/} {\bf 81} 062321

\bibitem{Schumacher1996}
Schumacher B 1996 {\em Phys. Rev. A\/} {\bf 54} 2614

\bibitem{Blanes2009}
Blanes S, Cases F, Oteo J~A and Ros J 2009 {\em Phys. Rep.\/} {\bf 470} 151

\bibitem{Magnus1954}
Magnus W 1954 {\em Commun. Pure and Appl. Math.\/} {\bf 7} 649

\bibitem{Waugh1968}
Haeberlen U and Waugh J~S 1968 {\em Phys. Rev.\/} {\bf 175} 453

\bibitem{Ernst1987}
Ernst R~R, Bodenhausen G~B and Wokaun A 1987 {\em Principles of Nuclear
  Magnetic Resonance in One and Two Dimensions\/} (New York: Oxford University
  Press)

\bibitem{SuterPRA2011}
Ajoy A, \'Alvarez G~A and Suter D 2011 {\em Phys. Rev. A\/} {\bf 83} 032303

\bibitem{Beauchamp1975}
Beauchamp K~G 1975 {\em Walsh Functions and their Applications\/} (London:
  Academic Press)

\bibitem{Walsh1923}
Walsh J~L 1923 {\em Amer. J. Math.\/} {\bf 45} 5

\bibitem{Tzafestas1985}
Tzafestas S~G 1985 {\em Walsh Functions in Signal and Systems Analysis and
  Design\/} (New York: Van Nostrand Reinhold)

\bibitem{Cooper14}
Cooper A, Magesan E, Yum H~N and Cappellaro P 2014 {\em Nat. Commun.\/} {\bf 5}

\bibitem{Harmuth1969a}
Harmuth H~F 1969 {\em Transmission of Information by Orthogonal Functions\/}
  (New York: Springer)

\bibitem{Harmuth1969b}
Harmuth H~F 1969 {\em IEEE Spectrum\/} {\bf 6} 82

\bibitem{Paley1932}
Paley R 1932 {\em Proc. London Math. Soc.\/} {\bf 2} 241

\bibitem{Rademacher1922}
Rademacher H 1922 {\em Math. Ann.\/} {\bf 87} 241

\bibitem{Horadam2007}
Horadam K~J 2007 {\em Hadamard Matrices and their Applications\/} (Princeton:
  Princeton University Press)

\bibitem{Khodjasteh2009}
Khodjasteh K and Viola L 2009 {\em Phys. Rev. A\/} {\bf 80} 032314

\bibitem{Souza2012}
Souza A~M, \'Alvarez G~A and Suter D 2012 {\em Phys. Rev. A\/} {\bf 85} 032306

\bibitem{Brown2004}
Brown K~R, Harrow A~W and Chuang I~L 2004 {\em Phys. Rev. A\/} {\bf 70} 052318

\bibitem{Brown2005}
Brown K~R, Harrow A~W and Chuang I~L 2005 {\em Phys. Rev. A\/} {\bf 72}(3)
  039905

\bibitem{WimperisBB11994}
Wimperis S 1994 {\em J. Magn. Reson., Series A\/} {\bf 109} 221

\bibitem{YanNature2013}
Yan F, Gustavsson S, Bylander J, Jin X, Yoshihara F, Cory D~G, Nakamura Y,
  Orlando T~P and Oliver W~D 2013 {\em Nat. Commun.\/} {\bf 4}

\bibitem{Gustavsson2012}
Gustavsson S, Bylander J, Yan F, Forn-D\'iaz P, Bolkhovsky V, Braje D, Fitch G,
  Harrabi K, Lennon D, Miloshi J, Murphy P, Slattery R, Spector S, Turek B,
  Weir T, Welander P~B, Yoshihara F, Cory D~G, Nakamura Y, Orlando T~P and
  Oliver W~D 2012 {\em Phys. Rev. Lett.\/} {\bf 108} 170503

\bibitem{Foletti_NP2009}
Foletti S, Bluhm H, Mahalu D, Umansky V and Yacoby A 2009 {\em Nat. Phys.\/}
  {\bf 5} 903

\bibitem{Britton2012}
Britton J~W, Sawyer B~C, Keith A~C, Wang C~C~J, Freericks J~K, Uys H, Biercuk
  M~J and Bollinger J~J 2012 {\em Nature\/} {\bf 484} 489

\bibitem{Lukin2013}
Grinolds M~S, Hong S, Maletinsky P, Luan L, Lukin M~D, Walsworth R~L and Yacoby
  A 2013 {\em Nat. Phys.\/} {\bf 9} 215

\bibitem{JessenPRL2013}
Smith A, Anderson B~E, Sosa-Martinez H, Riofrio C~A, Deutsch I~H and Jessen P~S
  2013 {\em Phys. Rev. Lett.\/} {\bf 111} 170502

\bibitem{Islam03052013}
Islam R, Senko C, Campbell W~C, Korenblit S, Smith J, Lee A, Edwards E~E, Wang
  C~C~J, Freericks J~K and Monroe C 2013 {\em Science\/} {\bf 340} 583

\bibitem{SoarePRA2014}
Soare A, Ball H, Hayes D, Zhen X, Jarratt M~C, Sastrawan J, Uys H and Biercuk
  M~J 2014 {\em Phys. Rev. A\/} {\bf 89} 042329

\bibitem{WienerKhinchin}
Miller S~L and Childers D 2012 {\em Probability and Random Processes with
  Applications to Signal Processing and Communications\/} (Boston: Academic
  Press)

\end{thebibliography}

%						__________________________________

%							     APPENDICES
%						__________________________________

\appendix

%__________________________________________________________________________________________

%					SECTION: DETAILED FF DERIVATION 
%__________________________________________________________________________________________

\section{\\Detailed filter-transfer Function Derivation}\label{App:DetailedFFDerivation}%\label{FirstOrderInfidelityAppendix}

In this appendix we derive the computational form of the first-order infidelity $\langle a_1^2\rangle = \langle\EV_1\EV_1^T\rangle$ expressed in Eq. \ref{FirstOrderInfidelityComputational}.  Recall, we write the total evolution operator $\Utot(t) = \Uc(t)\Uerr(t)$ where the \emph{error propagator} $\Uerr(t)$ satisfies the Schrodinger equation $i\dot{\Uerr}(t) = \Htog(t)\Uerr(t)$ in a frame co-rotating with the control, defined by the \emph{toggling frame Hamiltonian} $\Htog(t) :=\UcD(t)\Herr(t)\Uc(t)$. We may obtain an arbitrarily accurate, unitary estimate of the error propagator by performing a Magnus expansion, whereby $\Uerr(\tau) = \exp[-i\Phi(\tau)]$ and the effective error operator $\Phi(\tau) = \sum_{\mu = 1}^\infty\Phi_\mu(\tau)$ at the end of the interaction has Magnus expansion terms 
\begin{equation}
\begin{aligned}
\label{Magnus1}\Phi_1(\tau) &= \int_0^\tau dt \Htog(t)\\
\Phi_2(\tau) &= -\frac{i}{2}\int_0^\tau dt_1\int_0^{t_1} dt_2 \big[\Htog(t_1),\Htog(t_2)\big]\\
\Phi_3(\tau) &= \frac{1}{6}\int_0^\tau dt_1\int_0^{t_1} dt_2 \int_0^{t_2} dt_3
\Big\{\Big[\Htog(t_1),\big[\Htog(t_2),\Htog(t_3)\big] \Big]\\&+\Big[\Htog(t_3),\big[\Htog(t_2),\Htog(t_1)\big]\Big] \Big\}\\
&...
\end{aligned}
\end{equation}
These generally take the form of time-ordered integrals over nested commutators in $\Htog(t)$. 

We define the \emph{error vector} $\EV(\tau)$ by re-expressing the operator $\Phi(\tau)\equiv\EV(\tau)\cdot\sigvec$ in the basis of Pauli operators \footnote{This is valid since $\Phi(\tau)$ belongs to the Lie algebra of $SU(2)$, inheriting this property from the toggling frame Hamiltonians from which it is derived.}. Assuming unitary processes, one may then employ vector identities to expand $\EV(\tau) = \sum_{\mu = 1}^\infty\EV_\mu(\tau)$ in an infinite power series such that $\Phi_\mu(\tau) = \EV_\mu(\tau)\cdot\sigvec$, $\forall\mu\in\mathbb{N}$~\cite{GreenNJP2013}. The control propagator is therefore written $\Uerr(\tau) = \exp\big[-i\sum_{\mu = 1}^\infty\EV_\mu(\tau)\cdot\sigvec\big]$ and may be approximated,  with arbitrary accuracy, as a unitary operator in simple exponential form. In this paper we consider the first-order approximation $\EV(\tau)\approx\EV_1(\tau)$. Hence we restrict attention to deriving the form of $\EV_1(\tau)$ which, using Eq. \ref{Magnus1}, satisfies 
\begin{align}
\EV_1(\tau)\cdot\sigvec = \Phi_1(\tau) = \int_0^\tau dt \Htog(t).
\end{align}
Our first task is therefore to derive a computationally useful forms for the toggling frame Hamiltonian $\Htog(t)$. This is done in the following section.

%				-------------------------------------------------
%						sub:SECTION: Toggling Frame Hamiltonian 
%				-------------------------------------------------

\subsection*{Toggling Frame Hamiltonian $\Htog(t)$: Computational Form} 

The noise Hamiltonian $\Herr(t) = \Hd(t) + \Ha(t)$, presented in Sec. \ref{Sec:PhysicalSetting} is linear in the dephasing and amplitude contributions. Since the $\Htog(t)$ is linear in $\Herr(t)$ we may write
\begin{align}\label{TogHamPartitioned}
\Htog(t) =\Htogd(t)+\Htoga(t)
\end{align}
where we have defined the dephasing and amplitude toggling frame Hamiltonians by 
\begin{align}
%\EV_1(\tau)\sigvec =\Big[\EVa(\tau)+\EVd(\tau)\Big]\sigvec, %\\
\label{DephasingTogHam}\Htogd(t)&:=\UcD(t)\Hd(t)\Uc(t)\\
\label{AmplitudeTogHam}\Htoga(t)&:=\UcD(t)\Ha(t)\Uc(t).
\end{align}
It is convenient to employ the definitions of the scalar functions $R_{ij}(t)$, $R_{ij}^{P_l}(t-t_{l-1})$ and $\Lambda_{ij}^{(l-1)}$, $i,j\in\{x,y,z\}$ introduced by Green \emph{et. al} ~\cite{GreenNJP2013}. These are defined as the Cartesian expansion coefficients, in the basis of Pauli matrices, of the following operators 
%Computational Expansions 
\begin{align}
\label{Exp1}
&\UcD(t)\sig_i\Uc(t)=\sum_{j = x,y,z}R_{ij}(t)\sig_j,% = \mathbf{R}_i(t)\sigvec,
\\
\label{Exp2}
&U_c^\dagger(t,t_{l-1})\sig_i U_c(t,t_{l-1}) = \sum_{j=x,y,z}R_{ij}^{P_l}(t-t_{l-1})\sig_j, %= \mathbf{R}^{Pl}_i(t-t_{l-1})\sigvec,
\\
\label{Exp3}
&Q^\dagger_{l-1}\sig_iQ_{l-1} = \sum_{j=x,y,z}\Lambda_{ij}^{(l-1)}\sig_j.% = \mathbf{\Lambda}^{(l-1)}_i\sigvec.
\end{align}
These functions then serve as matrix elements, defining the computational matrices
%Computational Matrices 
\begin{eqnarray}
\label{TotalControlMatrix}\hspace*{-2.25cm}&\text{\emph{Total Control Matrix}:}\hspace{1cm}&[\mathbf{R}(t)]_{ij}:=\frac{1}{2}\text{Tr}\Big[\UcD(t)\sig_i\Uc(t)\sig_j\Big]
\\
\label{LocalControlMatrix}\hspace*{-2.25cm}&\text{\emph{Local Control Matrix}:}\hspace{1cm}&[\mathbf{R}^{P_l}(t-t_{l-1})]_{ij}:= \frac{1}{2}\text{Tr}\Big[U_c^\dagger(t,t_{l-1})\sig_i U_c(t,t_{l-1})\sig_j\Big]\\
\label{PulseHistoryMatrix}\hspace*{-2.25cm}&\text{\emph{Control History Matrix}:}\hspace{1cm}&[\HistMat]_{ij} := \frac{1}{2}\text{Tr}\Big[Q^\dagger_{l-1}\sig_iQ_{l-1}\sig_j\Big]
\end{eqnarray}
where the above expressions follow from post-multiplying Eqs. \ref{Exp1}, \ref{Exp2} and \ref{Exp3} by $\sig_j$, taking the trace and using the linearity of the trace operation. The matrix $\mathbf{R}(t)$ captures the qubit dynamics, in the time-domain, due to the control Hamiltonian \emph{at any time}; $\mathbf{R}^{Pl}(t-t_{l-1})$ captures essentially the same information, but restricted to the time interval $t\in I_l$. That is, during the $l$th pulse. The $3\times3$ \emph{Control History Matrix} $\mathbf{\Lambda}^{(l-1)}$, on the other hand, captures the accumulated effect of the previous $l-1$ completed pulses, implemented via the cumulative operator $Q_{l-1}$. 

%				-------------------------------------------------
%				sub:SECTION: Dephsasing Toggling Frame Hamiltonian  
%				-------------------------------------------------
\subsection*{Dephasing Toggling Hamiltonian $\Htogd(t)\textnormal{:}$}\label{DephasingNoiseTerm}
Substituting Eq. \ref{DephasingNoiseHamiltonian} into Eq. \ref{DephasingTogHam} the dephasing noise component of $\Htog(t)$ takes the form \begin{align}
\Htogd(t) &= \Bd(t)\UcD(t)\sigz\Uc(t).
\end{align}
Using Eq. \ref{Exp1}, we may express $\UcD(t)\sigz\Uc(t) = \sum_{j}R_{zj}(t)\sig_j$, $j\in\{x,y,z\}$, yielding
\begin{align}\label{TogHamDephComputational}
\Htogd(t) =\Bd(t)\Rd(t)\sigvec
\end{align}
where the time-domain \emph{Dephasing Control Vector} $\Rd(t)$ is defined as the third row of $\mathbf{R}(t)$. Here, and in the following derivations, for notational simplicity $\sigvec$ is understood to be the column vector
\begin{align}
\sigvec \equiv \left(\sigx,\hspace{0.1cm}\sigy,\hspace{0.1cm}\sigz\right)^T.
\end{align}
The computational form of $\Rd(t)$, by inspection of the more general computational form for $\mathbf{R}(t)$ derived in \ref{App:CVsComputationalForms}, is 
%\emph{Control Matrix} Time Domain Matrix Notation
\begin{align}\label{TimeDomainDephasingControlVectorComputational}
\Rd(t) = \sum_{l=1}^{n}G^{(l)}(t)\mathbf{R}_z^{P_l}(t-t_{l-1})\mathbf{\Lambda}^{(l-1)}.
\end{align}
%
%				-------------------------------------------------
%				sub:SECTION: Amplitude Toggling Frame Hamiltonian  
%				-------------------------------------------------
%
\subsection*{Amplitude Toggling Hamiltonian $\Htoga(t)\textnormal{:}$}\label{AmplitudeNoiseTerm} 

Similarly, substituting Eq. \ref{AmplitudeNoiseHamiltonian} into Eq. \ref{AmplitudeTogHam} we find
\begin{align}
\Htoga(t)&=\Ba(t)
\sum_{l=1}^{n} G^{(l)}(t) \frac{\Omega_l}{2}\Big\{ \UcD(t)\sigphiL\bla\Uc(t)\Big\}\\
\label{TogHamAmpCumulative}&=
\Ba(t)\sum_{l=1}^{n} G^{(l)}(t) \frac{\Omega_l}{2} 
Q_{l-1}^\dagger U_c^\dagger(t,t_{l-1})
\sigphiL\bla
U_c(t,t_{l-1})Q_{l-1}
\end{align}
where in the second line we have substituted $\Uc(t) = \sum_{l=1}^n G^{(l)}(t)U_c(t,t_{l-1})Q_{l-1}$ using Eq. \ref{ControlPropagatorComputational}. From Eq. \ref{LocalUnitaryRotation} the control operator $U_c(t,t_{l-1}) = \exp[-i\frac{\Omega_l}{2}\sigphiL(t-t_{l-1})]$ commutes with $\sigphiL$, $\forall l\in\{1,...,n\}$. That is, coaxial amplitude noise always ``tracks'' the direction of control. Hence Eq. \ref{TogHamAmpCumulative} reduces to 
\begin{align}
\label{TogHamAmpCommuting}\Htoga(t)&=
\Ba(t)\sum_{l=1}^{n} G^{(l)}(t) \frac{\Omega_l}{2}
Q_{l-1}^\dagger 
\sigphiL\bla
Q_{l-1}.
\end{align}
Now, from Eq. \ref{Exp3}, we know $Q^\dagger_{l-1}\sig_iQ_{l-1} = \mathbf{\Lambda}^{(l-1)}_i\sigvec$, where $\mathbf{\Lambda}^{(l-1)}_i$ denotes the $i$th row of $\mathbf{\Lambda}^{(l-1)}$. We may therefore write  
\begin{align}
\frac{\Omega_l}{2}Q_{l-1}^\dagger, 
\sigphiL\bla
Q_{l-1} &=\frac{\Omega_l}{2}\Big[\cos(\phi_l)\mathbf{\Lambda}^{(l-1)}_x
+
\sin(\phi_l)\mathbf{\Lambda}^{(l-1)}_y\Big]=\ProjVec\mathbf{\Lambda}^{(l-1)}\sigvec
\end{align}
where $\ProjVec \equiv\frac{\Omega_l}{2}(\cos(\phi_l),\hspace{0.1cm}\sin(\phi_l),\hspace{0.1cm}0)$ is the \emph{Projection Vector} defined in Eq. \ref{ProjectionVector}. Consequently, substituting into Eq. \ref{TogHamAmpCommuting}, the amplitude toggling Hamiltonian is re-expressed 
\begin{align}\label{TogHamAmpComputational}
\Htoga(t)=\Ba(t)\Ra(t)\sigvec
\end{align}
where, for compactness, we have defined the time-domain \emph{Amplitude Control Vector}
%Total Amplitude Control Vector
\begin{align}\label{AmplitudeControlVectorComputational}
\Ra(t) := \sum_{l=1}^n G^{(l)}(t)\ProjVec\HistMat. 
\end{align}

%				-------------------------------------------------
%						sub:SECTION: first-order Error Vector
%				-------------------------------------------------
\subsection*{First-Order Error Vector}

To summarize of the previous sections, the error propagator is now written 
\begin{align}
\Uerr(\tau) &= \exp\Big[{-i\Phi(\tau)}\Big]\\
\label{FirstOrderErrorSummary}& \approx\exp\Big[{-i\EV_1(\tau)\sigvec}\Big]\\
&= \exp\Big[{-i\int_0^\tau dt\Htog(t)}\Big]\\
& =\exp\Big[{-i\Big(\int_0^\tau dt \Htogd(t)+\int_0^\tau dt \Htoga(t)\Big)}\Big]\\
&\label{ControlVectorForm} =\exp\Big[{-i\Big(\int_0^\tau dt \Bd(t)\Rd(t)+\int_0^\tau dt \Ba(t)\Ra(t)\Big)\sigvec}\Big].
\end{align}
The first-order error vector consequently takes the form
\begin{align}\label{FirstOrderErrorVector}
\EV_1(\tau) &= \int_0^\tau dt \Bd(t)\Rd(t)+\int_0^\tau dt \Ba(t)\Ra(t)\\
& = \EVd +\EVa
\end{align}
where we define components 
\begin{align}
\label{DephasingErrorVector}\EVd(\tau)&:= \int_0^\tau dt\Bd(t)\Rd(t)\hspace{1.5cm}&&\text{(\emph{Dephasing Error Vector})}\\
\label{AmplitudeErrorVector}\EVa(\tau)&:=\int_0^\tau dt \Ba(t)\Ra(t) \hspace{1.5cm}&&\text{(\emph{Amplitude Error Vector})}.
\end{align}

%				-------------------------------------------------
%						sub:SECTION: first-order Infidelity: Time Domain
%				-------------------------------------------------
\subsection*{First-Order Infidelity}\label{FirstOrderInfidelityAppendix}

%We calculate the first-order infidelity $\langle a_1^2\rangle = \langle\EV_1\EV_1^T\rangle$ by substituting Eq. \ref{FirstOrderErrorVector}, resulting in a sum of 4 double time-integrals over the cross-correlation functions $\langle\beta_i(t_1)\beta_j(t_2)\rangle$ for $i,j\in\{z,\Omega\}$. In this work we assume fluctuations in the amplitude and frequency of the driving field arise from random, independent physical processes, hence $\Bd(t)$ and $\Ba(t)$ are uncorrelated stochastic processes with zero mean. Cross terms in $\Bd(t)$ and $\Ba(t)$ therefore vanish under an ensemble average and only terms proportional to the dephasing and amplitude autocorrelation functions $\langle\Bd(t_1)\Bd(t_2)\rangle$ and $\langle\Ba(t_1)\Ba(t_2)\rangle$ survive. If we further assume the noise is \emph{wide-sense-stationary} a spectral representation is possible by expressing these autocorrelation functions as the inverse Fourier transform of the dephasing and amplitude noise PSDs, $\Sd(\omega)$ and $\Sa(\omega)$. The spectral response of the control sequence is also explicitly introduced via the integral transforms in Eq. \ref{SpectralIntegralTransformForControlVectors} defining the frequency-domain control vectors, $\Rd(\omega)$ and $\Ra(\omega)$. The details of the above procedure are presented in \ref{FirstOrderInfidelityAppendix} with the result that the first-order infidelity takes the form %%%SEE ABOVE SECTION FOR REPETITION. NEAR LINE 400-500.

The time-domain representation of the first-order infidelity $\langle a_1^2\rangle = \langle\EV_1\EV_1^T\rangle$ now follows directly from Eq. \ref{FirstOrderErrorVector}. For filtering time-dependent noise, however, it is more useful to transform to a spectral representation in which case $\langle a_1^2\rangle$ separates into dephasing and amplitude noise terms, each appearing as an overlap integral between the noise PSD and a frequency-domain filter-transfer function. In this section we summarize this derivation, and define the filter-transfer functions. In the following section we present the final forms for these filter-transfer functions. 
\\
\\
\noindent The modulus square of the first-order error vector, using Eq. \ref{FirstOrderErrorVector}, is given by 
\begin{align}
\EV_1\EV_1^T& = (\EVd+\EVa)(\EVd+\EVa)^T\\
& = \EVd\EVd^T+\EVd\EVa^T + \EVa\EVd^T + \EVa\EVa^T.
\end{align}
Substituting Eqs. \ref{DephasingErrorVector} and \ref{AmplitudeErrorVector} and taking the ensemble average over time, the first-order infidelity $\langle a_1^2\rangle = \langle\EV_1\EV_1^T\rangle$ takes the form 
%\begin{align}
%\EV_1\EV_1^T&= 
%\Big\{
%\int_0^\tau dt_1\Ba(t_1)\Ra(t_1)+\int_0^\tau dt_3\Bd(t_3)\Rd(t_3)
%\Big\}\times...\\
%&\Big\{
%\int_0^\tau dt_2 \Ba(t_2)\Ra(t_2)+\int_0^\tau dt_4\Bd(t_4)\Rd(t_4)
%\Big\}^T\\
%&= 
%\Big\{
%\int_0^\tau dt_1
%\int_0^\tau dt_2
% \Ba(t_1) \Ba(t_2)\Ra(t_1)\big[\Ra(t_2)\big]^T
%\Big\}\\
%&+
%\Big\{
%\int_0^\tau dt_1
%\int_0^\tau dt_4
%\Ba(t_1)\Bd(t_4)\Ra(t_1)\big[\Rd(t_4)\big]^T
%\Big\}\\
%&+
%\Big\{
%\int_0^\tau dt_3
%\int_0^\tau dt_2
%\Bd(t_3)\Ba(t_2)\Rd(t_3) \big[\Ra(t_2)\big]^T
%\Big\}\\
%&+
%\Big\{
%\int_0^\tau dt_3
%\int_0^\tau dt_4
%\Bd(t_3)\Bd(t_4)\Rd(t_3)\big[\Rd(t_4)\big]^T
%\Big\}
%\end{align}
\begin{align}
\langle a_1^2\rangle &=
\int_0^\tau dt_1
\int_0^\tau dt_2
 \langle\Ba(t_1) \Ba(t_2)\rangle\Ra(t_1)\big[\Ra(t_2)\big]^T\\
&+
\int_0^\tau dt_1
\int_0^\tau dt_4
 \langle\Ba(t_1)\Bd(t_4)\rangle\Ra(t_1)\big[\Rd(t_4)\big]^T\\
&+
\int_0^\tau dt_3
\int_0^\tau dt_2
 \langle\Bd(t_3)\Ba(t_2)\rangle\Rd(t_3) \big[\Ra(t_2)\big]^T\\
&+
\int_0^\tau dt_3
\int_0^\tau dt_4
 \langle\Bd(t_3)\Bd(t_4)\rangle\Rd(t_3)\big[\Rd(t_4)\big]^T.
\end{align}
Here the time average only operates on the noise fields, not on the control vectors since these are deterministic. Assuming, as in Sec. \ref{Sec:PhysicalSetting}, $\Ba(t)$ and $\Bd(t)$ are uncorrelated, classical random variables with zero mean, the two-point cross-correlation functions  $\langle\Ba(t_i)\Bd(t_j)\rangle = \langle\Bd(t_k)\Ba(t_l)\rangle=0$. Hence the infidelity reduces to 
\begin{align}\label{AmplitudeAndDephasingInfidelityTimeDomain}
\langle a_1^2\rangle &=
\int_0^\tau dt_1
\int_0^\tau dt_2
 \langle\Ba(t_1) \Ba(t_2)\rangle\Ra(t_1)\big[\Ra(t_2)\big]^T\\
&+
\int_0^\tau dt_3
\int_0^\tau dt_4
 \langle\Bd(t_3)\Bd(t_4)\rangle\Rd(t_3)\big[\Rd(t_4)\big]^T.
\end{align}
We further assuming \emph{wide-sense-stationary}, so the remaining two-point correlation functions depend only on the time difference, and therefore reduce to \emph{auto-correlation functions}. In this case we may invoke the Wiener-Khinchin Theorem~\cite{WienerKhinchin}, 
\begin{align}
\big\langle \beta(t_1)\beta(t_2)\big\rangle = \frac{1}{2\pi}\int_{-\infty}^{\infty}S_\beta(\omega) e^{i\omega(t_2-t_1)}d\omega
\end{align} 
which relates the autocorrelation function of a signal $\beta(t)$ to the Fourier transform of its PSD $S_\beta(\omega)$. Denoting the dephasing and amplitude noise PSDs by $\Sd(\omega)$ and $\Sa(\omega)$, we may therefore re-express $\langle a_1^2\rangle$ in terms of the noise spectral properties, yielding
\begin{align}
\langle a_1^2\rangle %&=
%\frac{1}{2\pi}
%\int_{-\infty}^{\infty}d\omega
%\Sa(\omega) 
%\int_0^\tau dt_1
%\int_0^\tau dt_2
%e^{i\omega(t_2-t_1)}
%\Ra(t_1)\big[\Ra(t_2)\big]^T\\
%&+
%\frac{1}{2\pi}
%\int_{-\infty}^{\infty}d\omega'
%\Sd(\omega')
%\int_0^\tau dt_3
%\int_0^\tau dt_4
% e^{i\omega'(t_4-t_3)}
%\Rd(t_3)\big[\Rd(t_4)\big]^T\\
&=
\frac{1}{2\pi}
\int_{-\infty}^{\infty}d\omega
\Sa(\omega) 
\Big[
\int_0^\tau dt_1
e^{-i\omega t_1}
\Ra(t_1)
\int_0^\tau dt_2
e^{i\omega t_2}
\big[\Ra(t_2)\big]^T
\Big]\\
&+
\frac{1}{2\pi}
\int_{-\infty}^{\infty}d\omega'
\Sd(\omega')
\Big[
\int_0^\tau dt_3
e^{-i\omega' t_3}
\Rd(t_3)
\int_0^\tau dt_4
e^{i\omega' t_4}\big[\Rd(t_4)\big]^T
\Big].
\end{align}
Defining the frequency-domain control vectors via the integral transforms
\begin{align}\label{SpectralIntegralTransformForControlVectors}
\Ra(\omega) =-i\omega\int_{0}^{\tau}dte^{i\omega t}\Ra(t),\hspace{1cm}\Rd(\omega) =-i\omega\int_{0}^{\tau}dte^{i\omega t}\Rd(t)
%\Ra(\omega)^* &=i\omega\int_{0}^{\tau}dte^{-i\omega t}\Ra(t)\\
%\Rd(\omega)^* &=i\omega\int_{0}^{\tau}dte^{-i\omega t}\Rd(t)\\
\end{align}
it is then straightforeward to further re-express the infidelity in terms the spectral properties of the \emph{control}, yielding
%\begin{align}
%&\int_0^\tau dt_1
%e^{-i\omega t_1}
%\Ra(t_1) = \frac{1}{i\omega}
%\Big[
%\Ra(\omega)
%\Big]^*\\
%&\int_0^\tau dt_2
%e^{i\omega t_2}
%\big[\Ra(t_2)\big]^T=
%-\frac{1}{i\omega}
%\Big[
%\Ra(\omega)
%\Big]^T\\
%&\int_0^\tau dt_3
%e^{-i\omega' t_3}d\omega'
%\Rd(t_3)=
%\frac{1}{i\omega'}
%\Big[
%\Rd(\omega')
%\Big]^*\\
%&\int_0^\tau dt_4
%e^{i\omega' t_4}\big[\Rd(t_4)\big]^T=
%-\frac{1}{i\omega'}
%\Big[
%\Rd(\omega')
%\Big]^T
%\end{align}
\begin{align}
\langle a_1^2\rangle %&=
%\frac{1}{2\pi}
%\int_{-\infty}^{\infty}d\omega
%\Sa(\omega) 
%\Big\{
%\frac{1}{i\omega}
%\Big[
%\Ra(\omega)
%\Big]^*
%\Big\}
%\Big\{
%-\frac{1}{i\omega}
%\Big[
%\Ra(\omega)
%\Big]^T
%\Big\}\\
%&+
%\frac{1}{2\pi}
%\int_{-\infty}^{\infty}d\omega'
%\Sd(\omega')
%\Big\{
%\frac{1}{i\omega'}
%\Big[
%\Rd(\omega')
%\Big]^*
%\Big\}
%\Big\{
%-\frac{1}{i\omega'}
%\Big[
%\Rd(\omega')
%\Big]^T
%\Big\}\\
&=
\frac{1}{2\pi}
\int_{-\infty}^{\infty}
\frac{d\omega}{\omega^2}
\Sa(\omega) 
\Big[
\Ra(\omega)
\Big]^*
\Big[
\Ra(\omega)
\Big]^T\\
&+
\frac{1}{2\pi}
\int_{-\infty}^{\infty}
\frac{d\omega'}{\omega'^2}
\Sd(\omega')
\Big[
\Rd(\omega')
\Big]^*
\Big[
\Rd(\omega')
\Big]^T.
\end{align}
Defining the frequency-domain filter-transfer functions,
\begin{align}
\Fd(\omega) &:= \Big[
\Rd(\omega)
\Big]^*
\Big[
\Rd(\omega)
\Big]^T\hspace*{1cm}&&\text{(\emph{Dephasing Filter-Transfer Function})}\\
\Fa(\omega) &:=\Big[
\Ra(\omega)
\Big]^*
\Big[
\Ra(\omega)
\Big]^T\hspace*{1cm}&&\text{(\emph{Amplitude Filter-Transfer Function})}.
\end{align}
we therefore recover Eq. \ref{FirstOrderInfidelityComputational}
\begin{align}
\langle a_1^2\rangle &=
\frac{1}{2\pi}
\int_{-\infty}^{\infty}
\frac{d\omega}{\omega^2}
\Sd(\omega)
\Fd(\omega)+
\frac{1}{2\pi}
\int_{-\infty}^{\infty}
\frac{d\omega'}{\omega'^2}
\Sa(\omega') 
\Fa(\omega').
\end{align}

%__________________________________________________________________________________________

%					SECTION: Control Vectors: Computational Forms 
%__________________________________________________________________________________________

\section{\\Control Vectors: Computational Forms}\label{App:CVsComputationalForms}%\label{ComputationalFormsAppendix}

The dephasing and amplitude filter-transfer functions $\Fd(\omega)$ and $\Fa(\omega)$ are obtained by taking the modulus square respectively of the frequency-domain dephasing and amplitude control vectors $\Rd(\omega)$ and $\Ra(\omega)$, defined by Eq. \ref{SpectralIntegralTransformForControlVectors} in terms of a Fourier-type transform. In this section we derive the computationally useful forms of $\Rd(\omega)$ and $\Ra(\omega)$. 

%					-------------------------------------
%					sub-SECTION: Total Control Matrix
%					-------------------------------------
\subsection*{Total Control Matrix $\mathbf{R}(t)$ Computational Form}\label{TotalControlMatrixAppendix}
The time-domain \emph{Total Control Matrix} $\mathbf{R}(t)$ is defined by Eq. \ref{TotalControlMatrix} with elements $R_{ij}(t):=\frac{1}{2}\text{Tr}\Big[\UcD(t)\sig_i\Uc(t)\sig_j\Big]$. Substituting in Eq.  \ref{ControlPropagatorComputational} we then obtain 
%Matrix Elements of Lth Pulse \emph{Control Matrix}
\begin{align}
R_{ij}(t) &= \frac{1}{2}\sum_{l=1}^{n}G^{(l)}(t)\text{Tr}\Big[Q^\dagger_{l-1}\Big\{U^\dagger_c(t,t_{l-1})\sig_i U_c(t,t_{l-1})\Big\}Q_{l-1}\bla\sig_j\Big]\\
&= \frac{1}{2}\sum_{l=1}^{n}G^{(l)}(t)\text{Tr}\Big[Q^\dagger_{l-1}\Big\{\sum_{k=x,y,z}R_{ik}^{P_l}(t-t_{l-1})\sig_k\Big\}Q_{l-1}\bla\sig_j\Big]
\end{align}
where we have used Eq. \ref{Exp2} to re-express $U^\dagger_c(t,t_{l-1})\sig_i U_c(t,t_{l-1})$ in terms of the $R^{P_l}_{ij}$. Using the linearity of the trace operation and recalling the definition of the of the \emph{Control History Matrix} elements $\Lambda^{l-1}_{ij}$ from Eq. \ref{PulseHistoryMatrix}, we then obtain 
\begin{align}
R_{ij}(t)&= \frac{1}{2}\sum_{k=x,y,z}\sum_{l=1}^{n}G^{(l)}(t)\text{Tr}\Big[R_{ik}^{P_l}(t-t_{l-1})Q^\dagger_{l-1}\sig_kQ_{l-1}\bla\sig_j\Big]\\
&= \sum_{k=x,y,z}\sum_{l=1}^{n}G^{(l)}(t)R_{ik}^{P_l}(t-t_{l-1})\Big\{\frac{1}{2}\text{Tr}\Big[Q^\dagger_{l-1}\sig_kQ_{l-1}\bla\sig_j\Big]\Big\}\\
&\equiv \sum_{l=1}^{n}G^{(l)}(t)\sum_{k=x,y,z}\Big\{R_{ik}^{P_l}(t-t_{l-1})\Big\}\Big\{\Lambda^{(l-1)}_{kj}\Big\}. 
\end{align}
Hence, by definition of matrix multiplication, the time domain total control matrix takes the form
\begin{align}\label{TimeDomainTotalControlMatrixComputational}
\mathbf{R}(t) &= \sum_{l=1}^{n}G^{(l)}(t)\mathbf{R}^{P_l}(t-t_{l-1})\mathbf{\Lambda}^{(l-1)}.
\end{align}
We move to the frequency domain by performing the integral transform on $\mathbf{R}(t)$ defined by 
\begin{align}
\label{FormOfIntegralTransform}\mathbf{R}(\omega) &:=-i\omega\int_{0}^{\tau}dte^{i\omega t}\mathbf{R}(t)\\
&= -i\omega\sum_{l=1}^{n}\Big\{\int_{0}^{\tau}dte^{i\omega t}G^{(l)}(t)\mathbf{R}^{P_l}(t-t_{l-1})\Big\}\mathbf{\Lambda}^{(l-1)}\\
&= -i\omega\sum_{l=1}^{n}\Big\{\int_{t_{l-1}}^{t_l}dte^{i\omega t}\mathbf{R}^{P_l}(t-t_{l-1})\Big\}\mathbf{\Lambda}^{(l-1)}\\
&= \sum_{l=1}^{n}e^{i\omega t_{l-1}}\Big\{-i\omega\int_0^{\tau_l}dt'e^{i\omega t'}\mathbf{R}^{P_l}(t')\Big\}\mathbf{\Lambda}^{(l-1)}.
\end{align}
In the last line we have performed a change of variables using $t' = t-t_{l-1}$, $\forall l\in\{1,...n\}$. 
The frequency-domain \emph{Total Control Matrix} thus takes the computational form 
\begin{align}\label{FrequencyDomainTotalControlMatrixComputational}
\mathbf{R}(\omega)&=\sum_{l=1}^{n}e^{i\omega t_{l-1}}\mathbf{R}^{P_l}\bla(\omega) \mathbf{\Lambda}^{(l-1)}
\end{align}
where we have defined the frequency-domain \emph{Local Control Matrix} by
\begin{align}\label{LocalControlMatrixFrequencyDomain}
\mathbf{R}^{P_l}\bla(\omega) :=-i\omega\int_{0}^{\tau_l}dt'e^{i\omega t'} \mathbf{R}^{P_l}\bla(t').
\end{align}
The matrix elements of $\mathbf{R}^{P_l}\bla(\omega)$ are derived as functions of the control sequence in the section bellow. 
%					-------------------------------------
%					sub SECTION: Local Control Matrix Elements
%					-------------------------------------
\subsection*{Local Control Matrix Elements}\label{LocalControlMatrixElements}

As defined in  Eq. \ref{LocalControlMatrix}, the matrix elements of the time-domain \emph{Local Control Matrix} are given by $R_{ij}^{P_l}(t-t_{l-1}) = \frac{1}{2}\text{Tr}\Big[U_c^\dagger(t,t_{l-1})\sig_i U_c(t,t_{l-1})\sig_j\Big]$, where $\Uc(t,t_{l-1}) \equiv\exp[-i\frac{\Omega_l}{2}\sigphiL(t-t_{l-1})]$. The frequency domain representation $\mathbf{R}^{P_l}\bla(\omega)$ then follows from the integral transform defined by Eq. \ref{LocalControlMatrixFrequencyDomain}, with matrix elements expressed as functions of the control parameters $\{\Omega_l,\tau_l,\phi_l\}$. These matrix elements take the form

\begin{align}
%EXPRESSION FOR R_xx(\omega)
\label{RxPlx(omega)}
&\RPlxx=\cos^2(\phi_l)
\Big[
1-e^{i\omega \tau_l}
\Big]
+
\frac{\omega\sin^2(\phi_l)}{\omega^2-\Omega_l^2}
\VofL\\
%EXPRESSION FOR R_xy(\omega)
\label{RxPly(omega)}
&\RPlxy=\frac{1}{2}\sin(2\phi_l)
\Big[
1-e^{i\omega \tau_l}
\Big]
-
\frac{\omega\sin(2\phi_l)}{2(\omega^2-\Omega_l^2)}
\VofL\\
%EXPRESSION FOR R_xz(\omega)
\label{RxPlz(omega)}
&\RPlxz=-\frac{\omega}{\omega^2-\Omega_l^2}\sin(\phi_l)
\BofL\\
%EXPRESSION FOR R_yx(\omega)
\label{RyPlx(omega)}
&\RPlyx =\frac{1}{2}\sin(2\phi_l)
\Big[
1-e^{i\omega \tau_l}
\Big]
-
\frac{\omega\sin(2\phi_l)}{2(\omega^2-\Omega_l^2)}
\VofL\\
%EXPRESSION FOR R_yy(\omega)
\label{RyPly(omega)}
&\RPlyy=
\sin^2(\phi_l)
\Big[
1-e^{i\omega \tau_l}
\Big]
+
\frac{\omega\cos^2(\phi_l)}{\omega^2-\Omega_l^2}
\VofL\\
%EXPRESSION FOR R_yz(\omega)
\label{RyPlz(omega)}
&\RPlyz= \frac{\omega}{\omega^2-\Omega_l^2}\cos(\phi_l)
\BofL\\
%EXPRESSION FOR R_zx(\omega)
\label{RzPlx(omega)}
&\RPlzx=\frac{\omega}{\omega^2-\Omega_l^2}\sin(\phi_l)
\BofL\\
%EXPRESSION FOR R_zy(\omega)
\label{RzPly(omega)}
&\RPlzy =-\frac{\omega}{\omega^2-\Omega_l^2}\cos(\phi_l)
\BofL\\
%EXPRESSION FOR R_zz(\omega)
\label{RzPlz(omega)}
&\RPlzz=\frac{\omega}{\omega^2-\Omega_l^2}
\VofL
\end{align}
where we have defined
\begin{align}
\VofL &:= 
\Big[\violet
i\Omega_le^{i\omega\tau_l}
\sin(\Omega_l\tau_l)
-\omega e^{i\omega\tau_l}\cos(\Omega_l\tau_l)
+\omega\bla
\Big],\\
\BofL&:=
\Big[\brown
i\Omega_le^{i\omega\tau_l}\cos(\Omega_l\tau_l)
+\omega e^{i\omega\tau_l}\sin(\Omega_l\tau_l) 
-i\Omega_l
\bla\Big].
\end{align}

%					-------------------------------------
%					sub-SECTION: Amplitude Control Vector
%					-------------------------------------
\subsection*{Amplitude Control Vector}
The computational form for the frequency-domain \emph{Amplitude Control Vector} $\Ra(\omega)$ follows from substituting Eq. \ref{AmplitudeControlVectorComputational} into Eq. \ref{SpectralIntegralTransformForControlVectors}, yielding
\begin{align}
\Ra(\omega) &:=-i\omega\int_{0}^{\tau}dte^{i\omega t}\Ra(t)\\
&=-i\omega\Big\{\int_{0}^{\tau}dte^{i\omega t}
\sum_{l=1}^n G^{(l)}(t)\Big\}\ProjVec\HistMat\\
&=-i\omega
\sum_{l=1}^n
\Big\{
\int_{t_{l-1}}^{t_l}dte^{i\omega t}
\Big\}\ProjVec\HistMat\\
&=-i\omega
\sum_{l=1}^n
\frac{1}{i\omega}
\Big[
e^{i\omega t_l}
-e^{i\omega t_{l-1}}
\Big]\ProjVec\HistMat\\
&=
\sum_{l=1}^n
\Big[
e^{i\omega t_{l-1}} - e^{i\omega t_l}
\Big]\ProjVec\HistMat\label{Eq:AmplitudeControlVector}
\end{align}

\section{\\Walsh Rotary Spin Echo Derivations}\label{AppSec:WRSEDerivations}

The WRSE$_k$ sequence is defined by the phase modulation  $\phi(t) = \phi_0+\frac{\pi}{2}(1-y(t))$, where  $y(t)\in\{\pm1\}$.  Referring to Eq. \ref{SpinOperatorDefinition}, however, the spin operator $\sig_{\phi(t)}$ satisfies 
\begin{align}\label{Eq:PiPhaseShift}
\sig_{\phi_0+\pi} = -\sig_{\phi_0}\hspace{0.5cm}\iff\hspace{0.5cm}\sig_{\phi_0+\frac{\pi}{2}(1-y(t))} = y(t)\sig_{\phi_0}.
\end{align}
Consequently the sign-inversion may be absorbed into into a modulated Rabi-rate defined by $\Omega_y(t) := y(t)\Omega_0$. The sequence is then conveniently recast as \emph{amplitude} modulation with constant phase $\phi_0$. Defining $y(t):= \PAL_k(t/\tau)$ we therefore obtain the Walsh synthesis $\Omega_k(t) = \RabiWPMF\PAL_k(t/\tau)$ consisting of a single Walsh function. 
Referring to Eq. \ref{Eq:WAMFGeneralForm} the associated amplitude modulation is thus given by 
\begin{align}
\Rabivec = H_\MinDim\boldsymbol{\WAMampH} = \RabiWPMF\bP^{(k)}_\MinDim, \hspace{1cm}\MinDim=2^{m(k)}
\end{align}
where $\WAMampH_{i(k)} = \RabiWPMF$ is the only nonzero element of $\boldsymbol{\WAMampH}$ and, as in Sec. \ref{SubSec:PaleyOrdering},  $\bP^{(k)}_\MinDim:=(\Pk_1,\hspace{0.1cm}\Pk_2,...,\Pk_{\MinDim})^T$ defines the $i(k)$th column of $H_\MinDim$. Eq. \ref{Eq:WRSEPulseForm} is then more conveniently re-expressed
\begin{align}\label{Eq:WRSE_AM_PulseForm}
\text{WRSE}_k \equiv \prod_{l=1}^{\MinDim(k)}\exp\Big(i\frac{P_l^{(k)}\RabiWPMF}{2}\tauMIN\sig_{\phi_0}\Big)
\end{align}
In the next section we use this form to analyze the amplitude noise filtering properties. The subsequent section treats the filtering properties in the dephasing quadrature.

%				-------------------------------------------------
%				subsub:SECTION: WRSE as Amp Filters
%				-------------------------------------------------

\subsection*{WRSE$_k$ as Amplitude Noise Filters}\label{AppSubSec:WRSEAmpFF}

Referring to Eq. \ref{Eq:WRSE_AM_PulseForm}, the rotation operator $\sig_{\phi_0}$ for WRSE$_k$ is treated as fixed across all pulses. Thus it commutes with $P_l$ and, consequently, with $Q_l = P_lP_{l-1}...P_0$ $\forall l\in\{1,...,\MinDim(k)\}$. It folows $Q^\dagger_{l-1}\sig_{\phi_0}Q_{l-1}= \sig_{\phi_0}$ which, post-multiplying by $\sig_k$ and taking the trace of both sides, yields the identity 
\begin{align}
\frac{1}{2}\text{Tr}\Big[\sig_{\phi_0}\sig_k\Big]=\frac{1}{2}\text{Tr}\Big[Q^\dagger_{l-1}\sig_{\phi_0}Q_{l-1}\sig_k\Big],\hspace{1cm}k\in\{x,y,z\}.
\end{align}
The LHS expands to $\delta_{xk}\cos(\phi_0)+\delta_{yk}\sin(\phi_0)$ (where $\delta_{lk}$ is the Kronecker delta), and the RHS expands to $\cos(\phi_0)\Lambda^{(l-1)}_{xk}+\sin(\phi_0)\Lambda^{(l-1)}_{yk}$ (using the definition of the \emph{Pulse History Matrix} in Eq. \ref{PulseHistoryMatrix}). We thus obtain the following three identities $\forall \phi_0$: 
\begin{align}
\label{Eq:xIdentity}k=x:\hspace{0.75cm}&\cos(\phi_0)\Lambda^{(l-1)}_{xx}+\sin(\phi_0)\Lambda^{(l-1)}_{yx} =\cos(\phi_0)\\
\label{Eq:yIdentity}k=y:\hspace{0.75cm}&\cos(\phi_0)\Lambda^{(l-1)}_{xy}+\sin(\phi_0)\Lambda^{(l-1)}_{yy} =\sin(\phi_0)\\
\label{Eq:zIdentity}k=z:\hspace{0.75cm}&\cos(\phi_0)\Lambda^{(l-1)}_{xz}+\sin(\phi_0)\Lambda^{(l-1)}_{yz} =0.
\end{align}
Now, setting $\Omega_l \equiv \Pk_l\RabiWPMF$ the \emph{Projection Vector} defined in Eq. \ref{ProjectionVector}  becomes $\ProjVec = (\RabiWPMF\Pk_l/2)\big[\cos(\phi_0),\hspace{0.1cm} \sin(\phi_0),\hspace{0.1cm}0\big]$, in which case 
%Local Amplitude Control Vector
\begin{align}
\ProjVec\HistMat&=\frac{\RabiWPMF}{2}\Pk_l\Big\{
\cos(\phi_0)\mathbf{\Lambda}_x^{(l-1)}+
\sin(\phi_0)\mathbf{\Lambda}_y^{(l-1)}\Big\}\\
&=\frac{\RabiWPMF}{2}\Pk_l\left[ \begin{array}{c}  \cos(\phi_0)\Lambda^{(l-1)}_{xx}+\sin(\phi_0)\Lambda^{(l-1)}_{yx}\\
\cos(\phi_0)\Lambda^{(l-1)}_{xy}+\sin(\phi_0)\Lambda^{(l-1)}_{yy}\\
\cos(\phi_0)\Lambda^{(l-1)}_{xz}+\sin(\phi_0)\Lambda^{(l-1)}_{yz}\end{array} \right]^T\\
&=\frac{\RabiWPMF}{2}\Pk_l\left[ \begin{array}{ccc}  \cos(\phi_0), & \sin(\phi_0), & 0\end{array} \right]
\end{align}
where in the last equality we have used the identities derived above in Eqs. \ref{Eq:xIdentity}, \ref{Eq:yIdentity} and \ref{Eq:zIdentity}. Using Eq. \ref{AmplitudeControlVector:ComputationalForm} we therefore obtain
\begin{align}
\Ra(\omega) =\frac{\RabiWPMF}{2}
\sum_{l=1}^\MinDim\Pk_l
\Big[
e^{i\omega t_{l-1}} - e^{i\omega t_l}
\Big]
&\left[\begin{array}{c} \cos(\phi_0)\\ \sin(\phi_0)\\0\end{array} \right]^T
\end{align}
where $t_l = l\tau/\MinDim(k)$. From Eq. \ref{AmplitudeFF} the amplitude filter-transfer function therefore becomes 
\begin{align}\label{FFampWRSE Preliminary}
\Fa(\omega) 
&=\frac{\Omega^2}{4}\Big|
\sum_{l=1}^\MinDim\Pk_l
\Big[
e^{i\omega t_{l-1}} - e^{i\omega t_l}
\Big]
\Big|^2,
\end{align}
where, on taking the modulus square, the $\phi_0$ dependence amounts to $\cos(\phi_0)^2+\sin(\phi_0)^2 = 1$ and hence vanishes.

%				-------------------------------------------------
%				subsub:SECTION: WRSE as Dephasing Filters
%				-------------------------------------------------

\subsection*{WRSE$_k$ as Dephasing Noise Filters}\label{AppSubSec:WRSEforDephasing}

For a general WRSE$_k$ sequence, one can show the first-order Taylor coefficient for $\Fd(\omega)$ takes the analytic form
\begin{align}
C^{(z)}_2(\RabiWPMF;k) = \sinc^2\Big(\frac{\RabiWPMF}{2^{\kappa(k)}}\Big),\hspace{1cm}\kappa(k) =\cases{{m(k)}&if\hspace{0.15cm} $r(k) \ne 1$\\{m(k)+1}&if\hspace{0.15cm} $r(k)=1$} 
\end{align}
yielding the family of zeros $Z^{(k)}_2 = \{2^{\kappa(k)}\pi q \hspace{0.1cm}|\hspace{0.1cm}q\in\mathbb{N}\}$. Hence it is always possible to produce a first-order filter with $(p-1) = 1$ by setting $\RabiWPMF\in Z^{(k)}_2$. Higher-order filters for dephasing noise -- that is, such that $(p-1)>1$ -- then correspond to some $\eta$ satisfying Eq. \ref{Eq:WRSEDephFitlerCondition} such that $\eta\in Z^{(k)}_2$. Although a general analytical form for these higher-order coefficients is not easy to express\footnote{The higher-order $C^{(k)}_{2j}$ involve terms oscillating at multiple frequencies and have nontrivial dependencies on $\RabiWPMF$. Their zeros must in general be determined numerically.} we may still make progress, however, by simply substituting in the candidate values $\RabiWPMF = 2^{\kappa(k)}\pi q$ and determining which $q\in\mathbb{N}$ produce concurrent zeros of the $C^{(z)}_{2j}(\RabiWPMF)$. As a representative example we study the particular case for WRSE$_3$, deriving the coefficients
\begin{align}
C^{(z)}_4(\RabiWPMF;3) = &\frac{1}{\RabiWPMF^4}
\Big[
(\RabiWPMF^2-16)\cos\frac{\RabiWPMF}{2}
-2\RabiWPMF^2\cos\frac{\RabiWPMF}{4}-8\RabiWPMF\sin\frac{\RabiWPMF}{4}
+(\RabiWPMF^2+16)
\Big],\\
C^{(z)}_6(\RabiWPMF;3) = &\frac{1}{48\RabiWPMF^6}\Big[
(\RabiWPMF^4-96\RabiWPMF^2+1152)\cos\frac{\RabiWPMF}{2}-(5\RabiWPMF^4-192\RabiWPMF^2)\cos\frac{\RabiWPMF}{4}\nonumber\\
&\qquad-(28\RabiWPMF^3-768\RabiWPMF)\sin\frac{\RabiWPMF}{4}+4(\RabiWPMF^4-36\RabiWPMF^2-288)
\Big].
\end{align}
From above, the choice $k = 3$ implies $\kappa(k) = 2$ and consequently the candidate zeros take the form $\RabiWPMF = 4\pi q$. Substituting into the above expressions yields
\begin{align}
C^{(z)}_4(4\pi q;3)& =  \frac{1-(-1)^{q}}{8q^2\pi^2}\hspace{2.975cm}q\in\mathbb{N}\\
C^{(z)}_6(4\pi q;3)& = \cases{\frac{1}{(8\pi q)^4}&if
$q\hspace{0.2cm}\text{even}$\\
\frac{27-10\pi^2(1+2q)^2}{7\pi^4(1+2q)^4}&if $q\hspace{0.2cm}\text{odd}$}
\end{align}
%\begin{align}
%C^{(3)}_4(4\pi q)& =  \frac{1-(-1)^{q}}{8q^2\pi^2},\hspace{1.cm}
%C^{(3)}_6(4\pi q)& = \cases{\frac{1}{(8\pi q)^4}&if
%$m\hspace{0.2cm}\text{even}$\\
%\frac{27-10\pi^2(1+2q)^2}{7\pi^4(1+2q)^4}&if $q\hspace{0.2cm}\text{odd}$}
%\end{align}
Thus $q$ must be even to ensure $C^{(z)}_4 = 0$. However this choice implies $C^{(z)}_6>0$  (in fact $C^{(z)}_6>0$ for any choice of $q$) and it follows WRSE$_3$ is at maximum a second-order filter. In Fig. \ref{WRSE_C246_PAL(3)}a we plot $C^{(z)}_{2j}(\Omega_0;3)$, $j\in\{1,2,3\}$ showing the existence of concurrent zeros only for $j\in\{1,2\}$. The inset shows, as a representative case, the behaviour near $\Omega_0 = 8\pi$. In Fig. \ref{WRSE_C246_PAL(3)}b we plot $\Fd(\omega)$ setting $\Omega_0 = 2\pi q$, $q\in\{1,...,8\}$, showing values for which first and second-order filtering is achieved.  
%%%%%NO GO THEOREM. MAKE CLEARER. 
Repeating this general procedure for other values of $k$ we find similar results and conclude the WRSE$_k$ family are capable of up to second-order filtering against dephasing noise. 

%Fig: Taylor Expansion Coefficients 2,4,6 for RSE PAL(3) 
%\begin{figure}[!ht]
%\centering
%\includegraphics[width=1\columnwidth]{Fig_WRSE_Dephasing3}\\
%\caption{WRSE$_3$ dephasing noise filter characteristics. a) Taylor expansion coefficients $C^{(z)}_{2j}(\Omega_0;3)$, $j\in\{1,2,3\}$ of $\Fd(\omega)$. The inset shows typical behaviour: $\Omega_0=8\pi$ is a concurrent zero for $j \in\{1,2\}$, but not for $j = 3$. Hence WRSE$_3$ can only filter up to second-order. b) Dephasing filter-transfer functions for WRSE$_3$ corresponding to $\Omega_0 = 2\pi q$, $q\in\{1,..,8\}$. For example, $p-1 = 2$ when we choose the concurrent zero $\Omega_0 = 8\pi$.} \label{WRSE_C246_PAL(3)}
%\end{figure}

%__________________________________________________________________________________________

%							SECTION:  SK1-WAMF Method 2
%__________________________________________________________________________________________

%\section{\\UWMF Concatenation}\label{AppSec:UWMFs}

%Here we present the details of two alternative methods of concatenating WPMF$^{(c)}_1$ within WAMF$_{0,3}^{(1)}$ considered in Sec. \ref{Sec:UWMFs}. 

%				-------------------------------------------------
%				subsub:SECTION: SK1-WAMF Method 1
%				-------------------------------------------------

\end{document}